\documentclass[12pt,preprint]{aastex}
\usepackage{ulem}
\newcommand{\caii}{\ion{Ca}{2}}
\newcommand{\caby}{$Ca\ by$}
\newcommand{\cas}{$Ca$-$s$}
\newcommand{\caw}{$Ca$-$w$}
\newcommand{\str}{Str\"omgren}
\newcommand{\kms}{km s$^{-1}$}
\newcommand{\dra}{$\Delta RA$}
\newcommand{\ddec}{$\Delta Dec$}
\newcommand{\vmax}{$V^\mathrm{max}_\mathrm{rot}$}

\newcommand{\vsigstar}{$\left(V_\mathrm{rot}^\mathrm{max}/\sigma_\mathrm{m}\right)^*$}

\newcommand{\nsgb}{$n$(bSGB):$n$(fSGB)}

\newcommand{\nrgb}{$n$(\caw):$n$(\cas)}

\newcommand{\nhb}{$n$(BHB):$n$(EBHB)}
\newcommand{\ebv}{$E(B-V)$}
\newcommand{\ehk}{$E(hk)$}
\newcommand{\eby}{$E(b-y)$}

\begin{document}

\title{MULTIPLE STELLAR POPULATIONS OF GLOBULAR CLUSTERS FROM 
HOMOGENOUS $Ca\ by$ PHOTOMETRY. 
I. M22 (NGC 6656).\altaffilmark{1, 2}}

\author{Jae-Woo Lee\altaffilmark{3, 4}}

\altaffiltext{1}{Based on observations made 
with the Cerro Tololo Inter-American Observatory (CTIO) 1m and 0.9m telescopes, 
which are operated by the SMARTS consortium.}
\altaffiltext{2}{Partially based on observations made with the NASA/ESA 
{\it Hubble Space Telescope}, obtained at the Space Telescope 
Science Institute, which is operated by AURA, Inc.,
under NASA contract NAS 5-26555.}
\altaffiltext{3}{Department of Physics and Astronomy,
Sejong University, 98 Gunja-Dong, Gwangjin-Gu, Seoul, 143-747, Korea;
jaewoolee@sejong.edu, jaewoolee@sejong.ac.kr}
\altaffiltext{4}{Visiting Astronomer, CTIO, 
National Optical Astronomy Observatories (NOAO), operated by the Association
of Universities for Research in Astronomy, Inc., under cooperative agreement
with the National Science Foundation.}

\begin{abstract}
We investigate the multiple stellar populations in one of peculiar globular clusters (GCs)
M22 using new ground-based wide-field \caby\  and HST/WFC3 photometry
with equivalent passbands, confirming our previous result that M22 has a distinctive
red-giant branch (RGB) split mainly due to difference in metal abundances.
We also make use of radial velocity measurements of the large number of 
cluster membership stars by other.
Our main results are followings.
(i) The RGB and the subgiant branch (SGB) number ratios show that the calcium-weak (\caw)
group is the dominant population of the cluster.
However, an irreconcilable difference can be seen in the rather simple 
two horizontal branch (HB) classification by other.
(ii) Each group has its own CN-CH anti-correlation.
However, the alleged CN-CH positive correlation is likely illusive.
(iii) The location of the RGB bump of the calcium-strong (\cas) group is 
significantly fainter, which may pose a challenge to the helium enhancement
scenario in the \cas\ group.
(iv) The positions of the center are similar.
(v) The \caw\ group is slightly more centrally concentrated, while the the \cas\
is more elongated at larger radii. 
(vi) The mean radial velocities for both groups are similar, 
but the \cas\ group has a larger velocity dispersion.
(vii) The \cas\ group rotates faster.
The plausible scenario for the formation of M22 is that it
formed via a merger of two GCs in a dwarf galaxy environment
and accreted later to our Galaxy.

\end{abstract}


\section{INTRODUCTION}
The last decade has witnessed a drastic paradigm shift on the true nature of 
Galactic GCs. 
In particular, the conventional wisdom of the simple stellar population
of GCs is no longer valid in almost all GCs in our Galaxy. 
In this regard, variations in lighter elements in GC RGB
stars have been known for more than three decades \citep{cohen78} and
the Na-O anti-correlation, for example,
seen in almost all normal GCs \citep{carretta09,dercole08}
is now generally explained by two star formation episodes 
\citep[see also,][]{lee10},
in which the proto-stellar clouds of the second generation of stars
must have been polluted by the gaseous ejecta from
the first generation of the intermediate-mass asymptotic 
giant branch (AGB) stars \citep{dantona07} or
the fast rotating massive stars \citep{decressin07}.

Contrary to normal GCs showing the lighter elemental abundance variations,
peculiar GCs such as  $\omega$ Cen \citep{ywlocen, bedin04} and 
M22 \citep{jwlnat, marino09, marino11}
show spreads in the heavy elemental abundances which synthesized
and supplied by the core-collapsed supernovae. 
To retain ejecta from energetic supernova explosions requires
much more massive systems than typical Galactic GCs \citep{baumgardt08}.
Furthermore, some of massive GCs in our Galaxy appear to have
different kinematics than the normal GCs \citep{ywlee07}.
As a consequence, these peculiar GCs are thought to have an extragalactic origin, 
such as being the remnant of dwarf galaxies \citep[for example, see][]{bekki03}, 
and they are building blocks of our Galaxy.

One of the early hints on the peculiarity of M22 can be found in
\citet{hesser77} and \citet{hesser79}. They studied M22 using
the DDO photometry and the low resolution spectroscopy, finding
heterogeneous light elemental abundances in M22 RGB stars.
\citeauthor{hesser79} also noted that
M22 may share the similar elemental abundance anomalies of $\omega$ Cen.
Later \citet{norris83} found a variation in calcium abundance 
as well as those of the CN band and the G band strength in M22 RGB stars.
They suggested that calcium abundances among M22 RGB stars
can vary up to $\Delta$[Ca/H] $\approx$ 0.3 dex and
it is positively correlated with the carbon abundance.

Regarding variations of heavy elemental abundances including calcium
in M22, more intriguing results have been emerged in recent years.
In our previous work, \citet{jwlnat} studied M22 using 
the extended \str\ photometry
and we found the distinctive RGB split in the $hk$ index
[= $(Ca - b) - (b -y)$]. 
The $hk$ index of RGB stars is a measure of the \ion{Ca}{2} H \& K 
absorption strength \citep{att91}
and our previous result indicated discrete calcium abundances
among M22 RGB stars \citep[see also][]{lim15}.
We also showed that two discrete RGB sequences have indeed different
iron and calcium abundances by using elemental abundance measurement
from the high resolution spectroscopy by \citet{brown92}.
\citet{dacosta09} also obtained the same result from observations 
of the infrared \ion{Ca}{2} triplet that M22 RGB stars
show a substantial intrinsic metallicity spread. 
Finally, \citet{marino09,marino11} performed high resolution
spectroscopic studies of RGB stars in M22.
In addition to the spread in metallicity and the calcium abundance
among RGB stars in M22,
\citet{marino11} found that the metal-rich group of stars have
enhanced CNO, which may account for the the split in the SGB,
and $s$-process-element abundances.
Furthermore, each stellar group in M22 appears to exhibit 
its own Na-O anti-correlation.
Spectroscopic studies of the M22 HB stars also showed that 
the the true nature of HB populations may be as complex as that
of RGB populations \citep[e.g.][]{marino14,gratton14}.
All the available observational evidences up to date
may suggest that M22 must have retained not only the material from SNe 
but also from low-mass and intermediate-mass AGBs\footnote{See also
\citet{decressin07} for the fast rotating massive stars (FRMSs) as a source
of the Na-O anti-correlation of GCs.},
reflecting that the formation history of M22 must have been very complex
compared to those of the bulk of normal GCs in our Galaxy.

This is the first of the series of papers addressing the multiple
stellar populations of Galactic GCs based on the homogenous \caby\ photometry.
In this study we present a new wide-field \caby\ photometry of M22.
As we have extensively demonstrated in our previous work \citep{jwlnat},
the narrow-band photometric study of GCs can provide an efficient means 
to investigate multiple stellar populations in the Galactic GC systems.
Here we present more evidences that make M22 one of peculiar GCs
in our Galaxy.

\section{SEJONG $Ca\ by$ SURVEY}
Sejong $Ca\ by$ survey program has been launched in July, 2006 after two years
of preparation. As an official major partner of the SMARTS consortium,
we acquired the guaranteed access to the small telescopes operated at the CTIO
since 2006. For 8 years, we used about 270 nights in total, about
120 of which were photometric, for this survey program.

The main goals of our survey program are three fold.
Using the extended \str\ system 
(\str\ $by$ filters plus the $Ca$ filter),
we aimed at obtaining followings;
i) Homogeneous photometry of individual GCs without 
prerequisite information of the cluster. 
We tried to maintain a flux limited blind survey for Galactic GCs.
The main purpose of our GC survey program is to study multiple stellar populations
of Galactic GC systems to shed more light on the formation of the GC systems
and furthermore on the formation of our Galaxy.
ii) Metallicity distribution of selected fields in the Galactic bulge, 
in particular we are very interested in the metal-poor regime
of the Galactic bulge.
iii) Metallicity distributions and multiple stellar populations of nearby dwarf galaxies.

Our survey program is consisted of three phases.
The Phase I was from July, 2006 to November, 2010 \citep[see][]{jwlnat,jwln1851}.
During this period, we used \str\ $uvby$ and $Ca$ filters provided by the CTIO.
The CN band at $\lambda$ 3885 \AA\ is often very strong in the RGB stars
and the CTIO $Ca$ filter was originally designed to avoid the CN band contamination.
However, it was suspected that the CTIO $Ca$ filter had undergone degradation
due to aging effect and the original transmission function had been altered
to the shorter wavelength.
As a result, the CTIO $Ca$ filter is suspected to be suffered from 
the CN band contamination.

\begin{figure}
\epsscale{1}
\figurenum{1}
\plotone{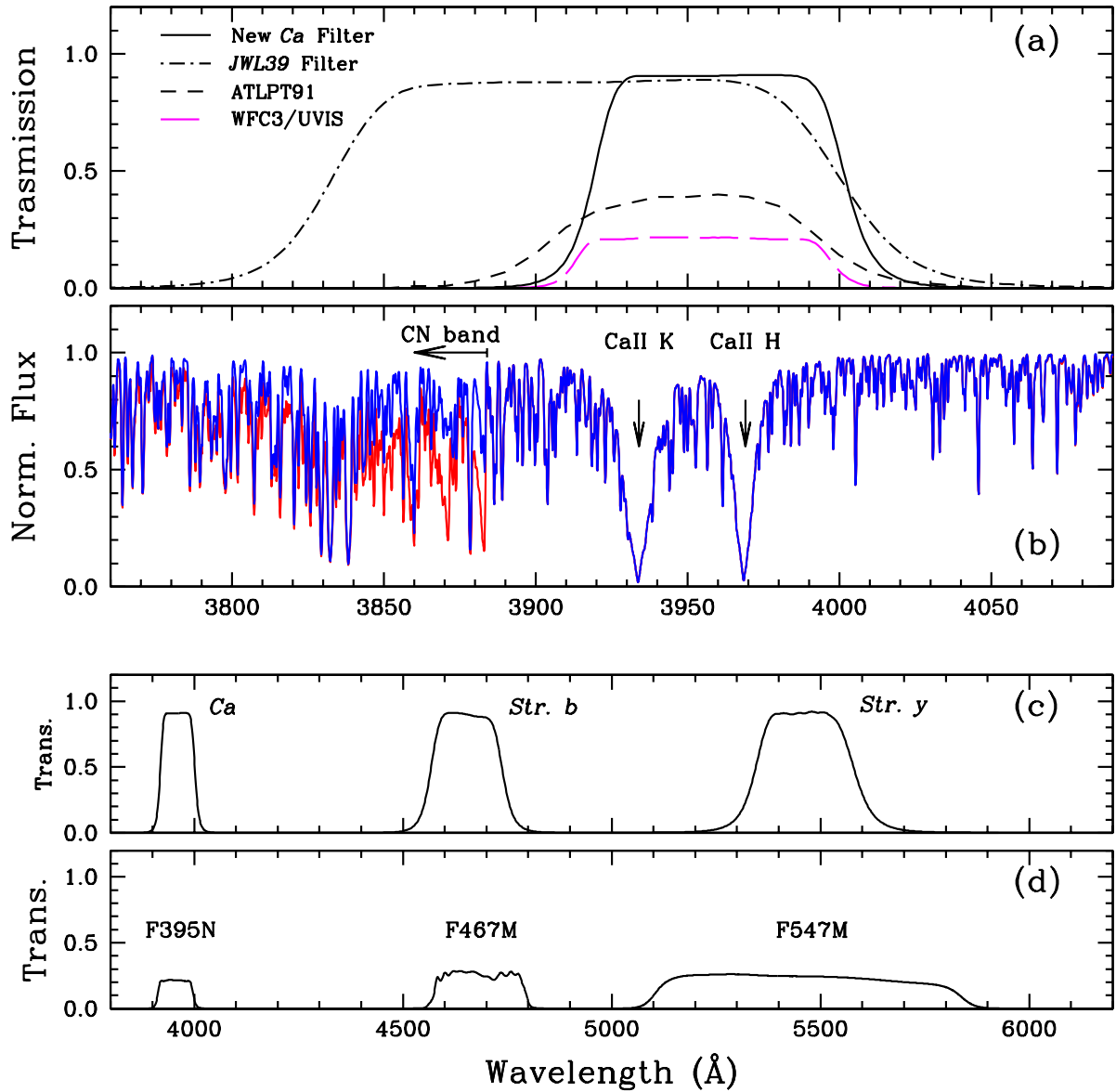}
\caption{(a) A comparison of $Ca$ filter transmission functions between
that in \citet{att91} (the dotted line) and that for our new $Ca$ filter
measured with collimated beam (the solid line).
Both filters have similar FWHMs, approximately 90 \AA\, but
our new $Ca$ filter has a more uniform and high transmission
across the passband, dropping more rapidly at both edges.
The transmission function for the new filter system, $JWL39$, 
is also shown in the plot.
(b) Synthetic spectra for the CN normal (the blue line) and the CN strong
(the red line) RGB stars. The CN band at $\lambda$ 3885 \AA\ lies outside
of the lower boundary of our new $Ca$ filter.
On the other hand, we can measure the CN band strength
by using our new filter system, $JWL39$. 
(c) Filter transmission functions for the ground-based extended \str\ photometry.
(d) Filter transmission functions for the HST WFC3/UIVS photometry.
Note that the bandwidth of the HST F547M is about twice as broad as
that of the \str\ $y$ filter.
}\label{fig:flt}
\end{figure}

Since March 2011 (Phase II), we used our own \str\ $by$ and new $Ca$ filters, 
all of which were manufactured by Asahi Spectra, Japan.
Our new $Ca$ filter was carefully designed to have very similar filter band width
and pivot wavelength as those of F395N filter in the Wide-Field Camera 3 (WFC3)
onboard the Hubble Space Telescope (HST).
In Figure~\ref{fig:flt}, we show a comparison of filter transmission functions 
between that in \citet{att91} and  that of our new $Ca$ filter 
measured with collimated beam by the manufacturer of the filter.
Both filter have similar full-width at the half maximums (FWHMs),
approximately 90 \AA\, but our new $Ca$ filter has a more uniform and 
high transmission across the passband, dropping more rapidly at both edges.
As shown in the figure, the CN band at $\lambda$ 3885 \AA\ lies
on the lower tail of the $Ca$ filter by \citet{att91} while our new $Ca$ filter
is designed to be completely free from the CN band contamination.

In Figure~\ref{fig:flt}, we also show the filter transmission functions for the HST WFC3
photometry system. The HST WFC3 F547M filter is supposed to be equivalent
to the ground-based \str\ $y$ filter. However, the bandwidth of the F547M
is twice as broad as the conventional \str\ $y$ filter.

The Phase III started in April 2013, we added a new filter system, $JWL39$, 
which allows us to measure the CN band at $\lambda$ 3885 \AA\
in combination with our new $Ca$ filter.
Note that the bandwidth of $JWL39$ is similar to that of DDO38 \citep{ddo},
but the pivot wavelength of the filter is at $\approx$ 3900 \AA.
We show the transmission function of our $JWL39$ filter in Figure~\ref{fig:flt}.

In our current work, we will present \caby\ photometry of M22 taken during
the Phases II and III only. 
We will present the results using the new $JWL39$  filter system in forthcoming papers.

There are at least two elements of strength in our survey program.
First, we used the same instrumental setups,
the CTIO 1m telescope and the Y4KCam,
for most of our observations except for two runs in 2012,
due to the maintenance of the Y4KCam.
Second, almost all the data have been obtained by the author of the paper
and all the data have been reduced and analyzed by the author of the paper,
which guarantees homogeneity of the final results.

\section{OBSERVATIONS AND DATA REDUCTION}
\subsection{Observations}
New $Ca\ by$ observations for M22 were made in 44 nights, 
23 of which  were photometric, in 10 runs from 
March 2011 to May 2014 using the CTIO 1.0m and 0.9m telescopes.
The CTIO 1.0m telescope was equipped with a STA 4k $\times$ 4k CCD camera,
providing a plate scale of 0.289 arcsec pixel$^{-1}$ and 
a field of view of about 20 $\times$ 20 arcmin.
The CTIO 0.9m telescope was equipped with the Tektronix 2048 No. 3 CCD,
providing a plate scale of 0.40 arcsec pixel$^{-1}$
and a field of view of 13.5 $\times$ 13.5 arcmin.
For most of our observations, we used the CTIO 1.0m telescope
except for one semester.
The Y4KCam mounted on the CTIO 1.0m telescope had been under inspection 
at Steward Observatory, Arizona, USA during the spring semester of 2012, 
and we used the CTIO 0.9m telescope for two runs in April and July 2012.

The total exposure times for M22 were 90,900 s, 20,930 s
and 9,145 s for our new $Ca$, \str\ $b$ and $y$, respectively.
We show the journal of observations in Table~\ref{tab:obslog}.

\subsection{Data Reduction}
The raw data were processed through the standard
IRAF\footnotemark[2] \footnotetext[2]{IRAF (Image Reduction 
and Analysis Facility) is distributed by the NOAO, 
which are operated by the Association of
Universities for Research in Astronomy, Inc., under contract with the
National Science Foundation.} packages using
twilight or dawn sky flat images to calibrate science exposures.
Since both CCD cameras are equipped with an iris type shutter,
the illumination across the CCD chips is not uniform and 
the shutter shading correction is applied for the photometric standard frames with
the exposure times shorter than 10 s \citep[see][for detailed discussion]{lee14}.
Note that the difference in the integration time between the center
and the edge of the CCD is $\approx$ 70 msec for the Y4KCam.
The most of photometric standards by \citet{tat95} and \citet{att98}
are too bright even for the small telescope like the CTIO 1.0m telescope.
For most cases, we intended to have exposure times longer than 10 s. 
Although defocusing observing technique has been frequently used
to observe isolated bright stars in the field, such as photometric standards,
the most of the standard stars are still too bright to be observed
with a integration time of longer than $\approx$ 10 s.
Without any viable options, science frames with a short integration
time are often necessary in order to secure wide range in colors
of photometric standards, especially in $y$ passband.

The field of view of the final combined science image of M22 is 
about 55 $\times$ 55 arcmin, which is more than 7 times larger than
that of our previous study of the cluster \citep{jwlnat}.
The photometry of the cluster and photometric standard frames were analyzed
using DAOPHOTII, DAOGROW, ALLSTAR and ALLFRAME, and  
COLLECT-CCDAVE-NEWTRIAL packages  \citep{pbs87,pbs93,pbs94,pbs95,turner95}
following the method described in \citet{lc99} and \citet{n6723}.
In order to derive the transformation relations
from the instrumental system to the standard system, we adopt the
following equations:
\begin{eqnarray}
y_{\mathrm inst} &=& y_{\mathrm STD} + \alpha_y (b-y)_{\mathrm STD} + \beta_y X + \gamma_y, \\
b_{\mathrm inst} &=& b_{\mathrm STD} + \alpha_b (b-y)_{\mathrm STD} + \beta_b X + \gamma_b, \nonumber \\
Ca_{\mathrm inst} &=& Ca_{\mathrm STD} + \alpha_{Ca} (b-y)_{\mathrm STD} + \beta_{Ca} X +  \delta hk_{\mathrm STD} + \gamma_{Ca}, \nonumber
\end{eqnarray}
where $X$ is the airmass and $y_{\mathrm STD}$, $b_{\mathrm STD}$, 
$Ca_{\mathrm STD}$, ($b-y$)$_{\mathrm STD}$ and $hk_{\mathrm STD}$ 
are the magnitudes and color on the standard system.
The total number of stars measured in our M22 field
from our ALLFRAME run  was more than 518,000.
Most of stars measured in our field turned out to be Galactic bulge stars.

\begin{figure}
\epsscale{1}
\figurenum{2}
\plotone{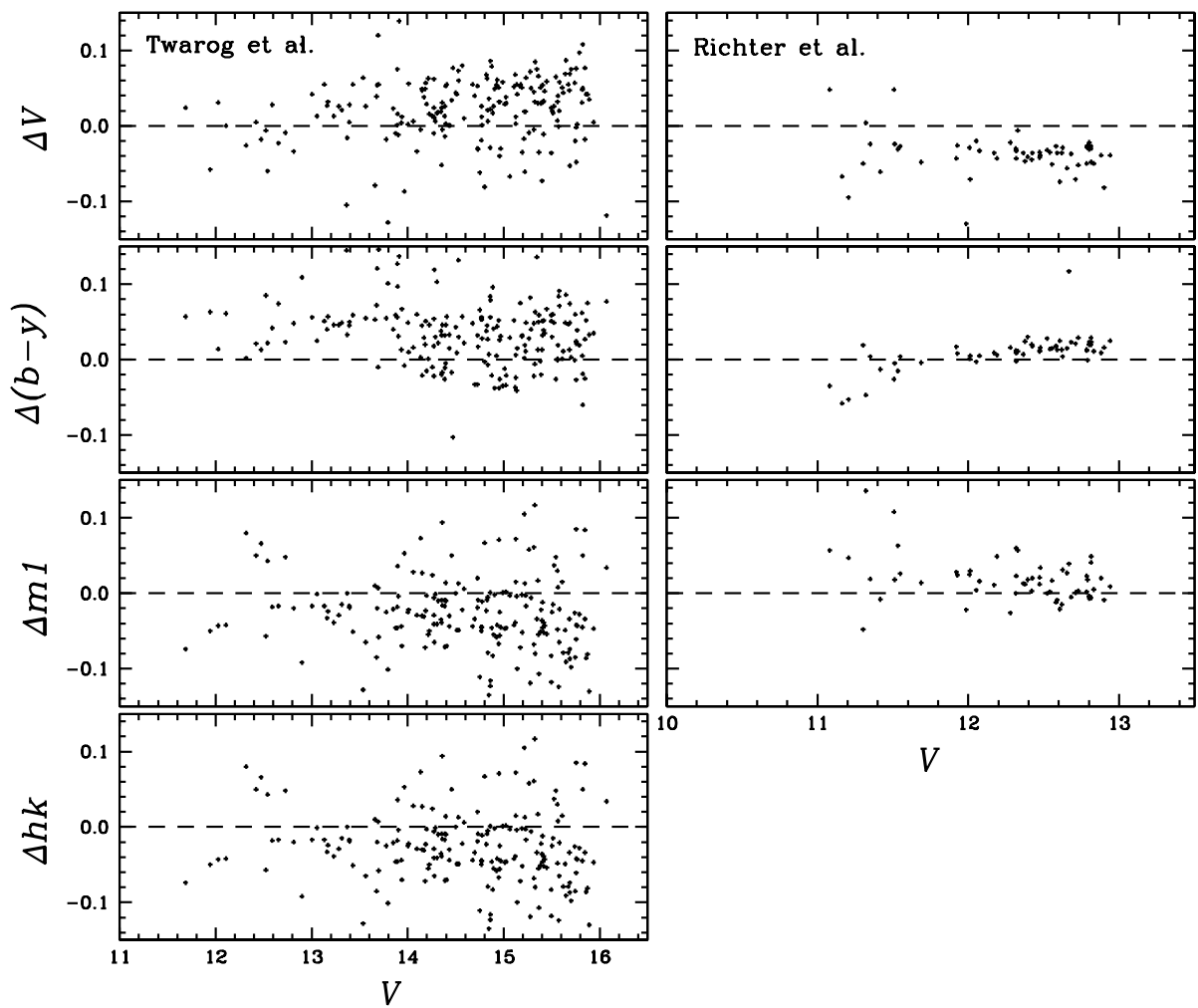}
\caption{Residuals of the CCD photometry of \citet{att95} and
\citet{richter99} as a function of the magnitude.
The differences are in the sense of Anthony-Twarog et al. or Richter et al.
minus our work.
}\label{fig:cmdcomp}
\end{figure}

In Figure~\ref{fig:cmdcomp} and Table~\ref{tab:comp}, 
we show comparisons of our photometry with those of previous studies 
by \citet{att95} and \citet{richter99}.
In the figure, the differences are given in the sense of other works minus
our work.
It can be seen that our results are in good agreement with
previous studies within measurement errors.
It should be emphasized that total exposure times of our observations
in each passband are much greater than those of \citet{att95} and
\citet{richter99}, and, therefore, the photometric quality of our work is 
believed to be superior to those of previous works.

Finally, astrometric solutions for individual stars in our field
have been derived using the data extracted from the Naval 
Observatory Merged Astrometric Dataset \citep[NOMAD,][]{nomad}.
We achieved the rms scatter in the residuals of less than 0.07 arcsec using 
the IRAF IMCOORS package.
Then the astrometric solution was applied to calculate the equatorial coordinates
for all stars measured in our science frames.

\subsection{Completeness tests}
To examine the completeness of our photometry as a function of $V$ magnitude,
we performed a series of artificial star experiments 
\citep[see for example,][]{sh88}.
We selected two square regions with a size of about 15 $\times$ 15 arcmin
from the central part and the outer part of the cluster.
We constructed a fortran program to distribute 1,000 stars in both fields
by adopting the observed radial stellar number density and 
the luminosity profiles in our study.
Note that the number of artificial stars used in our experiments
corresponds to about  1.3\% and 5.6\% of the total measured number of stars 
in the central and the outer part of the cluster, respectively.
This number of artificial stars added to the observed images is 
carefully chosen not to dramatically change the crowdedness characteristics
between our data reduction procedures and our artificial star experiments.

We used the observed point-spread functions (PSFs) and 
the DAOPHOT's ADDSTAR task  
to generate 20 artificial star images in each field.
Following the same data reduction procedure that we described above, 
we derived the completeness fractions as a function of $V$ magnitude.
We show our results in Figure~\ref{fig:complete}.
As shown in the figure our experiments suggest that our photometry
is complete down to $V$ $\approx$ 18.5 mag both in the central
and in the outer field of the cluster for the ground-based observations.
As will be discussed below, it should be emphasized that 
incompleteness at the fainter magnitude regime
does not affect our results presented in this work.

\section{RESULTS}
\subsection{Color-Magnitude Diagrams}\label{s:cmd}
\subsubsection{Ground-based observations: field star contamination}
In Table~\ref{tab:cmd}, we provide our photometric data for stars brighter
than $V$ = 19 mag.
Figure~\ref{fig:cmd} shows the color-magnitude diagrams (CMDs) of 
bright stars in the M22 field. 
Note that, for the sake of clarity, we show stars within 10 arcmin from
the center of the cluster.
As can be seen in the figure, the double RGB sequences of M22
can be clearly seen, especially the distinctive split in the RGB stars
in the $hk$ versus $V$ CMD.
Note that the $m1$ versus $V$ CMD is from our previous 
study of the cluster \citep{jwlnat} taken during the Phase I period.

Since M22 is located towards the Galactic bulge ($l = 9.9^\circ$, $b = -7.6^\circ$),
the contribution from the field star contamination is naturally expected to be very large 
\citep[see][for detailed discussions]{jwlnat}.
We attempted to remove the off-cluster field stars in our bright RGB sample
by using multi-color CMDs following the method described in \citet{jwlnat,jwln1851},
which turned out to be very successful as discussed below.

In Figure~\ref{fig:rgbsel}--(a) and (b), we show CMDs of the M22 field
using all the stars measured from  $V$ = 11.5 mag to 17.5 mag.
In the figures, blue solid lines indicate the RGB boundary grids
in the $(b-y)$ versus $V$ and in the $hk$ versus $V$ CMDs.
In order to prepare these RGB boundary grids, 
we made use of the fiducial sequence of the model isochrones 
by \citet{joo13}, which were kindly provided by S.-J. Joo,
and we adopted the distance modulus and the foreground reddening
value for M22 by \citet{harris96}. 
Then we added small amount of color offset values, 
$\Delta (b-y) \approx \pm$ 0.04 mag and $\Delta hk \approx \pm$ 0.07 mag,
in the transformed model isochrone sequences to make the final RGB boundary grids.
Although our definitions of the RGB boundary grids for $(b-y)$
and $hk$ indexes appear to be arbitrary, our approach
appears to work with great satisfaction as shown below.

The total number of RGB stars detected in the enclosed region 
in Figure~\ref{fig:rgbsel}--(c), is 3571. 
When plotted in the $hk$ versus $V$ plane,
it can be clearly seen that about 42\% of RGB stars selected 
in the $(b-y)$ versus $V$ plane does not belong to M22, 
i.e.\ either the disk or the bulge populations with large $hk$ index values.
A comparison with the radial velocity measurements by \citet{lane09} 
shows that 12 stars out of 307 stars in common turn out to be 
radial velocity non-member stars.
If we take this as a face value, the fraction of the inclusion of
the off-cluster field population in our M22 RGB selection procedure is
3.9 $\pm$ 1.4\%, which appears to be too low to dramatically change
the results presented in this paper.
Without 12 radial velocity non-member stars, we have,
2,054 RGB stars, from $V$ = 12.0 mag to 16.5 mag, in total.

The validity of our RGB selection procedure can be confirmed by
the surface-brightness profile (SBP) of the RGB stars in M22.
In Figure~\ref{fig:surfmag}-(d), we show the SBP of  the red-clump stars 
in the Galactic bulge 
[the stars lie inside a cyan box in Figure~\ref{fig:rgbsel}--(a)].
As expected, the SBP of the bulge population is almost flat against 
the radial distance from the center of M22.
Note that the mild fluctuation seen in the small $r$ region,
$\log (r$/arcsec) $\lesssim$ 2, is due to the small number of 
the bulge red-clump stars ($n \leq 15$) detected in this region. 

Figure~\ref{fig:surfmag}-(a)--(c) show
the SBPs of the \caw\ group, the \cas\ group and all RGB stars in M22
(see below for details on the definition of the \caw\ and the \cas\ groups).
With slightly different zero-point offset values for each group of stars,
our SBPs for M22 RGB stars are in excellent agreement with 
the Chebyshev polynomial fit of M22 by \citet{trager95}
up to more than 10$^3$ arcsec from the center of the cluster.
Our investigation of the SBPs of M22 RGB stars strongly suggests that
the contamination of the off-cluster field star in our result
is not severe, consistent with our previous result from the comparison of
the radial velocity measurements of M22 RGB stars.

\subsubsection{Ground-based observations: \caw\ versus \cas}
As shown in Figures~\ref{fig:cmd} and \ref{fig:rgbsel}, M22 exhibits 
the distinctive double RGB sequences in the $hk$ versus $V$ CMD,
which is mainly due to the difference in the heavy elemental abundances
between the two groups \citep[see also][]{jwlnat,marino09,marino11}.
As in our previous works \citep{jwlnat,jwln1851},
we define the calcium-weak (\caw) group of stars 
with smaller $hk$ index and the calcium-strong (\cas) group of stars
with larger $hk$ index at a given $V$ magnitude.

In the left panel of Figure~\ref{fig:rgbpop}, we show 2,054 RGB stars 
from  Figure~\ref{fig:rgbsel}--(d) on the $\Delta hk$ versus $V$ plane, 
where $\Delta hk$ is defined to be the difference in the $hk$ index of
the individual RGB stars with respect to the $hk$ index of 
the fiducial sequence of the \caw\ RGB group.
As shown, the differences in the $hk$ index between the mean values of 
the \caw\ and the \cas\ depend slightly on $V$ magnitude,
which is naturally expected if the difference in the absorption 
strengths of \caii\ H and K lines is mainly responsible for 
the separation in the $hk$ index of the RGB stars.
Due to this magnitude (more precisely, surface gravity)
dependency on the $\Delta hk$,
we divide M22 RGB stars into four different magnitude bins
and we show histograms of the $\Delta hk$ distributions of RGB stars of 
each magnitude bin in the right panels of Figure~\ref{fig:rgbpop},
all of which appear to show double peaks.
In order to separate the two different groups of RGB stars 
based on the $\Delta hk$ distribution, 
we adopt the expectation maximization (EM) algorithm 
for the two-component Gaussian mixture distribution model.
In an iterative manner, we derive the probability of individual RGB stars
for being the \caw\ and the \cas\ groups.
In the left panel of  Figure~\ref{fig:rgbpop}, stars with $P(w|x_i)$ $\geq$ 0.5 
from the EM estimator are denoted with blue dots which corresponds to 
the \caw\ group, 
while $P(s|x_i)$ $>$ 0.5 with red dots which corresponds to the \cas\ group.
In the right panel of the figure, 
we show the distributions of the \caw\ and the \cas\ groups with blue 
and red solid lines, respectively, and we show the number ratio
\nrgb\ for each bin.
In total the number ratio between the \caw\ and the \cas\ groups is
\nrgb\ = 68.8:31.2 ($\pm$3.3).

Our observed number ratio between the two groups does not necessarily 
reflect the initial stellar number ratio  or the initial mass ratio 
between the two groups, 
since both groups have different initial chemical compositions, 
different masses at a given luminosity, and furthermore slightly different ages 
\citep[e.g.,][]{jwlnat,marino09,marino11,joo13}.
In order to take care of the heterogeneous evolutionary effects between
the two groups, we constructed an evolutionary population synthesis model 
using the theoretical model isochrones by \citet{joo13}.
We populated 10$^7$ artificial stars in each group using 
the Salpeter's initial mass function (IMF)
and we generated 50 different sets for each group.
We, then, compared the number ratios between the two groups
in the magnitude range of 12.0 $\leq$ $V$ $\leq$ 16.5 mag, 
finding \nrgb\ = 48.3:51.7 ($\pm$ 0.3), in the sense that
RGB stars in the \cas\ group are slightly more numerous than 
those  in the \caw\ group at a fixed total number of stars in both groups.
Note that using other types of IMF does not affect 
our results presented here, which was also pointed out  by \citet{joo13}.
As an exercise, we calculated the number ratio using the universal IMF by \citet{kroupa01}
and we obtained \nrgb\ = 48.4:51.6 ($\pm$ 0.3), in excellent agreement
with that from the Salpeter's IMF within the 0.1\% level.

Figure~\ref{fig:LF} shows the mean theoretical luminosity functions (LFs)
for the \caw\ and the \cas\ groups. Note that the abscissa of the figure
is $M_V$ and the grey shaded area is corresponding to 
M22 RGB stars with 12.0 $\leq$ $V$ $\leq$ 16.5 mag.
As can be seen, the theoretical LFs for the RGB stars in both groups are very similar.
If we consider the evolutionary effect, the {\em intrinsic} or the {\em true} number ratio between
the two groups becomes \nrgb\ = 70.2:29.8 ($\pm$3.3).
Note that our new RGB number ratio is in good agreement with our previous
rough estimate of the number ratio between the two groups,
$\approx$ 0.70:0.30 \citep{jwlnat}.

As discussed by \citet{jwlnat} and \citet{marino09,marino11}, 
the double RGB sequences in M22 is due to
the difference in chemical abundances between the \caw\ and the \cas\ groups,
in particular the difference in the calcium abundance
measured with the $Ca$ filter and in heavy elements,
which was first hinted by the high-resolution spectroscopy of seven RGB stars
by \citet{brown92} and confirmed later by \citet{marino09,marino11} 
with the expanded sample of RGB stars of the cluster.
More detailed discussion on the difference in the chemical abundances between
the two groups will be given below.
Also worth emphasizing is the fact that the differential reddening effect and 
the contamination from the off-cluster populations cannot explain 
the double RGB sequences in M22 \citep{jwlnat}.
Assuming that the conventional interstellar reddening law 
for the extended \str\ photometry system by \cite{att91},
\ehk\ = $-$0.16$\times$\eby\ $\approx$ $-$0.12$\times$\ebv,
the mean $hk$ difference between the two RGB groups, $\Delta hk$ = 0.09 mag,
can be translated into $\Delta (b-y) = -0.56$ mag, $\Delta (B-V) = -0.75$ mag
and $\Delta V = -2.33$ mag, in the sense that the metal-rich \cas\ RGB stars
becomes bluer and brighter due to the foreground reddening.
Indeed, this is not the case. 

\subsubsection{Ground-based observations: $m1$ versus $V$}
The \str\ $v$ passband includes the CN band at $\lambda$ 4215 \AA\ and, 
as a consequence, $m1$ index [= $(v-b) - (b-y)$] is capable of distinguish 
the difference in the CN absorption strengths. 
In fact, the bimodal distribution in the $m1$ index of M22 RGB stars 
was known for decades and it has been frequently attributed 
to the bimodal distribution in the CN abundances 
\citep{norris83,att95,richter99,jwlnat}.
It is widely accepted that the variation of lighter elemental abundances, 
such as C, N and O, is thought to be resulted from chemical pollution 
by the intermediate-mass  AGB stars 
\citep[see, for example,][]{ventura01}.
The potential trouble with the $m1$ index is that the split between 
the double populations at the lower RGB sequence is not readily distinguishable 
in the $m1$ versus $V$ CMD, most likely due to the strong surface 
gravity sensitivity on the CN band strength, in the sense that 
the CN band strength is inversely correlated with  the surface gravity of stars 
\citep[see for example,][]{tripicco91,cannon98},
as shown in Figure~\ref{fig:cmd}.

On the other hand, as \citet{jwlnat} discussed extensively, 
the $hk$ index is a measure of calcium abundances of individual stars 
and, therefore, the difference in the $hk$ index at a given luminosity
indicates the heterogeneous calcium and other heavy elemental abundances 
between the double RGB populations in M22, which is confirmed
by the high-resolution spectroscopic study of RGB stars 
in the cluster \citep{brown92,marino09,marino11}.
Calcium and heavy elements can only be supplied through supernovae
explosions of massive stars and the origin of the split in the $hk$ index
is intrinsically different from that in the $m1$ index in M22,
if the $m1$ index is the most sensitively depends on the CN abundances
as previous investigators perceived frequently.
It is important to note that, however, since the \str\ $v$ passband includes 
numerous metallic lines, the $m1$ index also sensitively behaves 
in the difference in the heavy metal  abundances.

In the top panels of Figure~\ref{fig:fakecmd}, we show the CMDs of M22 again.
For the $m1$ versus $V$ CMD, we show stars with $(b-y)$ $\geq$ 0.45 mag
to prevent the degeneracy of the location of the HB and the RGB stars 
in the $m1$ versus $V$ CMD.
We also show the Padova model isochrones with [Fe/H] = $-$1.8 dex, comparable
to that of the \caw\ group, and $-$1.5 dex, comparable
to or slightly more metal rich than that of the \cas\ group, 
with a fixed [CNO/Fe] ratio \citep{padova}.
From the figure, it is palpable that the bifurcation in the RGB sequences
on the $m1$ versus $V$ CMD is mainly due to the difference in the metal abundance.
In addition, the higher mean [CNO/Fe] ratio in the \cas\ group\footnote{The 
$s$-process-rich group by \citet{marino11}.} may intensify the difference
in the $m1$ index between the two groups.

A comparison of M22 with two Galactic GCs, M55 and NGC~6752 
(Lee 2015 in preparation), may help to elucidate 
the nature of the double RGB sequences.
Note that the metallicity of M55 is [Fe/H] = $-$1.9 dex, 
equivalent to the \caw\ group in M22, while
that of NGC~6752 is $-$1.5 dex, equivalent to the \cas\ group in M22.
It is well known that the HB morphologies of both clusters are different.
M55 contains only the blue HB (BHB) population, 
while NGC~6752 contains the BHB and the extreme blue HB (EBHB) populations.
The CN distributions of both clusters are also different.
M55 appears to show an unimodal CN distribution \citep{kayser08},
in sharp contrast to the bimodal CN distribution of NGC~6752 \citep{norris81}.

In the bottom panels of Figure~\ref{fig:fakecmd}, we show the composite CMDs 
of M55 and NGC~6752. 
In order to make these composite CMDs, we tweaked the color indexes and 
the visual magnitude of individual stars in NGC~6752 to match with those in M55;
$\Delta(b-y)$ = 0.034 mag, $\Delta m1$ = $-$0.008 mag, $\Delta hk$ = $-$0.001 mag
and $\Delta V$ = 0.654 mag.
In the $m1$ versus $V$ CMD, we do not plot the HB stars in M55 and NGC~6752
to avoid the confusion with the RGB stars.
We emphasize that these composite CMDs can successfully reproduce 
the photometric characteristics of M22.
Especially, the CMDs of M22 and the composite GCs show remarkable resemblance
in the HB morphology and the double RGB sequences both in the $m1$
and the $hk$ indexes, suggesting that the double RGB sequence in the $m1$ index
of M22 is likely due to the difference in the mean heavy metal abundance.
If so, the origin of the RGB split in the $m1$ index is the same
as that of the $hk$ index in M22.

Nonetheless, using the $hk$ index as a metallicity indicator has advantages.
As \citet{att91} noted, the $hk$ index is about three times more sensitive to metallicity 
than the $m1$ index is, for stars more metal-poor than the Sun, and it has half the 
sensitivity of the $m1$ index to interstellar reddening.
The annoying trouble with the $m1$ index is that it loses sensitivity to metallicity
changes in the regime below [Fe/H] $\approx$ $-$2 dex, especially for stars with
higher surface gravity.

\subsubsection{Hubble Space Telescope Observations}
We also carried out the HST WFC3 photometry for the central part of M22.
In Figure~\ref{fig:cmdHST}, we show the preliminary results from 
HST WFC3 observations of the cluster (PI : J.-W.\ Lee and PID : 12193)
using the similar passbands that we used in our ground-based observations.

The left panel of the figure shows the $(m_{\rm F467M} - m_{\rm F547M})$ versus
$m_{\rm F547M}$ CMD, which is equivalent to the ground-based 
$(b-y)$ versus $V$ CMD,  near the the subgiant branch (SGB) and the faint RGB 
regions of M22. Similar to the ground-based $(b-y)$ versus $V$ CMD,
the split in the RGB or SGB can not be readily seen in the figure.
 
On the other hand, the $hk_{\rm STMAG}$
[ = $(m_{\rm F395N} - m_{\rm F467M}) - (m_{\rm F467M} - m_{\rm F547M})$]
versus $m_{\rm F547M}$ CMD shows the distinctive split in the
the SGB and the RGB sequences of M22,
confirming the results from our new ground-based observations discussed above.
Note that the SGB split in M22 has been already known in other HST passbands
by \citet{marino09,marino12} and the presence of the SGB split is 
likely due to variations in the CNO abundance \citep{cassisi08} . 
Especially, \citet{marino12} performed
a low-resolution spectroscopic study for a large number of SGB stars in M22,
finding that the elemental abundance trend of the faint SBG (fSGB) group is 
consistent with that of the $s$-process-rich group in their early study 
\citep[][i.e.\ \cas\ in this study]{marino11}, 
while the elemental abundance trend of the bright SGB (bSGB) is consistent
with that of the $s$-process-poor group (i.e.\ the \caw\ group),
demonstrating that the observed difference in the CNO abundance
can account for the observed SGB split.

\citet{milone09} devised a rather clever but complicated method to estimate 
the number ratio between the fSGB and the bSGB in NGC~1851.
However, there are at least three aspects that make difficult for M22 
to be analyzed using the method described by \citet{milone09}.
First, the spatial stellar number density in a fixed area on the sky 
(i.e.\ the number of stars/$\Box\arcsec$) of M22 is  about 20 times 
smaller than that of NGC~1851.
Therefore, the number of the SGB stars detected in our HST observations for M22
are much smaller than that in NGC~1851 by \cite{milone09}.
The small number of stars makes it difficult to delineate 
a clear SGB fiducial sequence for M22.
Secondly, the foreground interstellar reddening value of M22, \ebv\ = 0.34 , 
is much larger than that of NGC~1851, \ebv\ = 0.02 \citep{harris96}.
Furthermore, there may exist the differential foreground reddening
effect across M22, although the differential reddening is not the main reason
for showing the double RGB or SGB sequences in M22 as we have already discussed.
Lastly, the curvature of the SGB stars and the location of the transition
region between the SGB and the base of the RGB are different between
the HST $hk_{\mathrm STMAG}$ and the $F606W - F814W$ photometric systems.
The essential idea of the method proposed by \citet{milone09} is to estimate
the {\em intrinsic} or the {\em true}  number ratio between the two SGB populations,
derived from the {\em apparent} number ratio by taking care of 
the evolutionary effect as we have done for the RGB stars.
It also should be important to point out that their method is arbitrary
and model dependent, as they have noted.

Before turning to the observed number ratio between the two SGB groups in M22, 
we investigate the methodological aspect by employing the evolutionary population 
synthesis models that we constructed for the RGB stars above.
In the left panels of Figure~\ref{fig:SynSGB}, we show artificial CMDs 
around the SGB region of M22 for different photometric systems.
In order to construct the evolutionary population synthesis models, 
we use the model isochrones by \citet{joo13} with the Salpeter's IMF.
The photometric errors are estimated from the measurement errors 
around the SGB region in our HST observations.
We populate 10$^8$ artificial stars in each group.
In the figure, blue dots denote the bSGB group while
red dots the fSGB in M22. 
For the sake of clarity, we only show a small fraction of stars in the figure.
We then count the numbers of SGB stars for each group 
at a fixed color range (the vertically shaded areas) and 
at a fixed magnitude range (the horizontally shaded areas)
returned from our calculations.

The middle panels of Figure~\ref{fig:SynSGB} show the distributions
of SGB stars in the fixed color ranges for the different photometric systems,
while the right panels show those in the fixed magnitude ranges.
Similar to our RGB number ratio estimates,
we calculate the differences 
in the magnitudes or in the color indexes against those of the fiducial
sequences of the bSGB group.
The first number ratios in each plot are those returned from the EM estimator 
assuming the two-component gaussian mixture model 
while the number ratios inside the parenthesis are the input values 
used in our evolutionary population models.
Note that these number ratios in the parenthesis reflect the evolutionary effect
between the two SGB groups at a given color or magnitude ranges.
In the figure, our adopted ranges in each color index may be rather arbitrary
but we intend to maximize the available number of stars maintaining
the clear separation between the two SGB groups.

First, the SGB number ratio returned from the EM estimator between 
the two groups at the fixed (F606W $-$ F814W) color is \nsgb\ = 42:58, 
while that of the input value from the evolutionary population synthesis models
is 39:61.
On the other hand, the SGB number ratio determined at the fixed F606W magnitude 
appears to be more uncertain. 
We obtained \nsgb\ = 63:37 with the input value of 56:43.

Next, the SGB number ratios determined from the $b-y$ versus $V$ CMD 
appear to be the least reliable in our simulations.
We obtain \nsgb\ = 44:56 for the fixed color and 72:28 for the fixed magnitude
ranges, which number ratios deviate largely from the input values 
of 38:62 and 65:35, respectively.

Lastly, the SGB number ratios determined from the $hk$ versus $V$ CMD 
appear to be the most reliable. 
Due to the combined effect of the mean heavy metal and the CNO abundances
on the location of the SGB sequence on the $hk$ versus $V$ CMD,
the separation in the $hk$ index between the two SGB groups 
is the most conspicuous.
The EM estimator securely retrieve the input values within an uncertainty
of  $\leq \pm$ 0.2.
Since our simulations suggest that the total number of stars available 
in our simulations for the $hk$ versus $V$ CMD is about 3 times 
larger for the case with the fixed magnitude range, 
we estimate the observed SGB number ratio
from our HST observations of the cluster using the method with a fixed
magnitude range, i.e.\ using the distribution of individual stars
against $\Delta hk$, as for the RGB stars.
Also importantly, using the method with a fixed magnitude range has 
an advantage in correcting the evolutionary effect.
The SGB number ratio of \nsgb\ = 58:42 is, in fact, the evolutionary
effect correction factor shown in Figure~\ref{fig:LF} (the yellow shaded area)
and applying the correction factor of the differential evolutionary effect 
based on the $V$ magnitude is more direct.

In Figure~\ref{fig:sgbpop}-(a), we show the $hk_{\rm STMAG}$ versus 
$m_{\rm F547M}$ CMD around the SGB region of M22. 
Also shown is the fiducial sequence for the bSGB group.
Following the similar procedure that we used for the RGB stars,
we obtain the number ratio of \nsgb\ = 69:31 ($\pm$ 6).
If we apply the correction factor for the differential evolutionary effect
between the two group, the intrinsic number density becomes
\nsgb\ = 62:38 ($\pm$ 6). This SGB number ratio is in agreement
with the RGB number ratio, \nrgb\ = 70:30 ($\pm$ 3.3), 
within the measurement errors.

As mentioned above, \citet[][see references therein]{marino09} 
already discovered the SGB split in their HST Advanced Camera for 
Surveys Wide Field Channel (ACS/WFC) 
observations of M22 using F606W and F814W passbands, although
the separation between the fSGB and the bSGB populations 
in the $m_{\rm F606W} - m_{\rm F814W}$ versus $m_{\rm F814W}$ CMD is 
not as clear as that in $hk_{\rm STMAG}$ versus $m_{\rm F547M}$ CMD
as shown here.
Their SGB number ratio is \nsgb\ = 62:38 ($\pm$ 5),
which is in excellent agreement with our true SGB number ratio 
\nsgb\ = 62:38 ($\pm$ 6).

Finally, it should be worth mentioning that the $hk$ separation between 
the two SGB groups in the observed CMD (Figure~\ref{fig:cmdHST})
is more ambiguous than that of the simulated CMD with a proper treatment
of the photometric measurement errors (Figure~\ref{fig:SynSGB}).
This may indicate that not only the differential reddening 
but also the metallicity spreads in the double SGB populations 
are necessary to perform more realistic simulations.
In fact, \citet{marino11} reported rather large metallicity spreads
for the two groups of RGB stars; 0.10 dex for the $s$-process-poor group
and 0.05 dex for the $s$-process-rich group.
As shown in Figure~\ref{fig:fakecmd}, the $(b-y)$ width of M22 is compatible
with that of the composite GC (i.e.\ NGC~6752 + M55), 
where the $(b-y)$ locus of the RGB sequence is less sensitive to 
the variation in the heavy metal abundance, 
while the $hk$ width of M22 is broader than that of the composite GC.
Given the current interstellar reddening law (see below for more detail),
the reasonable explanation for the broader RGB sequence of M22 
in the $hk$ index is likely due to the spread in the heavy metal abundances 
of both stellar groups in M22.

\subsection{Differences in Chemical Abundances between Two RGB Groups}
\citet{jwlnat} showed that the distinctive RGB split 
in the $hk$ index in M22 is originated
due to the heterogeneous heavy elemental abundances, in particular calcium.
In our previous work, we demonstrated that the RGB stars in the \cas\ group 
in M22 are in fact Ca-rich, based on seven RGB stars studied by \citet{brown92}.

Here, using the expanded sample of RGB stars studied by \citet{norris83},
\citet{dacosta09}, and \citet{marino11},
we compared the differences in the elemental abundances between
the \caw\ and \cas\ groups in M22, providing more observational evidences
that the two populations of stars are intrinsically different.

\subsubsection{Comparisons with \citet{norris83}}
\citet{hesser77} and \citet{hesser79} first noticed that M22 have RGB stars with
heterogeneous light elemental abundances using the DDO photometry and
the low resolution spectroscopy, respectively, leading them to suggest that
M22 may share the similar elemental abundance anomalies of $\omega$ Cen.
Later \citet{norris83} performed a spectroscopic survey study
for about 100 RGB stars in the cluster and 
they showed that the calcium abundance is correlated with
the CN band and the G band strengths in RGB stars in M22. 

In Figure~\ref{fig:norris}, we show a plot of the calcium index $A$(Ca)
as a function of $V$ magnitude [see Equation (2) of \citeauthor{norris83}
for the definition of the $A$(Ca) index].
In the Figure,  we also show the least square fit to the \caw\ RGB stars,
which can provide a baseline, 
and we calculate the deviations of individual stars from the fitted line.
As discussed by \citet{norris83}, the residuals of individual stars
from this fitted line against the $V$ magnitude may correct 
the temperature and the surface gravity effects 
on the absorption strengths to the first order.
The figure shows that the $A$(Ca) index of \citet{norris83}
is correlated well with our $hk$ index, which should not be a surprise
because both quantities measure 
the same \caii\ H and K line strength of RGB stars
photometrically and spectroscopically. 

\subsubsection{Comparisons with \citet{dacosta09}}
\citet{dacosta09} studied M22 RGB stars using intermediate resolution spectra
and they obtained infrared \caii\ triplet strengths $\Sigma$(CaT) 
(= $W_{8542} + W_{8662}$) for membership RGB stars, 
finding a substantial intrinsic metallicity spread in M22 RGB stars.
The advantage of using the infrared \caii\ triplet is that
these lines arises from a lower energy level of 1.70 eV, while
the lower energy level of \caii\ H and K lines is 0 eV.
Therefore, the infrared \caii\ triplet lines
do not suffer from interstellar reddening contamination.\footnote{
\cite{jwlnat} showed that the differential reddening effect on the \caii\ H and K lines
across the cluster does not affect on the RGB split in M22. 
We demonstrated that, for example,
the differential reddening effect of $\Delta E(B-V) \approx$ 0.3 mag
results in $\Delta Ca$ $\leq$ 0.01 mag, too small value to fully explain
the RGB split in the $hk$ index in M22.}
Also, since the absorption lines from other elements around 
the infrared \ion{Ca}{2} triplet lines are sparsely populated,
contamination from absorption lines of other elements would be less severe.

In the upper panel of Figure~\ref{fig:dacosta}, we show the plot of 
$\Sigma$(CaT) against $V-V_{\rm HB}$, the $V$ magnitude difference 
from the HB.
In the figure, we also show the least square fit
to the data with a dotted line.
The residuals around this fitted line will correct the temperature
and the surface gravity effects on $\Sigma$(CaT) to the first order.
As can be seen, the RGB stars in the \cas\ group have larger
$\Sigma$(CaT) values at a given $V-V_{\rm HB}$ magnitude
than those in the \caw\ group,
consistent with our results that the \cas\ RGB stars are
more calcium rich than the \caw\ RGB stars in M22.

\subsubsection{Comparisons with \citet{marino09,marino11}}
Finally, we make use of results from high resolution spectroscopy by
\citet{marino11}. In Figure~\ref{fig:marino}, we show plots of the
[Ca/H] versus [Fe/H] and the [Na/Fe] versus [O/Fe] for RGB stars
in the \caw\ and the \cas\ groups.
Similar to the results in \citet{jwlnat,jwln1851},
the plot of [Ca/H] versus [Fe/H] shows that the RGB stars in the \caw\ group 
have lower calcium and iron abundances than those in the \cas\ group,
confirming that our $hk$ index measures the calcium abundances
of RGB stars in M22.
The plot of [Na/Fe] versus [O/Fe] shows that each group based on our
$hk$ index exhibits its own Na-O anti-correlation.
This was also pointed out by \citet{marino11}, who found two separate
Na-O anti-correlations in M22 RGB stars 
with different $s$-process elemental abundances.
Note that the separation in the $hk$ index in our current work is equivalent 
to that in  $s$-process elemental abundances by \citet{marino11}.
These separate Na-O anti-correlations made \citet{marino11} to suggest
that M22 may be composed of the merger of two clusters.
In fact, NGC~1851 shows the similar behavior in the Na-O anti-correlations
between the two distinct RGB populations.
\citet{carretta10} suggested that the simple explanation 
for the observational evidences of NGC~1851 would be a merger scenario,
preferentially in a dwarf galaxy environment.

However, it is worth emphasizing is the fact that the mean values
of the \caw\ RGB stars appear to have a higher oxygen and 
a lower sodium abundances, representative of the primordial elemental
abundance pattern, than the \cas\ RGB stars,
consistent with the result by \citet{marino11}.
Also lighter elements C,  N and heavy elements Ca and Fe abundances 
are different between the two groups \citep{marino11}.
Therefore, the feedback on the overall chemical evolution between
the two groups can not be completely neglected.

\bigskip

From comparisons of our $hk$ photometry with three independent works by others,
we conclude that our $hk$ index measurements for RGB stars in M22 
are very consistent with
calcium abundance measurements by others from \ion{Ca}{2} H and K, 
infrared \ion{Ca}{2} triplet lines and visual \ion{Ca}{1} lines.
Therefore, the $hk$ index can be {\em securely} and {\em efficiently}
used to distinguish the difference in the calcium abundances 
at a given luminosity and furthermore the multiple RGB populations in GCs.

\subsubsection{On the internal $\delta CN$ vs $\delta CH$ positive correlation}
\citet{lim15} presented low resolution spectroscopy for M22 RGB stars and
confirmed the CN-CH positive correlation discovered by \citet{norris83}
\citep[see also][]{att95}.\footnote{It is worth mentioning that 
\citet{kayser08} did not find any CN-CH positive correlation in M22.
Examinations of their data led \citet{pancino10} to attribute
this to low S/N ratios of spectra by \citet{kayser08} and the differential 
reddening effect. 
However, there is other possibility that the measurements
by \citet{kayser08} may be not correct.
For example the CH strength measurements 
by \citet{kayser08} are unlikely very large for their adopted definition
of the CH strength from \citet{harbeck03}, indicating that the results 
presented by \citet{kayser08} are suspected to be unreliable.
\citet{pancino10} examined the $\delta$CN versus the $\delta$CH distribution
using the data by \citet{kayser08}, finding neither a positive correlation 
nor an anti-correlation, contradictory to the results by \citet{norris83}
or by \citet{lim15}.}
Here, we discuss more about the metallicity effect on 
the CN-CH positive correlation between the two RGB groups in M22.

The variations in lighter elements in GC stars have been known
for more than three decades, pioneered by \citet{cohen78},
and it is now generally believed to be rooted in the primordial origin
\citep{cottrell81,kraft94}.
The CN-CH anti-correlation is one of  well-known
characteristics in the Galactic GC system.
One probable explanation for origin of this CN-CH anti-correlation 
can be found in \citet{grundahl02}.
They nicely demonstrated that there exist a CH-NH anti-correlation
and a CN-NH positive correlation among RGB stars in NGC~6752,
suggesting that the nitrogen abundance is inversely correlated with
the carbon abundance.

The CN-CH positive correlation shown in Figure~13 of \citet{lim15}
appears to show two separate CN-CH anti-correlations
among stars in each group at a first glance.
It would have been very surprising if there exists a single 
CN-CH anti-correlation between the \caw\ and the \cas\ groups in M22.
It is thought that a CN-CH positive correlation superposed on
two separate CN-CH anti-correlation in M22
can be expected naturally if M22 is composed of
two groups of stars with heterogeneous metallicities.
It is because the positive correlation 
in the $\delta$CN versus  $\delta$CH relation can occur
when one compares the $\delta$CN and the $\delta$CH of 
different groups of stars with  heterogeneous metallicities 
\citep[see][for the definition of $\delta$CN and $\delta$CH]{norris81}.

In order to investigate the metallicity effect on the positive correlation between
the CN and CH band strengths, we used the homogeneous measurements 
of the CN and CH band strengths for stars in 8 GCs observed 
in the course of Sloan Digital Sky Survey (SDSS) by \citet{smolinski11}.
Since they provided magnitudes of stars in the SDSS photometric system,
we converted their SDSS photometry to the Johnson system
using the relation given by \citet{jester05},
\begin{equation}
 V = g - 0.59(g-r) - 0.01,
\end{equation}
and we calculated the $V-V_{\rm HB}$ using the $V_{\rm HB}$ values
given by \citet{harris96}.

In Figure~\ref{fig:cnch}, we show distributions of the CN(3839) and 
the CH(4300) indexes against $V-V_{\rm HB}$ for 8 GCs.
As shown, the CH(4300) strengths are well correlated with the metallicity
of GCs at a given magnitude (for example, the RGB stars around
the level of the HB magnitude shown with vertical dashed lines), 
while the CN(3839) strengths exhibit rather weak correlation against
metallicity mainly due to the multi-modal distributions of the CN(3839) 
strengths of individual GCs
\citep[for example, see Figure~4 of][]{smolinski11}.
Also shown are M22 and NGC~288 from \citet{lim15}, whose measurements
do not agree with those of \citet{smolinski11} due to 
heterogeneous definition of the band strengths between the two works.
Note that \citet{smolinski11} used the definition of the CN(3839) band strength
by \citet{norris81} for RGB and SGB stars, 
that by \citet{harbeck03} for MS stars
and the definition of the CH(4300) band strength by \citet{leesg99}, 
while \citet{lim15} used the definitions by \citet{harbeck03} for both bands.
As a consequence of heterogeneous definition of the molecule band strengths,
the extent of the CN(3839) band strengths for M22 and NGC~288 by \citet{lim15}
is greater than those by \citet{smolinski11}.
Also, the CH(4300) values for M22 and NGC~288 by \citet{lim15} are 
$\gtrsim$ 1.0 and they are out of range in the figure.
Therefore, a direct comparison between the two works is not feasible.

In the figure, the thick solid lines are the common lower envelopes for 8 GCs.
Due to the difference in the mean metallicity among GCs,
these common lower envelopes may not correct both the temperature and 
the surface gravity effects  simultaneously on the CN(3839) and 
the CH(4300) band strengths for 8 GCs.
This is exactly what expected to happen when one compares two groups of RGB stars 
in M22 with different mean metallicities using a single fitted line, 
since the two groups of RGB stars in M22 have very different metallicities.
In Figure~\ref{fig:cnchgrad}, we show the distributions of the $\delta$CN
and the $\delta$CH against metallicity,
where the $\delta$CN and the $\delta$CH are defined to be the differences
in the CN and CH between individual stars and the common lower envelopes.
As expected from Figure~\ref{fig:cnch}, the $\delta$CN and the $\delta$CH
have substantial gradients against metallicity, 0.169 $\pm$ 0.031 and
0.077 $\pm$ 0.010 mag/dex, respectively.
It is believed that these positive slopes are the source of 
the CN-CH positive correlation when one compares the 
$\delta$CN and the $\delta$CH of stars with heterogeneous metallicities.

In Figure~\ref{fig:lim}, we show the CN(3839) and the CH(4300) band strengths
by \citet{lim15} against $V$ magnitude. Also shown are the lower envelops for
the \caw\ (the blue dotted line) and the \cas\ (the red dotted line) groups.
Stars with $-1.0 \leq V-V_{\rm HB} \leq 0.5$ mag are shown with plus signs
(\caw) and crosses (\cas).
Note that the lower envelope for the \caw\ group can be served as 
the common lower envelope for both groups.

Figure~\ref{fig:lim2}-(a) shows a distribution of the $\delta$CN versus
the $\delta$CH of M22 RGB stars, where the $\delta$CN and
the $\delta$CH are calculated with respect to the common lower envelope
(i.e.\ the blue dotted lines in Figure~\ref{fig:lim}).
This plot is similar to Figure~13 of \cite{lim15}, where the CN-CH
positive correlation can be seen. 
Noted that \citet{lim15} used the mean fitted line, not the lower envelope used 
in this analysis or those of others 
\citep[for example][]{norris81,kayser08,smolinski11}
and detailed distributions of individuals stars are slightly different
from those presented here.

In Figure~\ref{fig:lim2}-(b) and (c), we show distributions of 
the $\delta$CN versus $\delta$CH for the \caw\ and the \cas\ groups
using their own lower envelopes.
Each plot appears to show its own CN-CH anti-correlation,
a well-known characteristics of normal GCs.
This is also consistent with the presence of  the separate C-N anti-correlations 
in each of the two $s$-process groups \citep{marino11}.
Since its own lower envelope for each group can correct
the metallicity effects, the $\delta$CN and the $\delta$CH values
calculated from its own lower envelope reflect the relative
CN and CH band strengths at a given metallicity and surface gravity.
As a consequence, the CN-CH positive correlation with an apparently 
ambiguous origin disappears. 
In this regards, the most simplest explanation would be 
a merger of two GCs with different metallicities.
Each group of stars could have been chemically enriched
independently and exhibits its own CN-CH anti-correlation 
as observed in normal GCs in our Galaxy.

Finally, it is worth pointing out again that the extents in the $\delta$CN and
the $\delta$CH of M22 using the data  by \citet{lim15} are greater than
those can be expected from the slopes in the metallicity gradients 
as shown in Figure~\ref{fig:cnchgrad}.
This is partially due to the different definitions in the CN and the CH 
band strengths. It would be highly desirable to re-analyze spectra in
a homogeneous manner in the future.

\subsubsection{Spread in nitrogen abundances}
\citet{yong08} investigated nitrogen abundances of RGB stars in NGC~6752
using NH molecule lines.
They noted that \str\ $u$ passband contains the NH molecule band
at $\lambda$ 3360 \AA\ and they showed that the $c_1$ index
are correlated with the nitrogen abundances of individual stars.
They devised an arbitrary empirical index $cy\ [= c_1 - (b-y)]$
and derived a well correlated relation between the nitrogen abundance 
and the $cy$ index of RGB stars at about $V_{\mathrm HB}$ of NGC~6752,
[N/Fe] $\propto 16.11\times cy$.

In Figure~\ref{fig:cy}, we show $cy$ versus $V$ CMDs of M22 RGB stars,
showing broad spreads in the $cy$ index in both groups.
Assuming Gaussian distributions, we calculated the standard deviation
around the mean $cy$ values and obtained $\sigma (cy) \approx$ 0.03 mag 
at $V_{\mathrm HB}$  for both groups of stars in M22.
Since the metallicity of M22 is comparable to that of NGC~6752
and if the spread in the NH absorption strengths in individual RGB stars
is solely responsible for the spreads in the $cy$ index,
$\sigma (cy) \approx$ 0.03 mag corresponds to 
$\sigma$[N/Fe] $\approx$ 0.5 dex in each group of stars.

Note that our photometric estimate of the spread in the nitrogen
abundances in M22 RGB stars appear to be slightly larger than
those of \citet{marino11}, who obtained $\sigma$[N/Fe] $\approx$ 0.2 dex
for their $s$-poor and $s$-rich groups
based on the observations of 14 stars.

One of the remarkable features can be seen in Figure~\ref{fig:cy} is that
the RGB bump magnitudes of both populations appear to be different.
In the right panel of Figure~\ref{fig:cy},
we show the cumulative luminosity functions (LFs)
for the \caw\ and the \cas\ populations. 
We show the $V$ magnitude levels of the RGB bump, $V_{\mathrm bump}$, 
where the slope of the luminosity function changes abruptly,
for each population by arrows. 
We obtained $V_{\mathrm bump}$ = 13.91 mag
for the \caw\ population while 14.06 mag for the \cas\ population.

\subsubsection{RGB bump magnitude: $V_{\mathrm bump}$ versus $\Delta hk$}
Here, we explore more details on the RGB bump in M22.
As already shown in the left panel of Figure~\ref{fig:rgbpop},
it can be seen that the $V$ magnitude level of the RGB bump
appears to be positively 
correlated with the $\Delta hk$ values of RGB stars 
at around $V$ $\approx$ 14.0 mag.
Especially, due to its high stellar number density, 
the tilted $V_{\mathrm bump}$ magnitude
level against $\Delta hk$ for the \caw\ group is conspicuous,
while the tilted HB structure in the \caw\ group cannot be seen
in the $cy$ versus $V$ CMD as shown in Figure~\ref{fig:cy}.
It should be worth noting that neither this tilted $V$ magnitude against 
$\Delta hk$ in the \caw\ group nor the difference in the bump magnitude
between the two groups can be explained 
by the differential reddening effect \citep[see also,][]{jwlnat}.

The difference in the mean $\Delta hk$ values between the \caw\ stars 
with $\Delta hk$ $<$ 0 mag and those with $\Delta hk$ $\geq$ 0 mag 
at $V$ = 13.95 mag is $\Delta hk$ = 0.04 mag.
Again, assuming that the conventional interstellar reddening law 
by \cite{att91} is applicable,
\ehk\ = $-$0.16$\times$\eby\ $\approx$ $-$0.12$\times$\ebv,
and that the differential reddening effect across the cluster is 
the main reason for showing a broad RGB sequence in the $hk$ index,
this 0.04 mag difference in the $\Delta hk$ index can be translated into
$A_V$ = $-$1.05 mag, in the sense that the \caw\ RGB stars 
with $\Delta hk$ $\geq$ 0 mag should be about 1 mag brighter than
those with $\Delta hk$ $<$ 0 mag.
This is the opposite of what observed in M22. 
As shown in Figure~\ref{fig:bump}, the $V$ magnitude for the RGB bump
of the \caw\ stars with $\Delta hk$ $\geq$ 0 mag is about 0.06 mag fainter
than that of the \caw\ stars with $\Delta hk$ $<$ 0 mag 
(see also Figure~\ref{fig:cy}).
Also, if this 0.04 mag difference in $\Delta hk$ is originated solely by 
the differential reddening across the cluster, not by the difference
in the chemical compositions, the $(b-y)$ color of the \caw\ group
should have been broadened by $\approx$ 0.25 mag.
However the RGB sequence in M22 is not as broad as 0.25 mag
as shown in Figure~\ref{fig:cmd}.
Therefore, the difference in the RGB bump magnitude is not mainly due to 
differential reddening effect but due to the difference in chemical compositions.

It is well known that at a given age the RGB bump becomes fainter 
with increasing metallicity and with decreasing helium abundance,
due to changes in the envelope radiative opacity
\citep[see for example,][]{cassisi13}. 
The fainter $V_{\rm bump}$ magnitude with increasing $\Delta hk$
can be interpreted that $\Delta hk$ increases with either increasing
metallicity or decreasing helium abundance.
One can quantitatively estimate the effect of metallicity and
the helium abundance on the RGB bump luminosity
\citep{bjork06,valcarce12}.

In their Table~2, \citet{bjork06} presented the absolute $V$ magnitude
of the RGB bump with metallicity and we obtained 
$\Delta M_{V,{\rm bump}}/\Delta$[Fe/H] $\approx$ 0.93 mag/dex for 13 Gyr.
The effect of helium abundances on the RGB bump can be found 
in \citet{valcarce12}. From their Figure~9, we obtained 
$\Delta m_{\rm bol} \approx  2.5\times\Delta Y$ for the isochrones
with $Z = 1.6\times10^{-3}$ and 12.5 Gyr.
If the 0.06 mag spread in the RGB bump magnitude in the \caw\ group
is solely due to metallicity, one expects to see the spread in metallicity
by $\Delta$[Fe/H] $\approx$ 0.06 dex.
Similarly the 0.15 mag difference between the \caw\ and the \cas\ groups
can translate into $\Delta$[Fe/H] $\approx$ 0.16 dex that agrees well with
what \citet{marino11} obtained for the $s$-poor and the $s$-rich groups,
$\Delta$[Fe/H] $\approx$ 0.15 $\pm$ 0.02 dex.
However, if we take into account the enhanced helium abundance for 
the \cas\ group by $\Delta Y \approx$ 0.09 \citep{joo13},
the RGB bump magnitude for the \cas\ group should be 
$\approx$ 0.08 (= 0.23 $-$ 0.15) mag brighter than that for the \caw\ group,
which is not the case for our observation.
The formation of the EBHB stars, as progenies of the \cas\ RGB stars,
may require the enhanced helium abundance but the difference 
in the RGB bump magnitude is difficult to explain in this scenario.
Perhaps, may this suggest that only a fraction of the \cas\ RGB stars
have been evolved into the EBHB?

\subsection{Centers}
With the bright RGB stars selected above, we measured the centers of 
each populations using three different methods; 
the arithmetic mean, the half-sphere and the pie-wedge methods.

Using the coordinate of the center of the cluster measured by \citet{goldsbury}
as an initial value ($\alpha$ = 18:36:23.94 and 
$\delta$ = $-$23:54:17.1; J2000)
we chose RGB stars in each population within 300 arcsec from the center 
of the cluster and we calculated the mean values for each group.
The half-light radius of M22 is about 200 arcsec \citep{harris96} and
the circle with a radius of 300 arcsec can contain more than 70\% of RGB stars
in our sample. It is believed that using this radius is sufficient enough
to derive the coordinate of the center of each group with decent accuracy.

We obtained the offset values with respect to the coordinate of center
of the cluster by \citeauthor{goldsbury},  (\dra, \ddec) = (3\farcs8, 0\farcs4) 
for the \caw\ group and (0\farcs8, 5\farcs6) for the \cas\ group, 
resulted in an angular separation of 6 arcsec 
between the centers of both populations.
Note that the core radius of the cluster is about 80 arcsec \citep{harris96},
and the angular separation between the centers of the two populations
is relatively small.

Similar to the simple mean calculation, we use the coordinate of the cluster by
\citeauthor{goldsbury} as an initial value, we chose RGB stars in each population 
within 300 arcsec from the center of the cluster.
Then we divided the sphere into two halves by assuming 
the radial symmetry in the distribution of RGB stars in M22.
We compared the number of RGB stars between the two halves 
by rotating the position angle by 10 degree at a fixed coordinate of the center
and we obtained the differences in the number of RGB stars between both halves.
We repeated this calculation with varying coordinates of the center
and we derived the coordinates of the centers of each group 
with the minimum difference in the number of RGB stars between the two halves.
We found (7\farcs5, 1\farcs6) for the \caw\ group and
($-$1\farcs4, $-$1\farcs4) for the \cas\ group, slightly different from
those from the simple mean method.
The angular separation between the two groups is about 9 arcsec
and, again, it is relatively small compared the core radius of the cluster.

Finally, we applied the pie-wedge method. 
Using the coordinate by \citet{goldsbury}, we divided the sphere 
of a radius of 300 arcsec into 12 different slices. 
Then we compared the number of stars in the opposing distribution.
We repeated this calculation with varying coordinates of the center
and obtained the center of each population with the minimum differences.
We found (7\farcs7, $-$2\farcs8) for the \caw\ group and
($-$4\farcs9, 6\farcs5) for the \cas\ group.
Again, the angular separation between the two groups is about 16 arcsec
and it is still relatively small compared with the core ($\approx$  80 arcsec)
or the half-light ($\approx$  200 arcsec) radii of the cluster.

We show our results in Table~\ref{tab:cnt} and Figure~\ref{fig:cnt}.
We conclude that the coordinates of the centers of each population
are slightly different. However the differences in the coordinates of the center
from various methods do not appear to be substantially large 
to claim that the centers of the RGB distributions of two populations 
are distinctively different.

\subsection{Spatial Distributions}
\subsubsection{Red giant branch stars}
In Figure~\ref{fig:rgbdist}, we show the number ratios of RGB stars 
in the \cas\ group to those in the \caw\ group
against the radial distance from the center using the coordinate
of the center by \citet{goldsbury}.
It is worth noting that there appears to exist a weak gradient 
of the number ratio between the two populations against the radial distance 
from the center, indicating that the distribution of the \caw\ group is 
slightly more centrally concentrated. 
However, the difference in the mean number ratio are in agreement 
within the measurement errors from the central part of the cluster 
up to $\approx$ 800 arcsec from the center, about four times 
of the half-light radius of M22, similar to that can be seen in NGC~1851.
\citet{milone08} showed that the number ratio between the two distinctive 
populations in NGC~1851 does not appear to vary up to 1.7 arcmin from the center,
which is about 2 times of the half-light radius of the cluster.\footnote{
\citet{carretta11} claimed that the metal-poor RGB population in NGC~1851
is more centrally concentrated than the the metal-poor population 
from the incomplete sample of their spectroscopic study 
with larger radial distance from the center, $\approx$ 12 arcmin.}
The flat number ratio between the multiple populations in M22 and NGC~1851 is 
in sharp contrast with that in $\omega$ Cen.
\citet{bellini09} found that both intermediate-metallicity and the metal-rich 
RGB populations are more centrally concentrated than the metal-poor 
RGB population in $\omega$ Cen \citep[see also][]{sollima07}. 

The relaxation time at the half-mass radius for $\omega$ Cen is about 10 Gyr
\citep{harris96} and it is substantially large that the later generation
of stars in $\omega$ Cen may not have sufficient time to be homogenized completely.
As a consequence, the later generation of stars fails to achieve similar radial
distribution as the early generation of stars in the cluster.
On the other hand, the relaxation times at the half-mass radius
for M22 and NGC~1851 are about 1.7 Gyr and 0.7 Gyr, respectively,
significantly smaller than that of $\omega$ Cen.
If the second generation of stars in M22 and NGC~1851 formed from
gaseous ejecta expelled from the first generation of stars, 
they might have enough time to become relaxed systems.
In this regard, \citet{decressin08} claimed that
stars initially centrally concentrated toward the center
need about the two relaxation times to achieve a completely homogenization
throughout the cluster and any radial difference among multiple stellar
populations would be erased over the dynamical history of old GCs.\footnote{
The work of \citet{vesperini13} is also worth mentioning. They performed 
$N$-body numerical simulations and showed that the time required to have 
a flat number ratio between the first and the second generations 
can be larger than 45 half-mass relaxation time for their $r10$ system, 
where $r10$ refers the ratio between the two half-mass radii
(see their Figure 7).}
However, it is also possible that the merger of two GCs can maintain
a flat RGB number ratio in M22.

We show a comparison of cumulative distributions
of RGB stars in both groups in the lower panel of Figure~\ref{fig:rgbdist}. 
As can be seen, the radial distribution of RGB stars
in the \caw\ group is slightly more centrally concentrated. 
We performed a Kolmogorov-Smirnov (K-S) test and we found the probability
of being drawn from identical populations is 0.1\%, with a K-S discrepancy
of 0.10, indicating that they have different parent populations.
As we have demonstrated earlier, the difference in the radial distribution
is not likely due to the field star contamination.

Although small, the difference in the distributions of RGB stars is difficult 
to understand within the current theoretical framework of the GC formation.
If the second generation of stars formed out of gas 
expelled from the first generation of stars in the central part of the cluster 
\citep[][see for example]{bekki10}, the second generation of stars should
have a more centrally concentrated distribution in the early history
of the cluster. Also importantly, if the complete homogenization 
had been achieved within a couple of relaxation times,  
the two different populations in M22 are expected to have 
indistinguishable radial distributions.

We speculate that the difference in the radial distributions of RGB stars
may suggest that either (or perhaps both) of followings is correct;
(i) The \cas\ population did not form out of the gaseous ejecta
of the first generation of stars. Or,
(ii) the complete homogenization may require much longer time scale
than that proposed by \citet{decressin08}.
In this regard, the merger of two GCs can naturally explain the observed
radial distributions of RGB stars.

We turn our attention to the spatial distributions of RGB stars in both groups.
In the top panel of Figure~\ref{fig:density}, 
we show the projected spatial distributions 
of RGB stars in the \caw\ and the \cas\ groups in M22.
In the Figure, the offset values of the projected right ascension and 
the declination in the units of arcsec were calculated 
using the transformation relations in \citet{vandeven},
\begin{eqnarray}
\Delta RA &=& \frac{648000}{\pi}\cos\delta\sin\Delta\alpha, \\
\Delta Dec &=& \frac{648000}{\pi}(\cos\delta\cos\delta_0 - 
\cos\delta\sin\delta_0\cos\Delta\alpha), \nonumber
\end{eqnarray}
where $\Delta\alpha = \alpha - \alpha_0$ and $\Delta\delta = \delta - \delta_0$,
and $\alpha_0$ and $\delta_0$ are the coordinate of the center.
For our calculations, we adopted the coordinate for the cluster center
measured by \citet{goldsbury}.

In the lower panel of the Figure, we show smoothed density distributions of
RGB stars along with iso-density contours for each population.
For the smoothed density distribution of each population, 
we applied a fixed Gaussian kernel estimator algorithm 
with a FWHM of 70 arcsec \citep{silverman}.
We show the FWHM of our Gaussian kernel in the lower left panel of the figure.
To derive the iso-density contour for each population,
we applied the second moment analysis \citep{dodd,stone}.
In the Figure, we show the iso-density contour lines for 90, 70, 50, and 30\% 
of the peak values for both populations.

At a glance of the figure, the distribution of the \cas\ group appears
to be more spatially elongated than that of the \caw\ group does.
In Table~\ref{tab:fit}, we show the axial ratio, $b/a$,
and the ellipticity, $e$ $(= 1 - b/a)$, and  we show 
radial distributions of the axial ratio and the ellipticity 
of the \caw\ and the \cas\ groups in Figure~\ref{fig:axis}.
As shown, the radial distributions for both groups
agree rather well up to $\approx$ 130 arcsec 
($\approx$ 1.5 core radii of the cluster) from the center
but two distributions bifurcate at the larger radial distance, 
in the sense that the \cas\ group is more elongated.

\subsubsection{Horizontal branch stars}

In recent years, the nature of the multiple stellar populations of the HB stars 
in GCs has been emerged through the detailed spectroscopic studies
\citep[e.g.][]{marino13,marino14, gratton14}.
\citet{marino13} studied 7 HB stars in M22 and they found that
all their HB stars are Ba-poor and Na-poor, equivalent to the $s$-poor
RGB population, leading them to suggest
that the position of a HB star is strictly related to the chemical composition. 
More recently, \citet{gratton14} investigated the chemical compositions
of more than 90 HB stars in M22, using the multi-object spectroscopy facility.
Their results showed that there appears to exist 3 different groups of
HB stars in M22. However, they did not find any severely O-depleted HB stars.
They claimed that the progeny of the severely O-depleted RGB stars may
evolve into the hotter part of the HB, where the surface temperature
of these HB stars are beyond the temperature range of their sample.

In addition to metallicity and age,
it is generally believed that the HB morphology of GCs can be governed
by the helium abundances \citep[e.g.][]{dantona02,joo13,milone14}
and there also exist direct spectroscopic observational evidences
of helium enhancement in GC HB stars \citep{marino14,gratton14}.

In their recent work, \citet{joo13} investigated M22 HB populations
using the synthetic population models.
Although simplistic, they divided M22 HB populations into two groups, 
the BHB and the EBHB, and they delineated a connection between
the RGB and the HB of M22 (see their Figure~13). 
Here, we explore more details on the number ratio and the spatial distribution
of the HB populations in M22 using our wide field data. 
Note that the $hk$ index is not an useful tool to study the BHB
or the EBHB stars, since the \caii\ H and K lines become suppressed
and the absorption strength of H$_\epsilon$ at $\lambda$ 3970 \AA\
reaches its maximum at A0 and can contaminate the adjacent
\caii\ H line at the high effective surface temperature of BHB or EBHB stars.

In Figure~\ref{fig:hbcmd}, we shows CMDs of the HB region of M22.
As for the RGB populations, we used all available multi-color information
in order to remove the contamination from the off-cluster field star population 
in our study.
Also shown in the figure is the cumulative LF of the HB stars,
where the slope in the cumulative LF of HB stars
becomes briefly zero at the location of the HB gap at $V$ = 15.97 mag.
This magnitude is about the same magnitude level that \citet{joo13}
separated the BHB and the EBHB populations in M22.
Furthermore, \citet{joo13} claimed that the BHB stars are the first generation 
of stars (corresponding to the progeny of the \caw\ RGB population in our notation), 
while the EBHB stars the second generation of stars 
(the \cas\ population) of the cluster.

\begin{figure}
\epsscale{1}
\figurenum{3}
\plotone{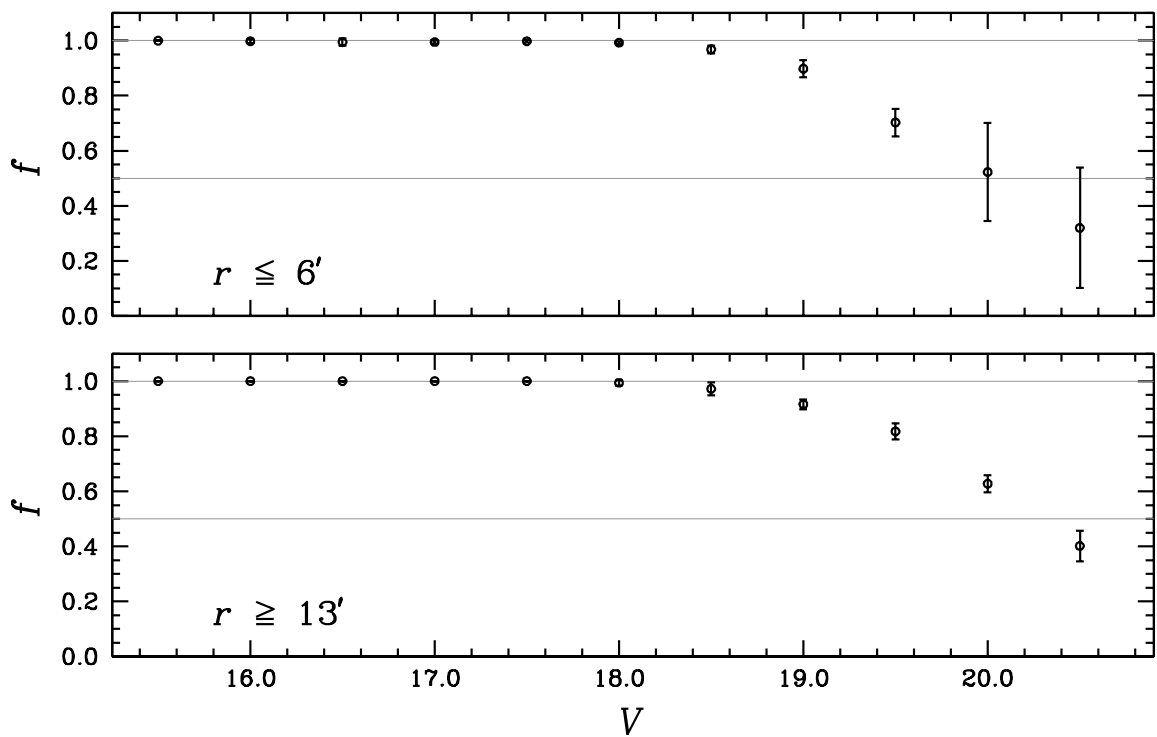}
\caption{Completeness fractions at each magnitude bin measured from
artificial star experiments for the inner and outer parts of M22.
The experiments suggest that our photometry is complete down to
$V \approx$ 18.5 mag.}\label{fig:complete}
\end{figure}

\begin{figure}
\epsscale{1}
\figurenum{4}
\plotone{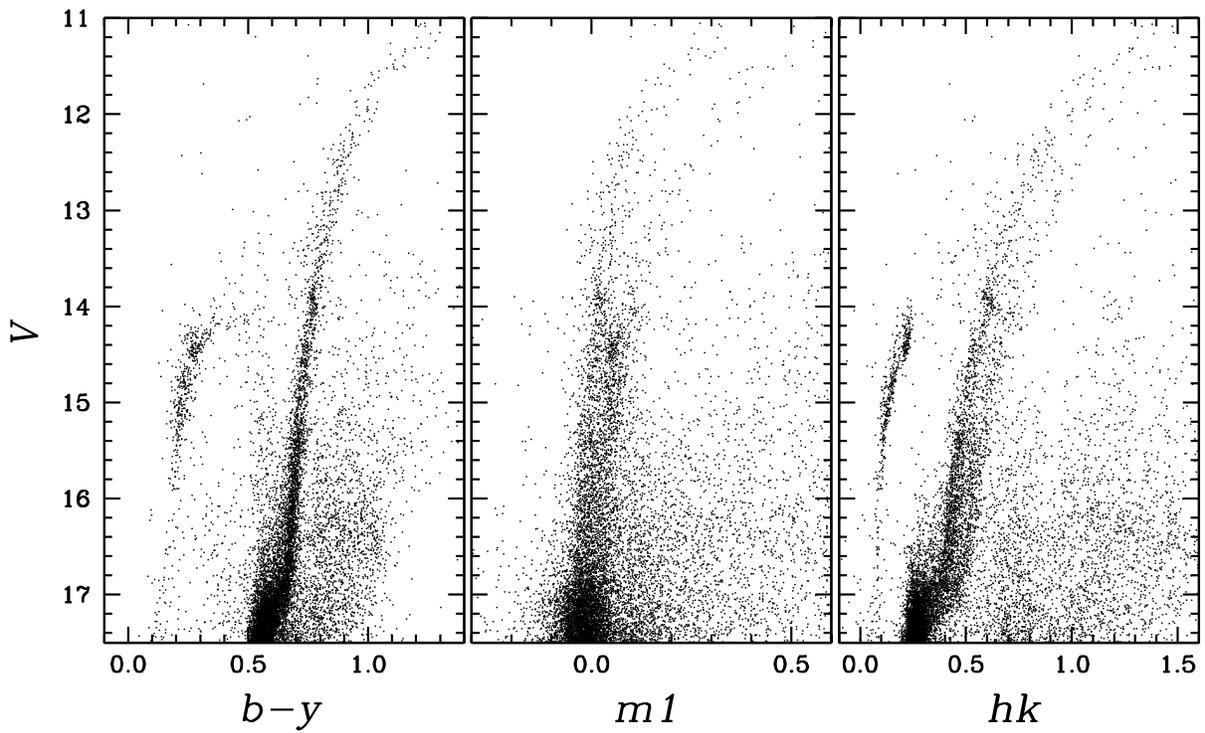}
\caption{Color-magnitude diagrams for bright stars 
located within 10 arcmin from the center of the cluster.
Note that discrete two RGB sequences can be clearly seen
in the $hk$ versus $V$ CMD.
In the figure, the $m1$ versus $V$ CMD is taken from \citet{jwlnat}.
}\label{fig:cmd}
\end{figure}

\begin{figure}
\epsscale{1}
\figurenum{5}
\plotone{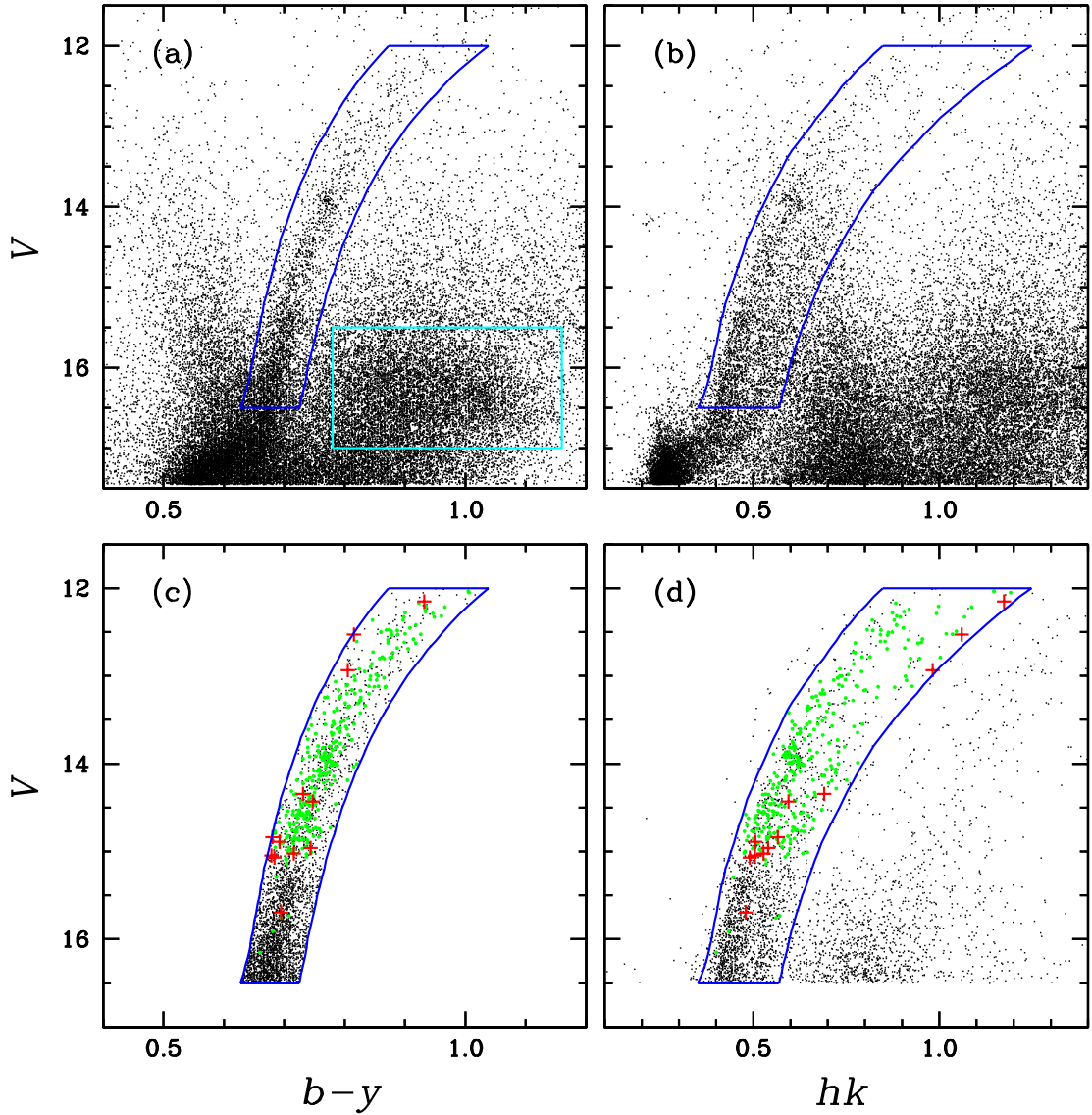}
\caption{(a) \& (b) CMDs of the M22 field. 
The enclosed regions by blue lines indicate the location of RGB stars in M22, 
while the enclosed region by cyan box in (a) 
mostly represents the bulge red-clump star population;
(c) The number of stars in the enclosed region is 3571.
Compared with \citet{lane09}, we have 307 stars in common;
295 stars are the radial velocity member star (green dots)
and 12 are radial velocity non-member star (red plus signs).
(d) The $hk$ versus $V$ CMD for 3571 stars from (c). The number of stars
in the enclosed region is 2054 without the radial-velocity non-member stars.
}\label{fig:rgbsel}
\end{figure}

\begin{figure}
\epsscale{1}
\figurenum{6}
\plotone{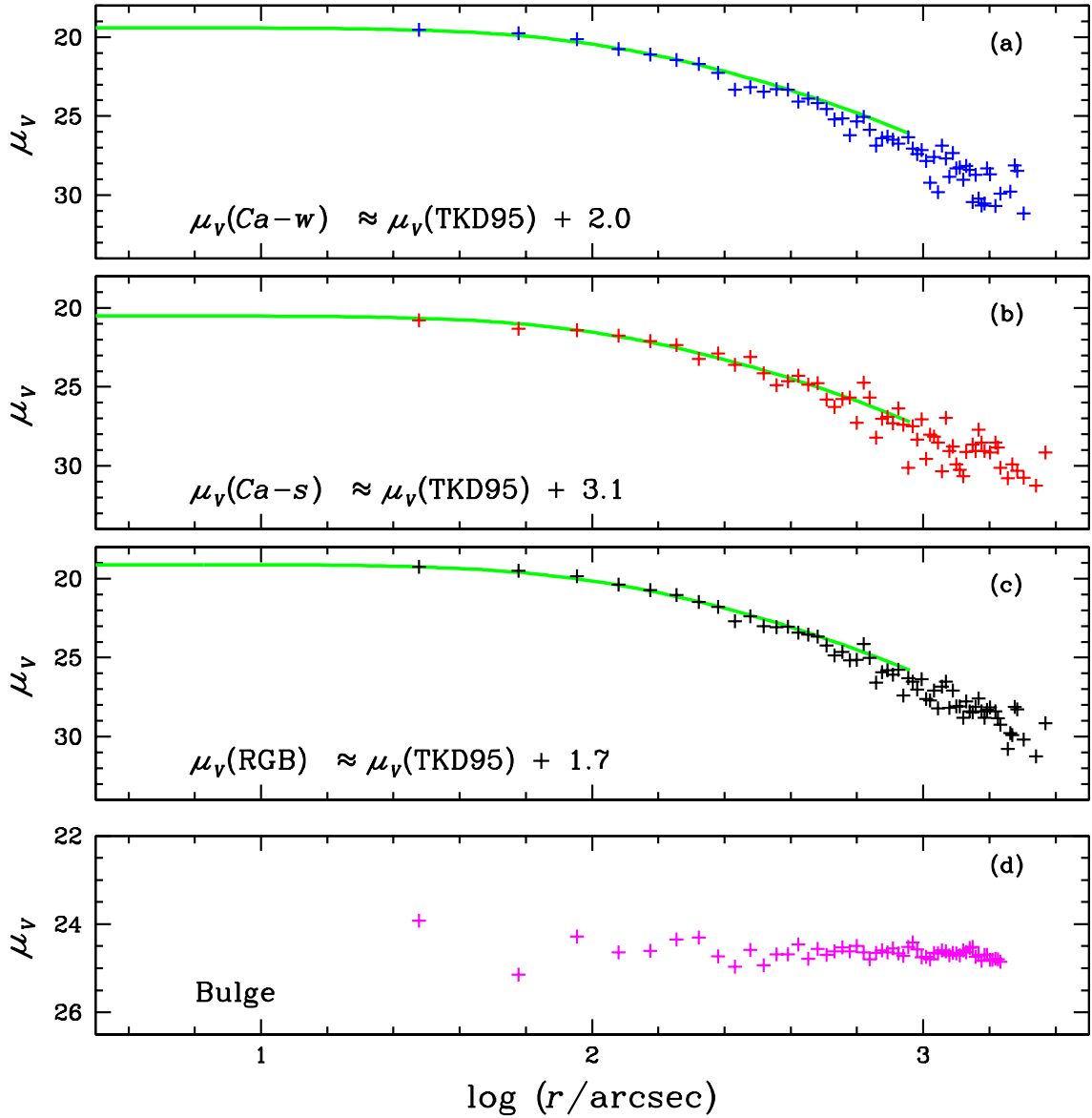}
\caption{The surface-brightness profiles for M22 RGB stars 
with 12.0 $\leq$ $V$ $\leq$ 16.5 mag. 
(a) \caw\ RGB stars only; (b) \cas\ RGB stars only; and
(c) All RGB stars in M22.
Our surface-brightness profiles for M22 RGB stars are in excellent
agreement with the Chebyshev polynomial fit of M22 by
\citet{trager95}, shown with green solid lines.
On the other hand, the surface-brightness profile for
the Galactic bulge red-clump stars is almost flat against
the radial distance from the center of the cluster.
Our results strongly suggest that the contamination from the off-cluster 
field star in our RGB selection procedure is not severe.
}\label{fig:surfmag}
\end{figure}

\begin{figure}
\epsscale{1}
\figurenum{7}
\plotone{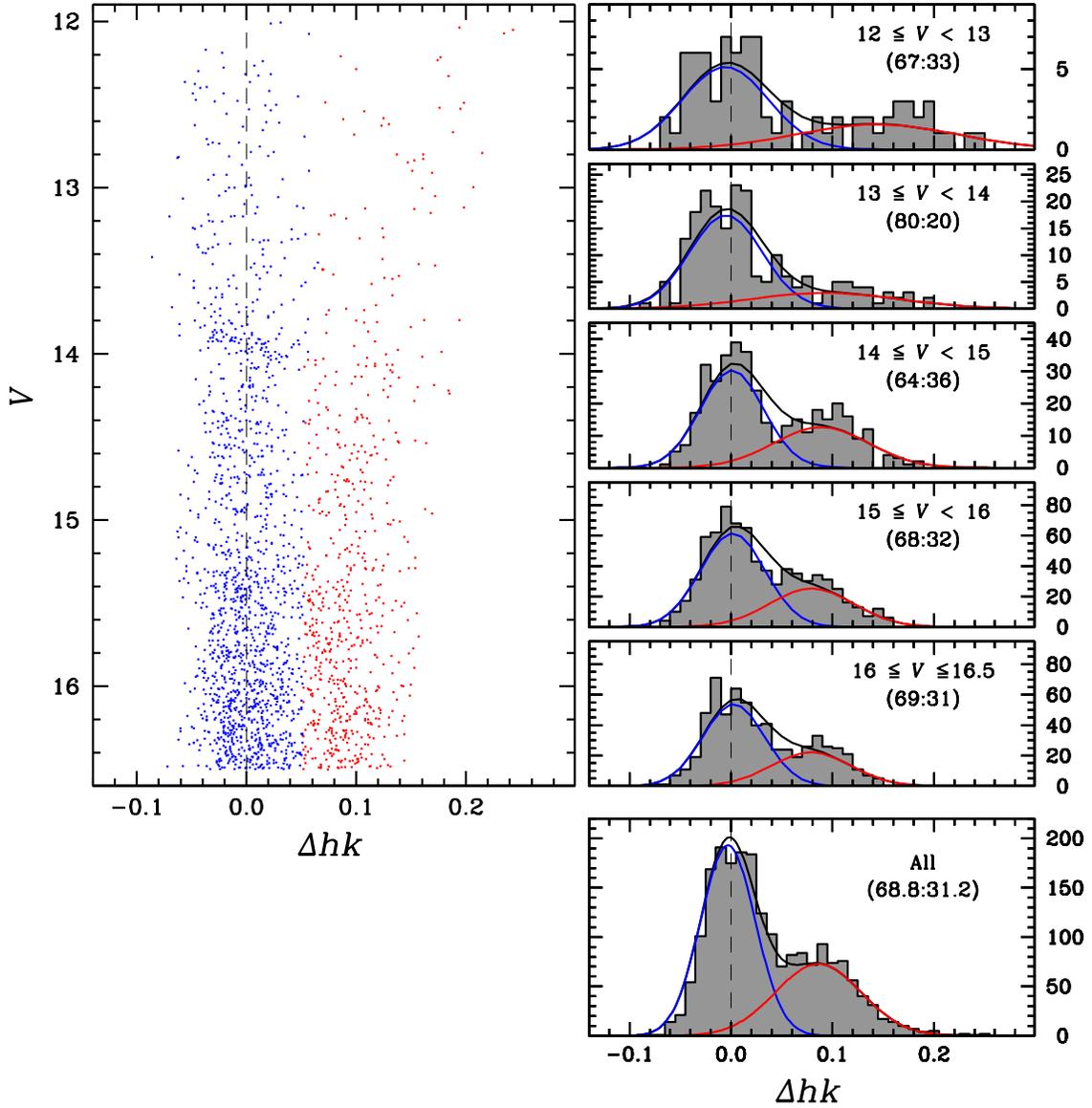}
\caption{(Left) The plot of $\Delta hk$ versus $V$ for M22 RGB stars,
where $\Delta hk$ is defined to be the difference in color from the mean
fiducial sequence of the \caw\ group.
Stars with the probability of being the \caw\ group $P(w)$
$\geq$ 0.5 from the EM estimator assuming the two gaussian mixture model
are denoted with blue dots,  while $P(s)$ $>$ 0.5 with red dots.
(Right) Histograms of M22 RGB stars with different magnitude bins. 
The two gaussian distribution was estimated by using the EM method
and the number ratio in the parenthesis is for \nrgb\ in each bin.
}\label{fig:rgbpop}
\end{figure}

\begin{figure}
\epsscale{1}
\figurenum{8}
\plotone{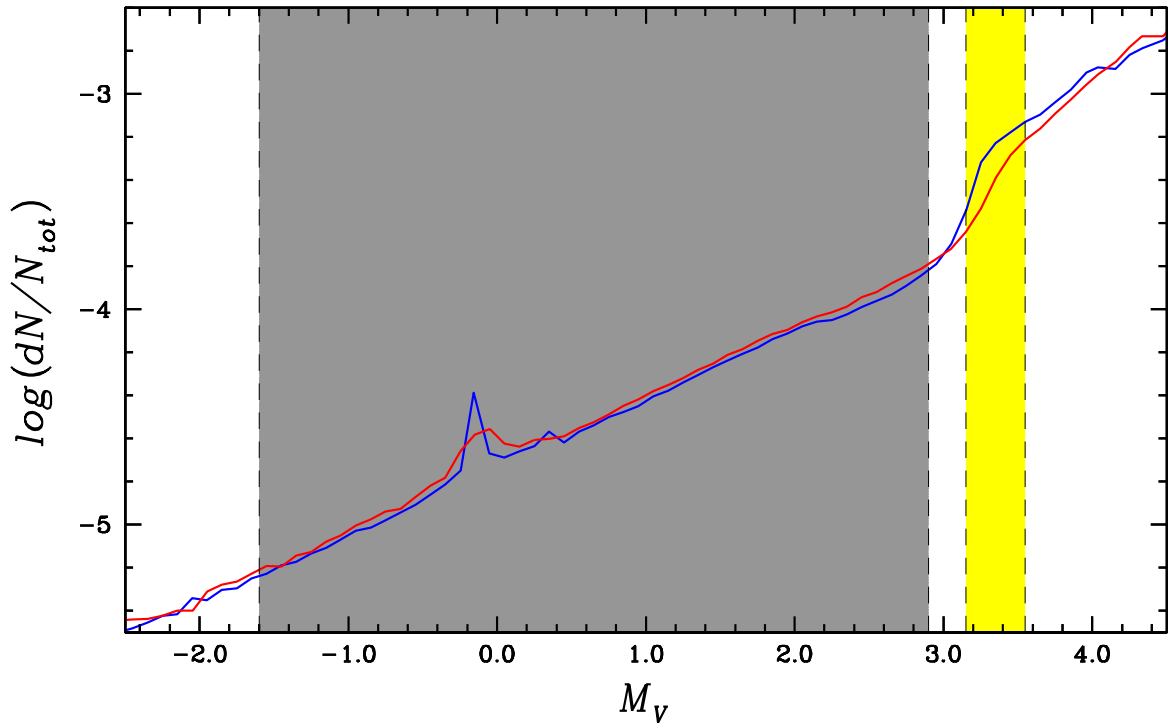}
\caption{The theoretical luminosity functions for the \caw\ group (the blue line)
and the \cas\ group (the red line). The grey shaded area corresponds 
to the RGB stars and the yellow shaded area to the SGB stars analyzed
in our current work.
Note that the RGB number ratio between the two groups at the fixed magnitude 
bin is very similar, \nrgb\ = 48.3:51.7 ($\pm$ 0.3),
while the number of SGB stars in the \caw\ (or bSGB) group is slightly larger than
that in the \cas\ (or fSGB) group, \nsgb\ = 58.0:42.0 ($\pm$ 0.2).
}\label{fig:LF}
\end{figure}

\begin{figure}
\epsscale{1}
\figurenum{9}
\plotone{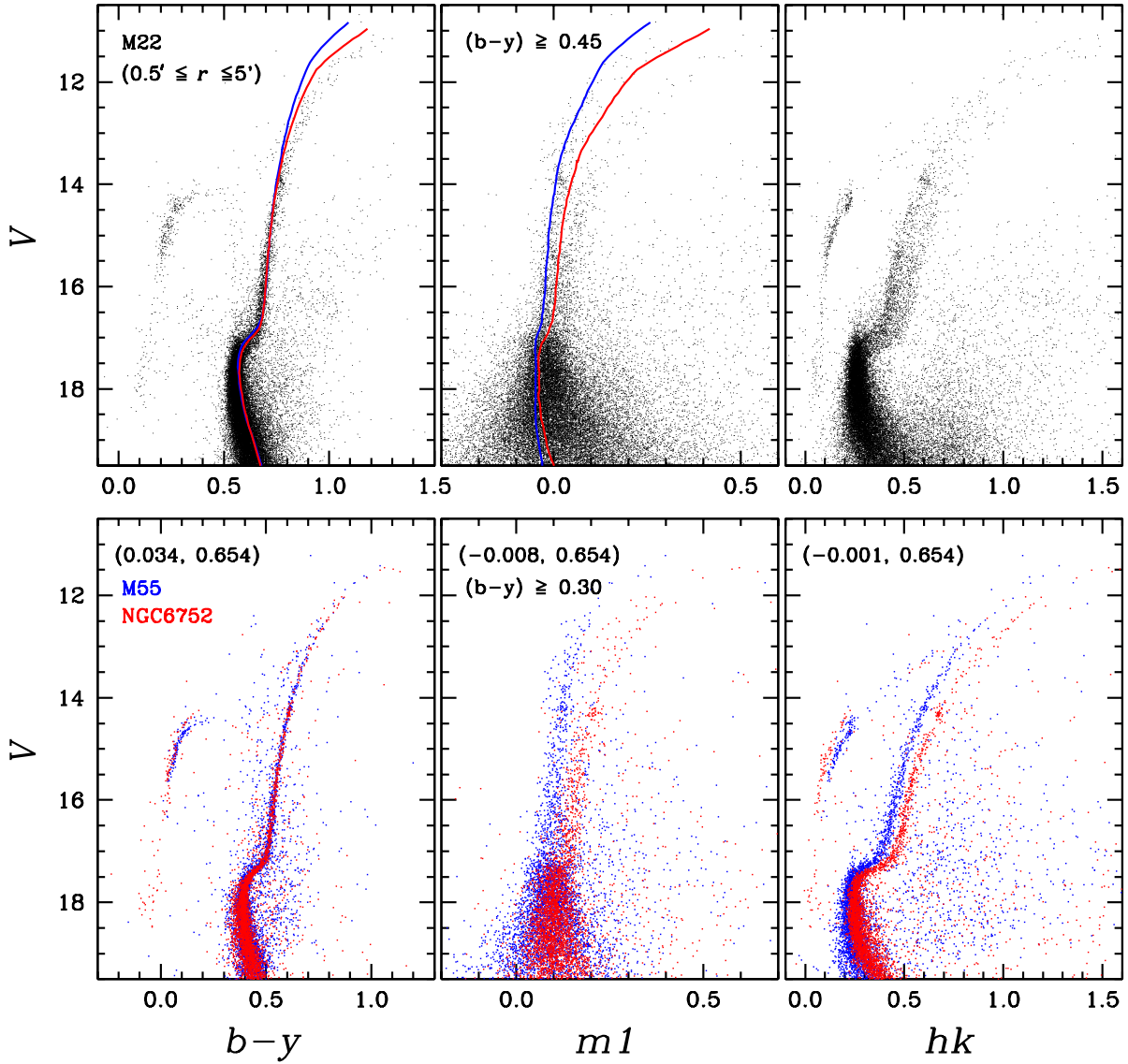}
\caption{(Upper panels) CMDs of M22 using stars with 
0.5 $\leq$ $r$ $\leq$ 5.0 arcmin. In the $m1$ versus $V$ CMD, we show
stars with $(b-y) \geq$ 0.45 mag, since the location of HB stars in the CMD
overlaps with that of RGB stars. Note that the blue solid line is for
the Padova isochrone with [Fe/H] = $-$1.8 and the red solid line
for $-$1.5 \citep{padova}.
(Bottom panels) Composite CMDs for globular clusters M55, which mimics
the \caw\ groups in M22, shown with blue dots and NGC~6752, which mimics
the \cas\ groups, with red dots.
The numbers in the parentheses indicate differential colors and 
the visual magnitude terms for NGC~6752 with respect to M55.
The composite CMDs using M55 and NGC~6752 reproduce very well the photometric
characteristics of M22 in the $m1$ versus $V$ and $hk$ versus $V$ CMDs.
}\label{fig:fakecmd}
\end{figure}

\begin{figure}
\epsscale{1}
\figurenum{10}
\plotone{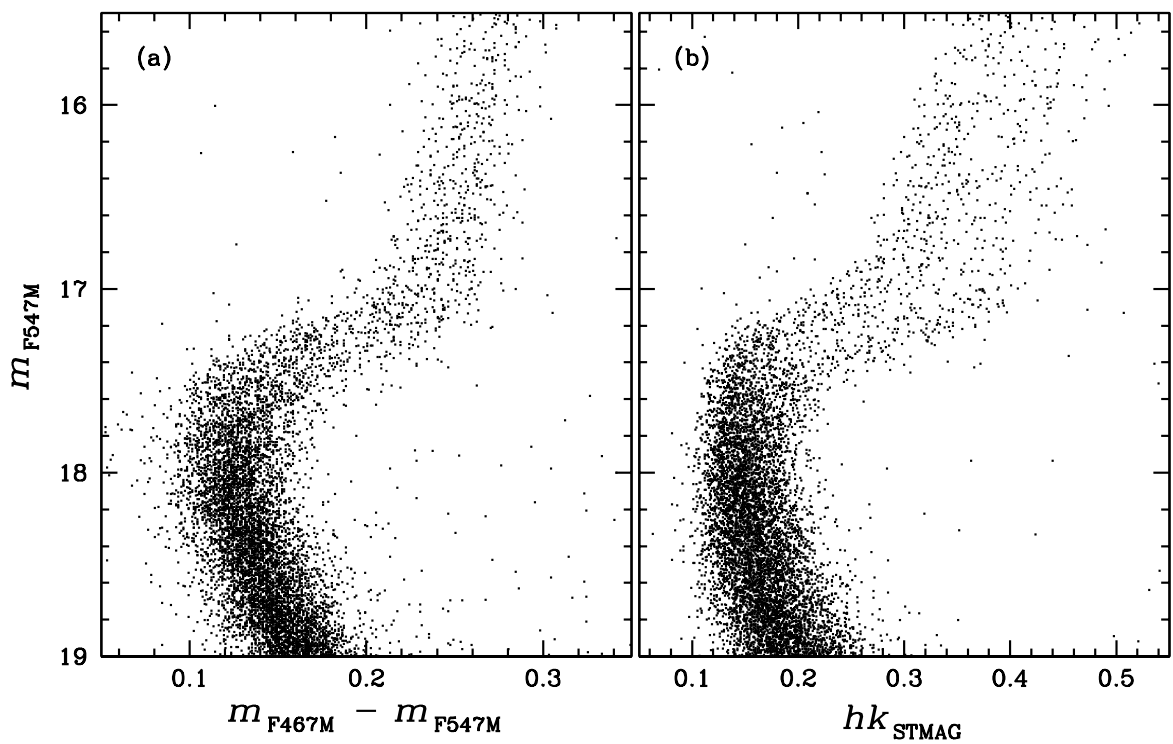}
\caption{Color-magnitude diagrams of M22 using the HST data.
The $hk_{\rm STMAG}$ denotes 
$(m_{\rm F395N} - m_{\rm F467M}) - (m_{\rm F467M} - m_{\rm F547M})$.
The split in the SGB and the RGB sequences in M22
can be clearly seen in the $hk_{\rm STMAG}$ versus $m_{\rm F547M}$ CMD.
}\label{fig:cmdHST}
\end{figure}

\begin{figure}
\epsscale{1}
\figurenum{11}
\plotone{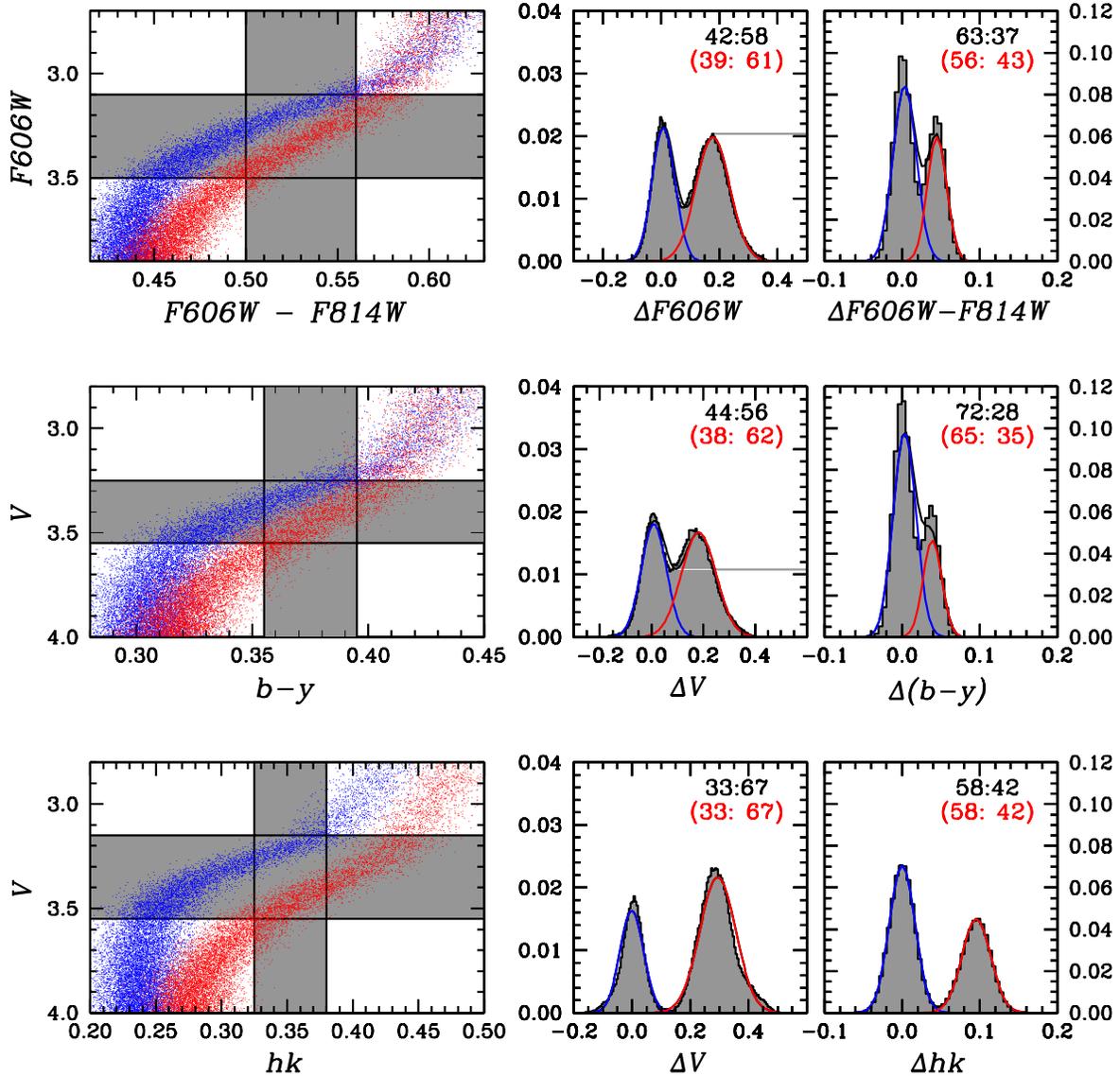}
\caption{(Left panels) Artificial CMDs around the SGB region of M22 
for different photometric systems. The blue dots are for the bSGB stars
and the red dots are for the fSGB stars.
The vertical and horizontal grey shaded regions indicate
ranges in the color indexes and in the magnitudes used to derive
the number ratios between the two groups.
We show only a small fraction of stars in each plot for the sake of clarity.
(Middle panels) The distributions of SGB stars at the fixed color indexes.
The first number ratios in each plot are those returned from
the EM estimator assuming the two gaussian mixture model 
while the number ratios in the parenthesis are the input values 
from our evolutionary population models.
(Right panels)  The distributions of SGB stars at the fixed magnitudes.
Note that the number ratios from the ($hk$, $V$) plane
are the most reliable, while those from the ($b-y$, $V$) plane are 
the least reliable.
}\label{fig:SynSGB}
\end{figure}

\begin{figure}
\epsscale{1}
\figurenum{12}
\plotone{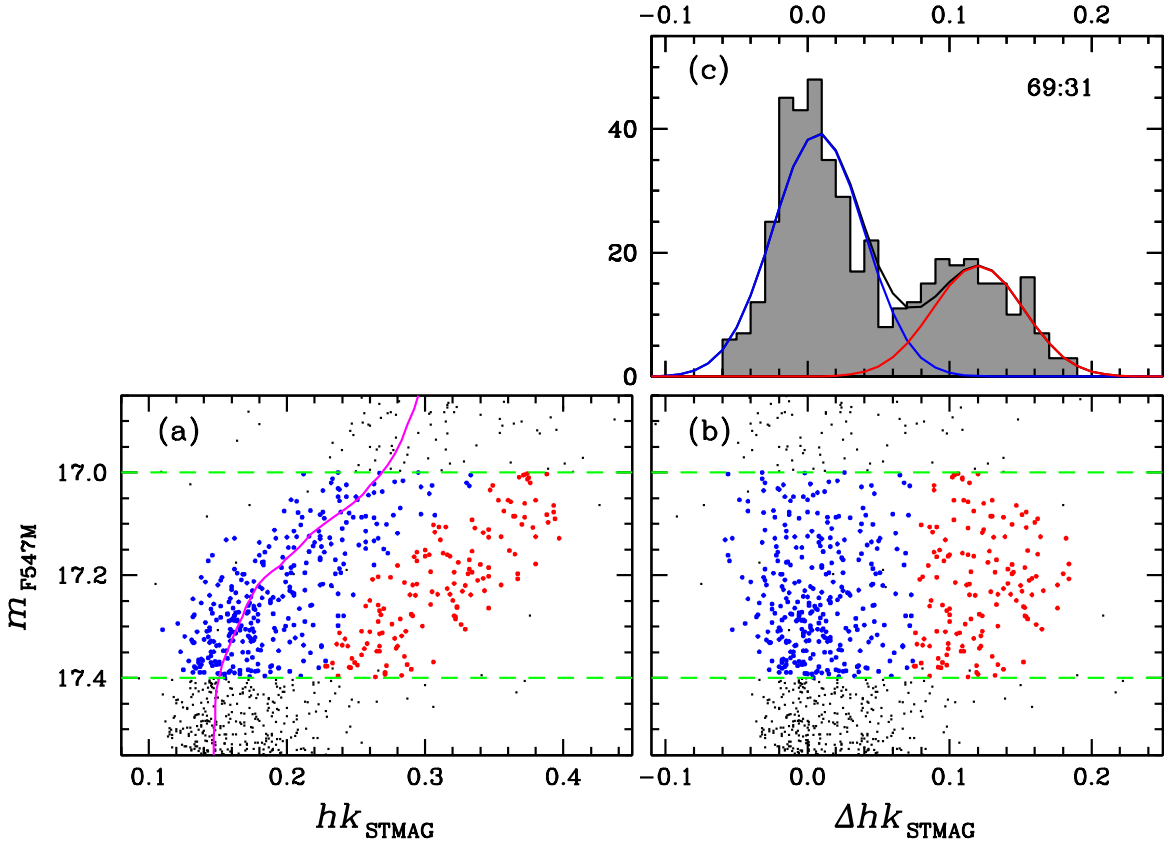}
\caption{(a) The $hk_{\rm STMAG}$ versus $m_{\rm F547M}$ CMD 
around the SGB region of M22. Also shown is the fiducial sequence 
for the bSGB group. We use the SGB stars with
17.0 mag $\leq$ $m_{\rm F547M}$ $\leq$ 17.4 mag to estimate
the number ratio between the two groups.
The stars with $P$(bSGB) $\geq$ 0.5 from the EM estimator are denoted
with blue dots, while $P$(fSGB) $>$ 0.5 with red dots.
(b) The plot of $\Delta hk_{\rm STMAG}$ versus $V$, 
where $\Delta hk_{\rm STMAG}$ is defined to be
the difference in the $hk_{\rm STMAG}$ index between the individual stars
and the fiducial sequence of the bSGB group.
(c) The distribution of $\Delta hk_{\rm STMAG}$ and the two gaussian mixture
model fit to the data.
We obtain the apparent number ratio of \nsgb\ = 69:31 ($\pm$ 6)
and the intrinsic number ratio of \nsgb\ = 62:38 ($\pm$ 6) after
applying the correction for the differential evolutionary effect.
This SGB number ratio is in agreement
with the RGB number ratio within measurement errors.
}\label{fig:sgbpop}
\end{figure}

\begin{figure}
\epsscale{1}
\figurenum{13}
\plotone{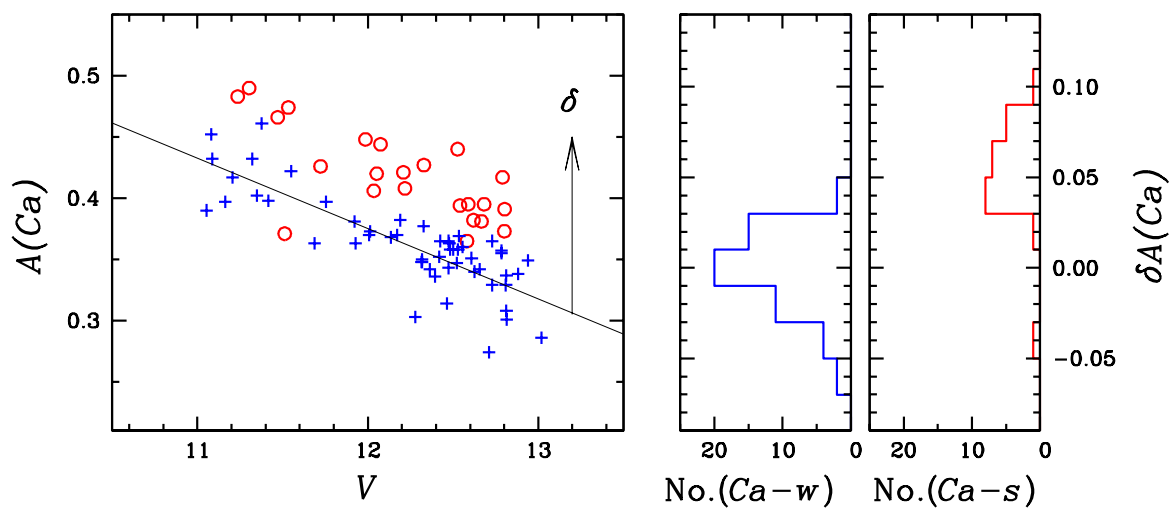}
\caption{A comparison of the Calcium index $A$(Ca) 
measured by \citet{norris83} between the RGB stars in the \caw\ (blue plus signs)
and  the \cas\ (red circles) groups in M22 as functions of $V$ magnitude.
We also show the least square fit to the \caw\ RGB stars
and the differences are measured from the fitted line
to correct the temperature and the surface gravity effects
on the absorption strengths.
The histograms for each group are shown in the right panel.
Note that the boundary between the two groups is rather sharp.}
\label{fig:norris}
\end{figure}

\begin{figure}
\epsscale{1}
\figurenum{14}
\plotone{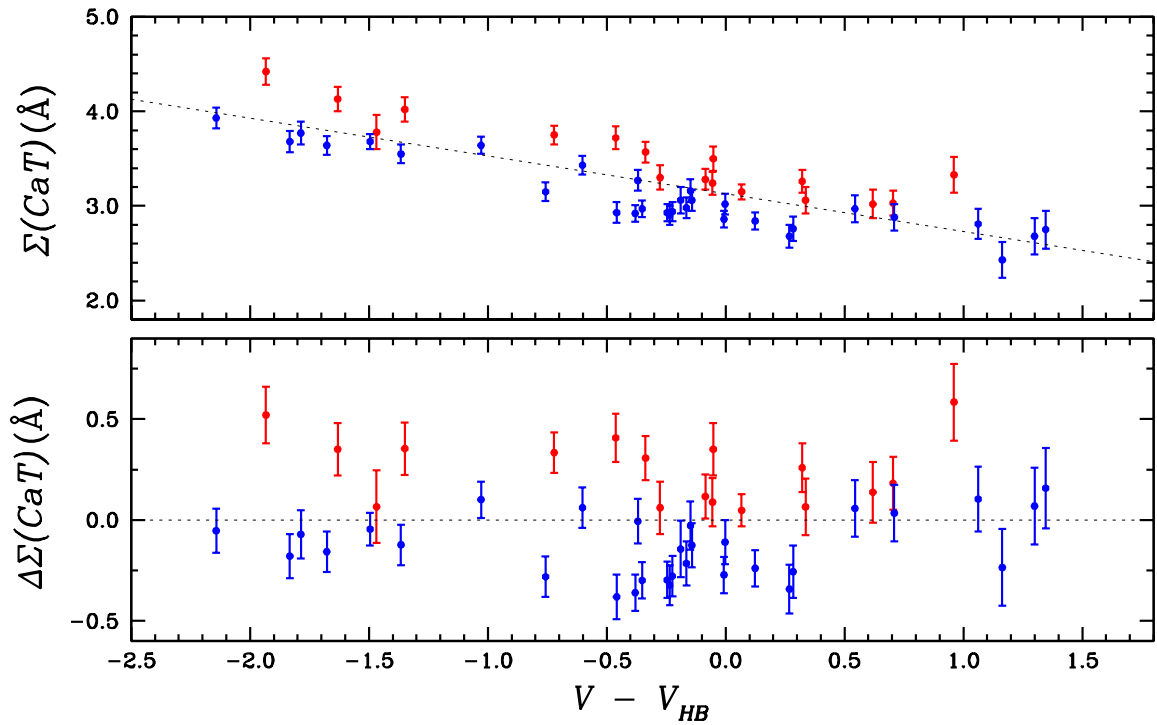}
\caption{Comparisons of \ion{Ca}{2} triplet strength $\Sigma$(CaT), 
(= $W_{8542} + W_{8662}$), measured by \citet{dacosta09} 
for the \caw\ (blue circles) and 
the \cas\ (red circles) RGB stars against magnitude difference
from the horizontal branch, $V-V_{\rm HB}$.
In the upper panel, the dotted line denotes the least square fit to the data.
The lower panel show the residuals around  the fitted line.
}\label{fig:dacosta}
\end{figure}

\begin{figure}
\epsscale{1}
\figurenum{15}
\plotone{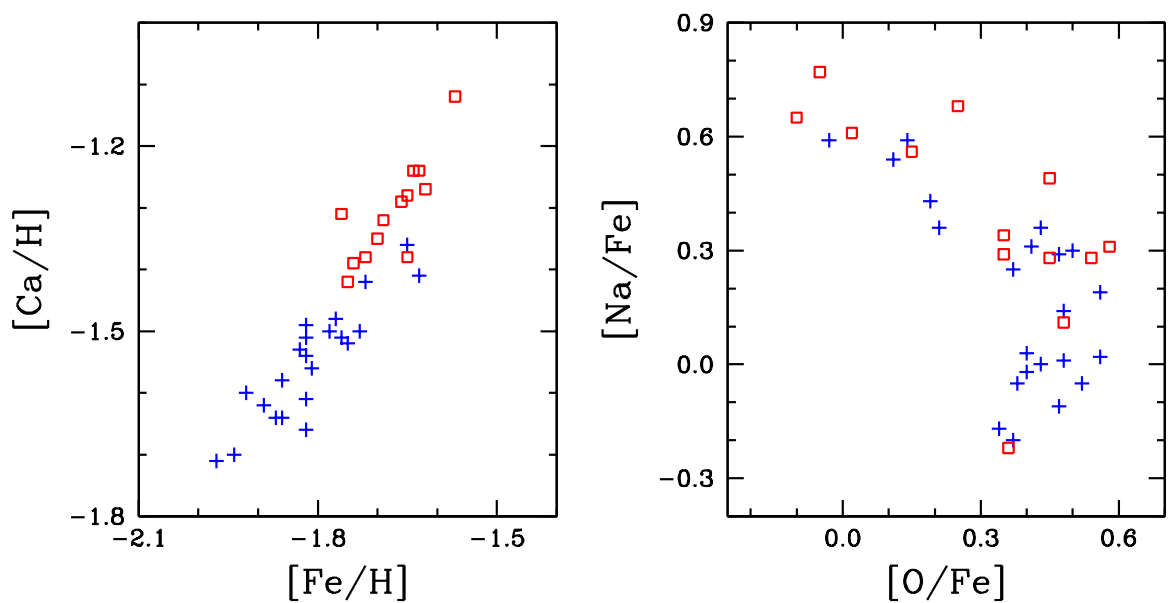}
\caption{Plots of [Ca/H] versus [Fe/H] and [Na/Fe] versus [O/Fe]
for the \caw\ (blue pluses) and the \cas\ (red squares) RGB stars
using high-resolution spectroscopic measurements by \citet{marino11}.
The plot of [Ca/H] versus [Fe/H] shows that the RGB stars in the \caw\ group 
have lower calcium and iron abundances than those in the \cas\ group.
The plot of [Na/Fe] versus [O/Fe] shows that each group has its own
Na-O anti-correlations. In the mean the \caw\ group appears to have a higher
oxygen and a lower sodium abundances than the \cas\ group.
}\label{fig:marino}
\end{figure}

\begin{figure}
\epsscale{1}
\figurenum{16}
\plotone{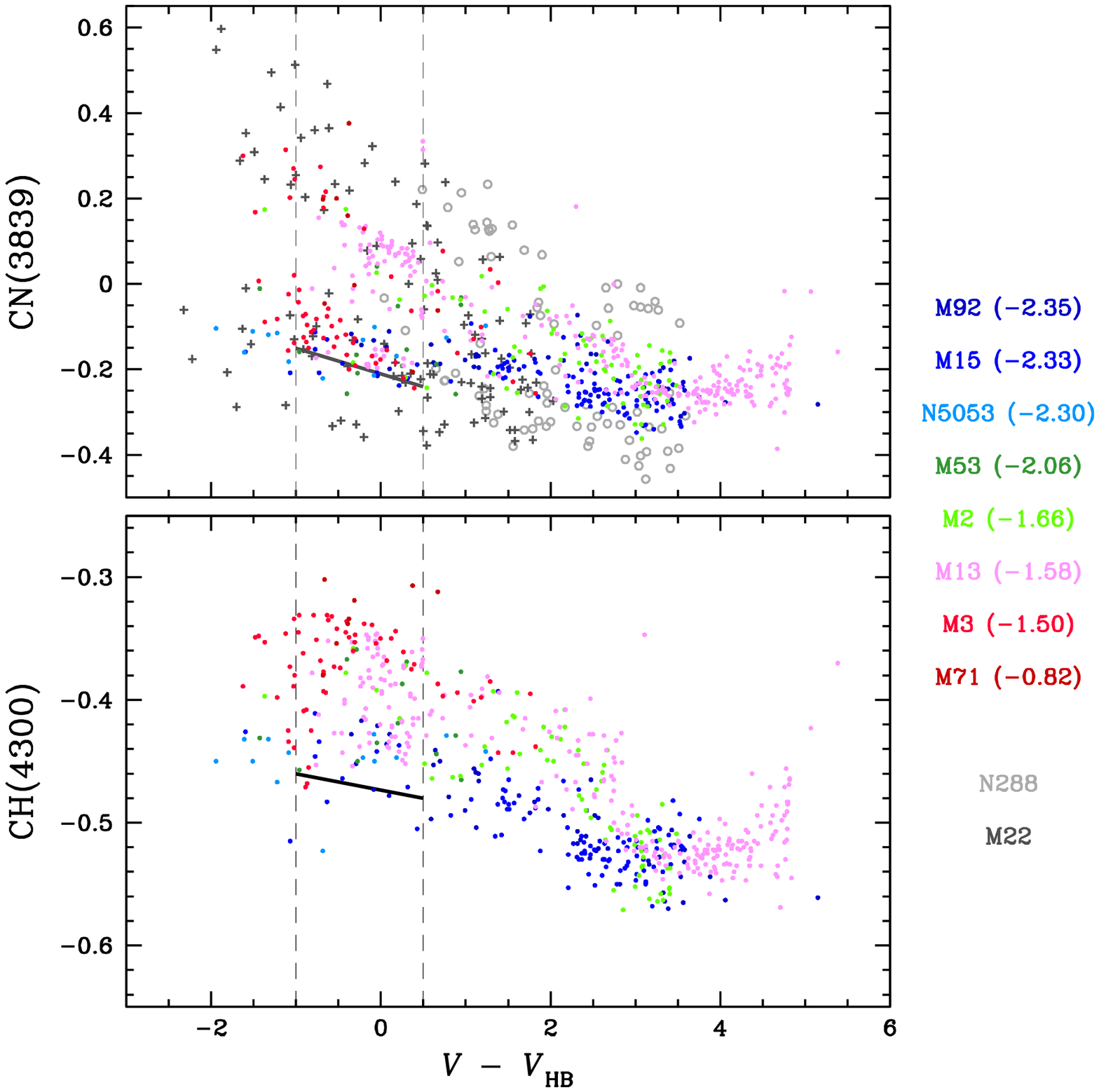}
\caption{Distributions of CN(3839) and CH(4300) against $V-V_{\rm HB}$
for 8 GCs by \citep{smolinski11}. 
The numbers inside the parentheses are the metallicity of each GC.
Also shown are two GCs, M22 and NGC~288, by
\cite{lim15}.
The thick solid lines denote the common lower envelopes for RGB stars around 
$V_{\rm HB}$,  $-1 \leq V - V_{\rm HB} \leq 0.5$, used in our analysis.
Note that the CN(3839) and CH(4300) index values by \citet{smolinski11} do not
agree with those by \citet{lim15} due to different definitions of the band strengths
between the two works.
}\label{fig:cnch}
\end{figure}

\begin{figure}
\epsscale{1}
\figurenum{17}
\plotone{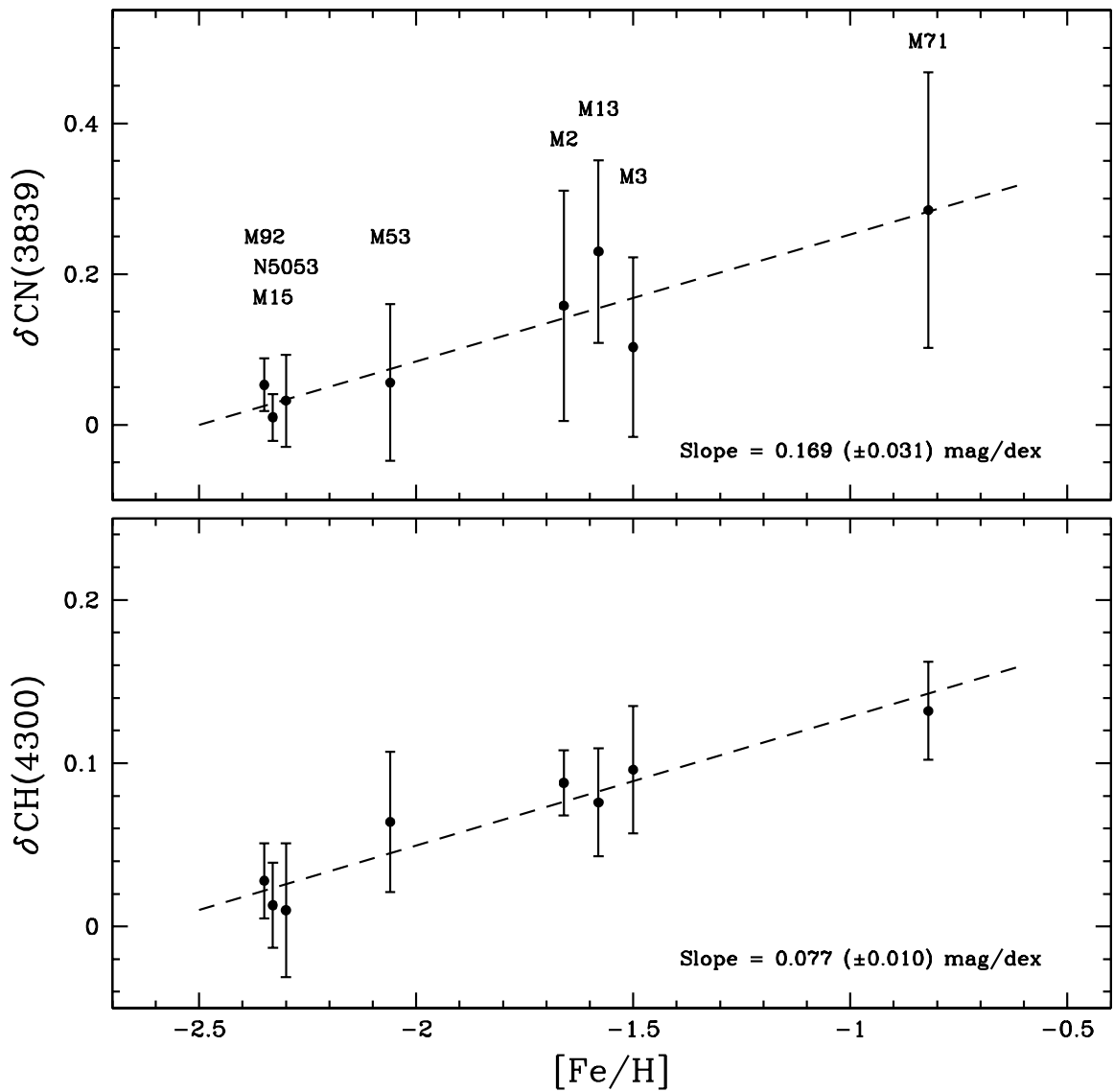}
\caption{The distributions of the $\delta$CN(3839) and the $\delta$CH(4300)
against [Fe/H] for 8 GCs by \citet{smolinski11}, where 
the $\delta$CN(3839) and the $\delta$CH(4300) are defined to be 
the differences in the band strengths from the common lower envelopes.
The least square fits to the data are shown with dashed lines.
As expected from Figure~\ref{fig:cnch}, both the $\delta$CN(3839) and 
the $\delta$CH(4300) have the gradients against metallicity.
}\label{fig:cnchgrad}
\end{figure}

\begin{figure}
\epsscale{1}
\figurenum{18}
\plotone{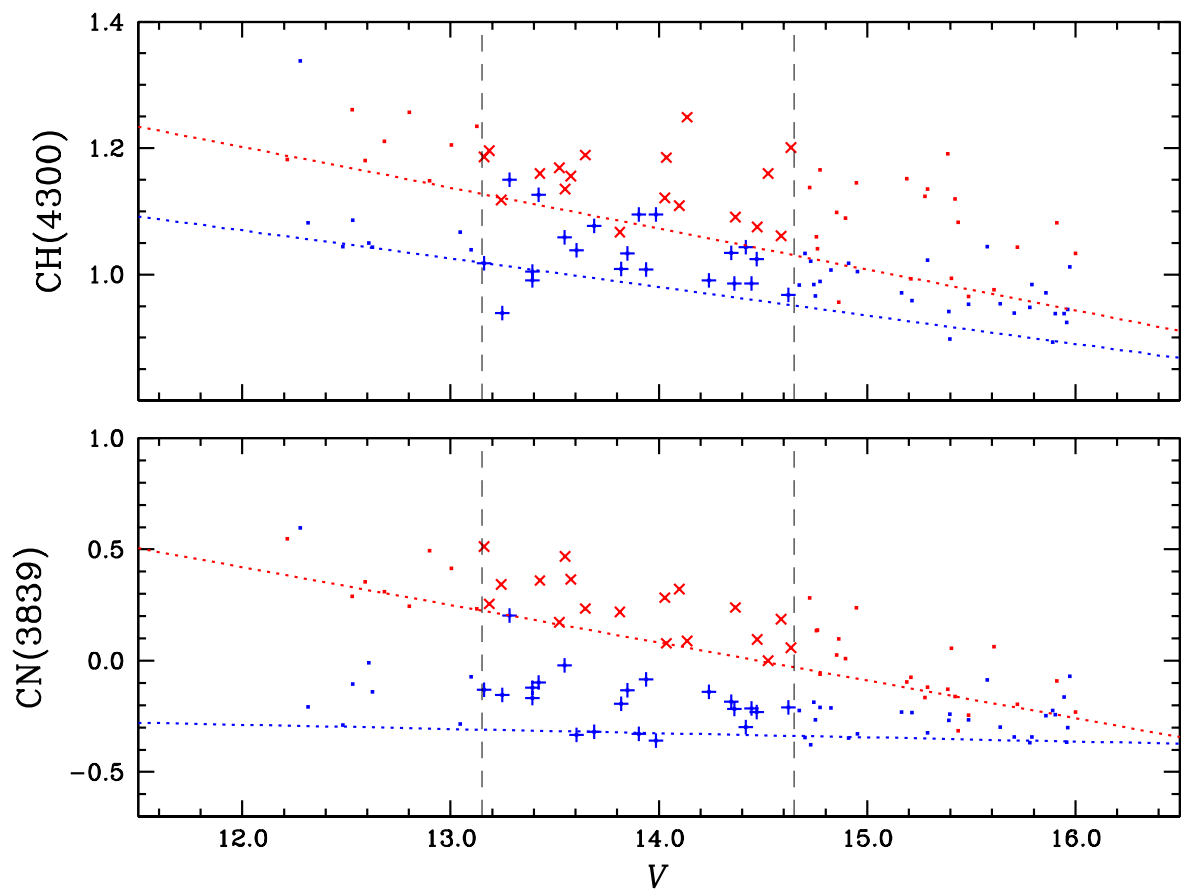}
\caption{Distributions of CN(3839) and CH(4300) against $V-V_{\rm HB}$
for M22 RGB stars by \cite{lim15}.
The blue color and the red color denote the \caw\ and the \cas\ RGB stars
in M22, respectively. The blue dotted lines denote the lower envelope for
the \caw\ stars, while the red dotted lines for the \cas\ stars.
The vertical dashed lines denote the region around the $V_{\rm HB}$ of M22,
$-1 \leq V - V_{\rm HB} \leq 0.5$.
}\label{fig:lim}
\end{figure}

\begin{figure}
\epsscale{1}
\figurenum{19}
\plotone{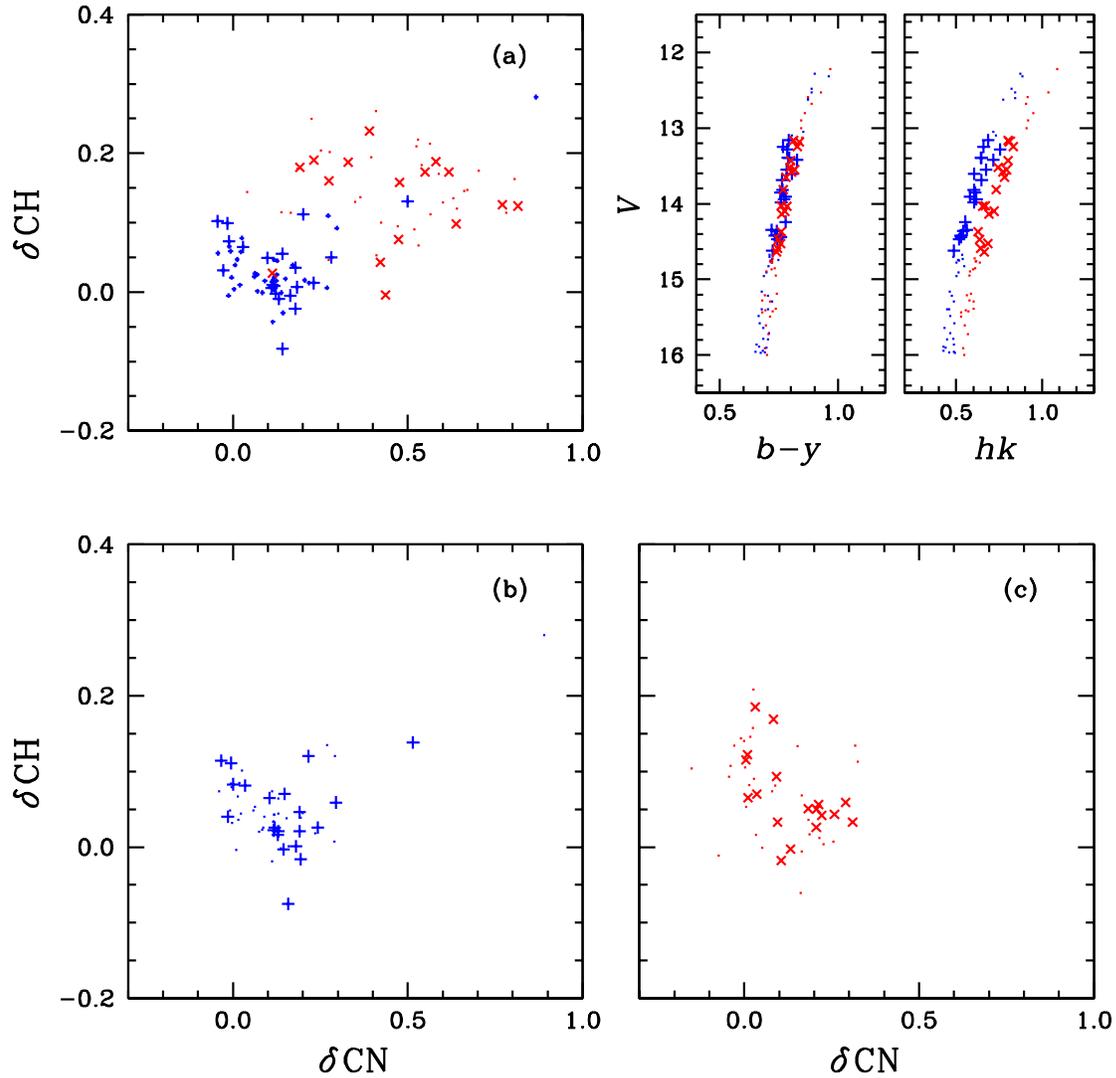}
\caption{The distributions of the $\delta$CN(3839) versus the $\delta$CH(4300);
(a) Using the common lower envelopes. 
(b) The \caw\ RGB stars using its own lower envelopes.
(c) The \cas\ RGB stars using its own lower envelopes.
The RGB stars around the $V_{\rm HB}$ are shown with the plus signs and crosses.
The $\delta$CN--$\delta$CH positive correlation in (a) is 
most likely due to the metallicity effect between the two groups
as shown in Figure~\ref{fig:cnchgrad}.
}\label{fig:lim2}
\end{figure}

\begin{figure}
\epsscale{1}
\figurenum{20}
\plotone{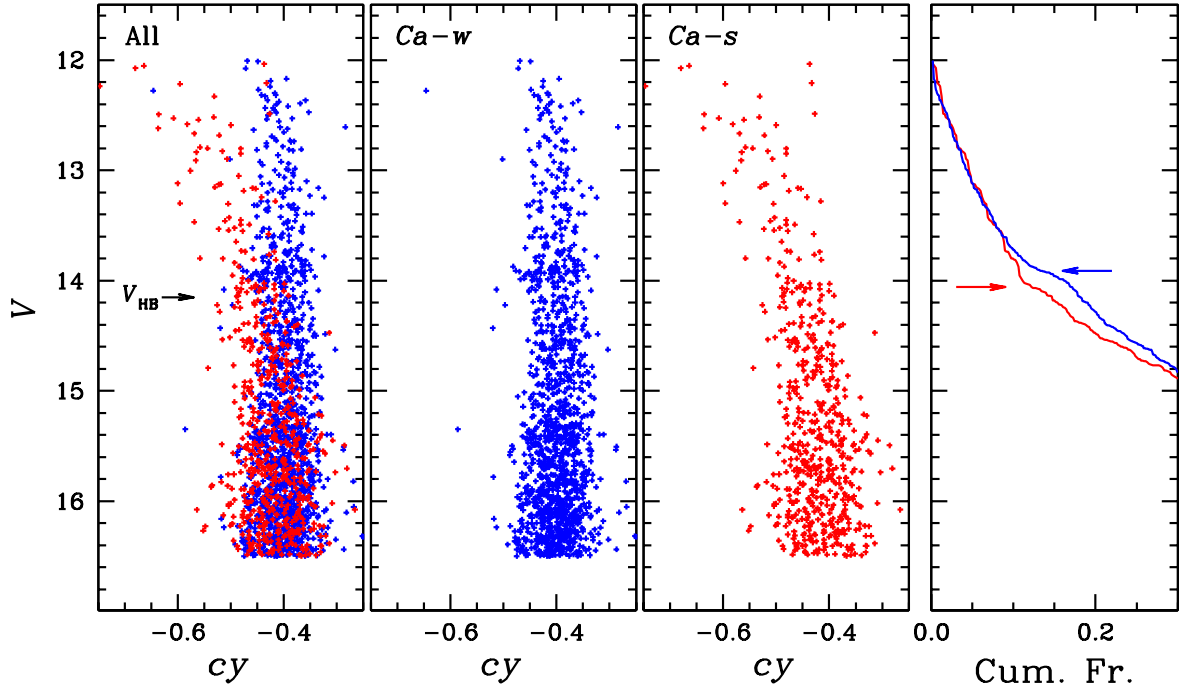}
\caption{Plots of $cy$ versus $V$ CMDs of M22 RGB stars.
The $V$ magnitude of the horizontal branch, $V_{\rm HB}$  = 14.15 mag,
is indicated by an arrow.
The $cy$ spread at $V_{\rm HB}$ is $\sigma (cy) \approx$ 0.03 mag
in each group, corresponding to $\sigma$[N/Fe] $\approx$ 0.5 dex,
if the spread in the NH absorption strengths are solely responsible
for the $cy$ spreads in M22 RGB stars.
In the right panel of the figure, we show the cumulative LFs
for the \caw\ and the \cas\ populations. We show the $V$ magnitude levels
of the RGB bump, $V_{\mathrm bump}$, 
at which the slope of the LF changes abruptly,
for each population by arrows. We obtained $V_{\mathrm bump}$ = 13.91 mag
for the \caw\ population while 14.06 mag for the \cas\ population.
The fainter bump magnitude in the \cas\ population suggests the metallicity
effect on the bump brightness is greater than 
the effect from the helium enrichment is.}\label{fig:cy}
\end{figure}

\begin{figure}
\epsscale{1}
\figurenum{21}
\plotone{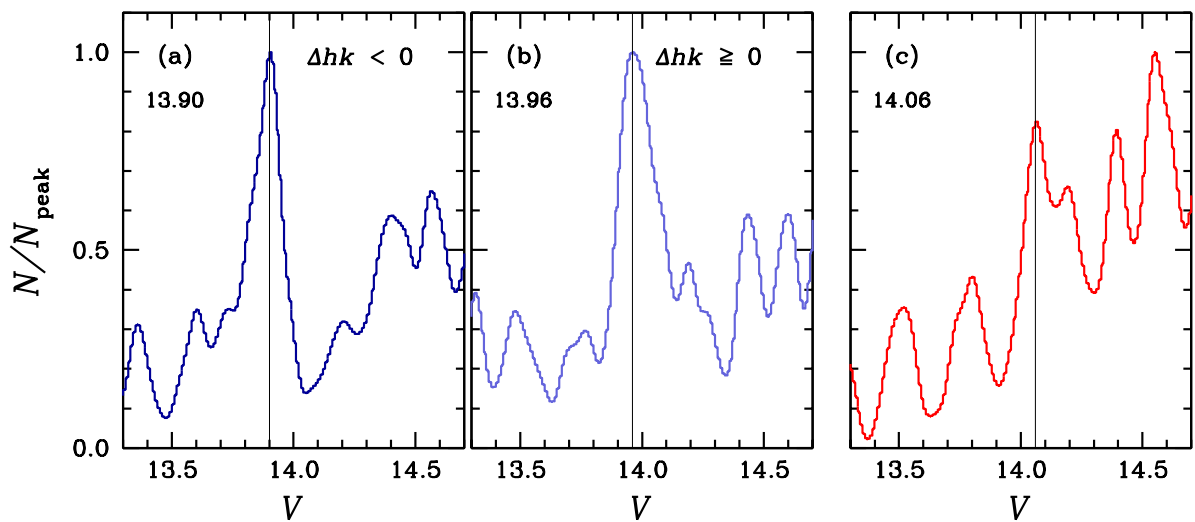}
\caption{Generalized histograms for RGB stars against $V$ magnitude.
(a) \caw\ RGB stars with $\Delta hk <$ 0 from Figure~\ref{fig:rgbpop}.
(b) \caw\ RGB stars with $\Delta hk \geq$ 0.
(c) \cas\ RGB stars.
The vertical thin lines represent $V_{\mathrm bump}$ for each subset,
13.90, 13.96 and  14.06 mag respectively.}\label{fig:bump}
\end{figure}

\begin{figure}
\epsscale{1}
\figurenum{22}
\plotone{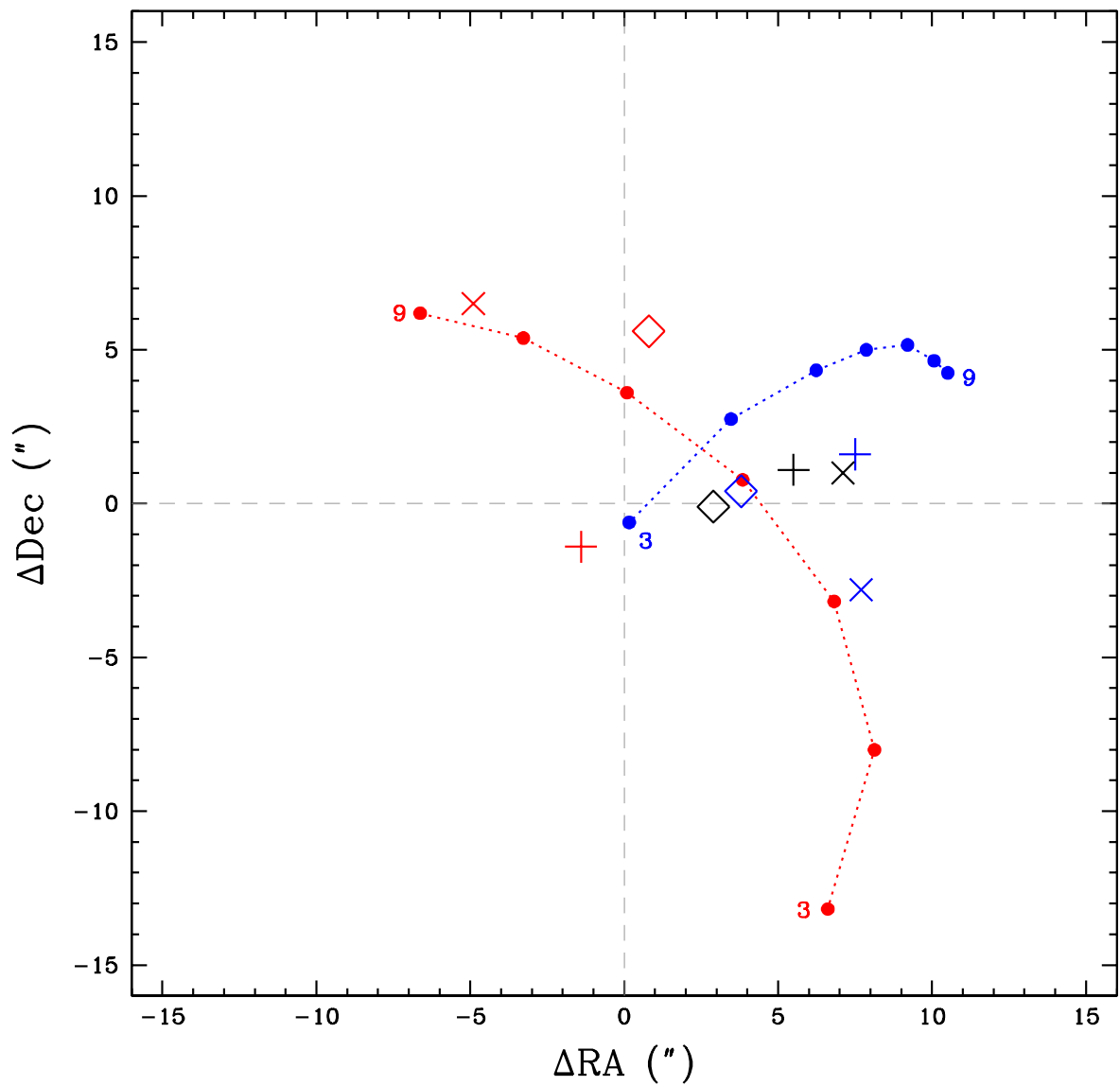}
\caption{Relative positions of centers with respect to that of \citet{goldsbury}.
The blue color is for the \caw\ group, the red color for the \cas\ group
and the black for the combined RGB stars.
Diamond signs denote for the centers from the simple mean,
plus signs for those from the half-sphere method and
crosses for those from the pie-slice method.
The filled circles are centers of each population from the iso-density contour
method from 90\% to 30\% levels.}\label{fig:cnt}
\end{figure}

\begin{figure}
\epsscale{1}
\figurenum{23}
\plotone{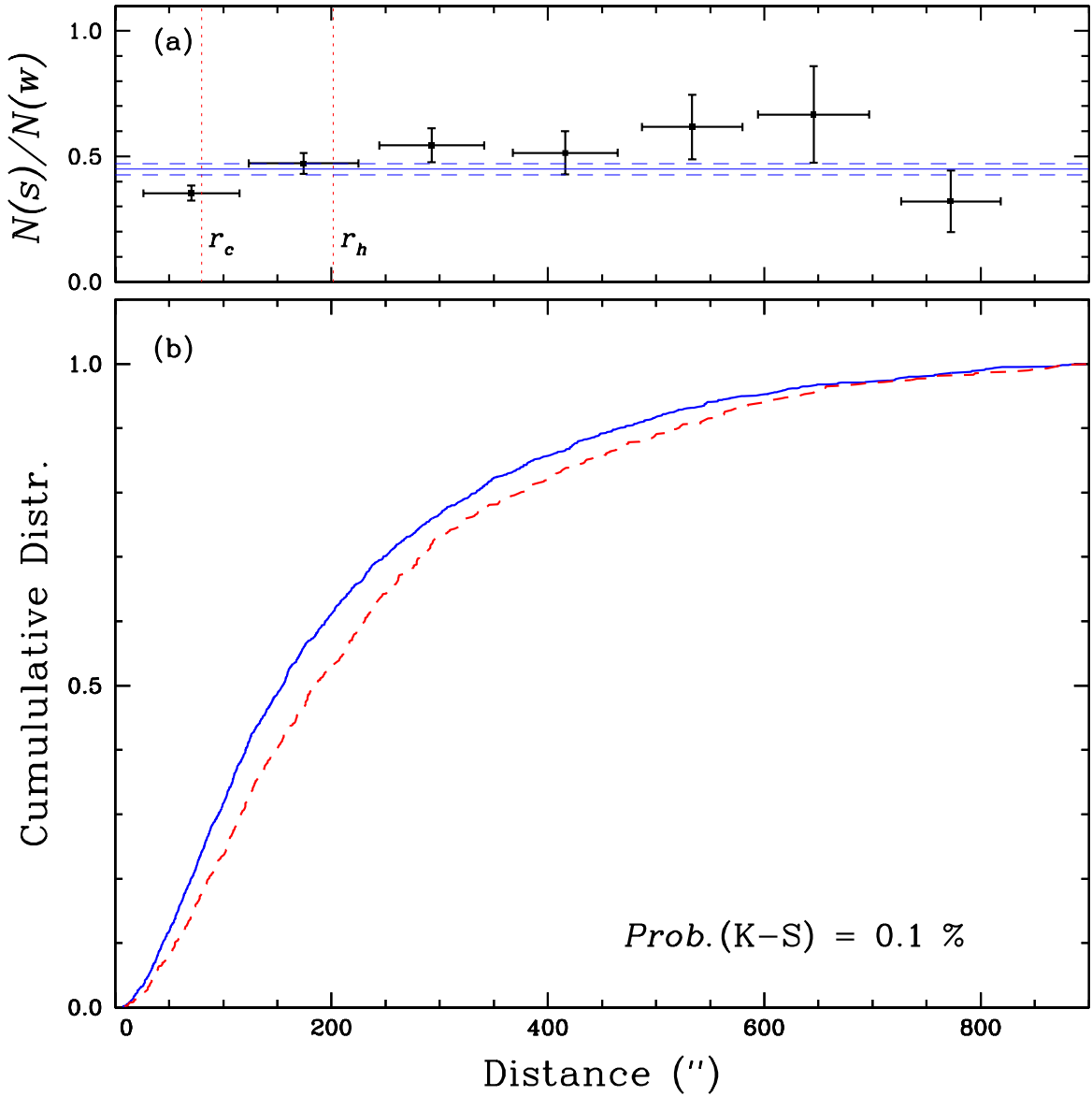}
\caption{(a)The number ratios of the \cas\ to the \caw\ groups against
the radial distance from the center. 
The RGB number ratio appears to have a weak radial gradient
up to $r \approx$ 700 arcsec from the center,
in the sense that the number of the \caw\ population is slightly
more centrally concentrated than that of the \cas\ population.
In the figure, the vertical lines denote 
the cluster's core and half-light radii and the horizontal lines denote
the mean number ratio and the standard deviations.
(b) Cumulative radial distributions of \caw\ (blue) and \cas\ (red)
RGB populations. The \caw\ RGB stars are slightly more centrally concentrated.
The K-S test indicates a probability of 0.1\% that two populations 
are drawn from the same parent population.}\label{fig:rgbdist}
\end{figure}

\begin{figure}
\epsscale{1}
\figurenum{24}
\plotone{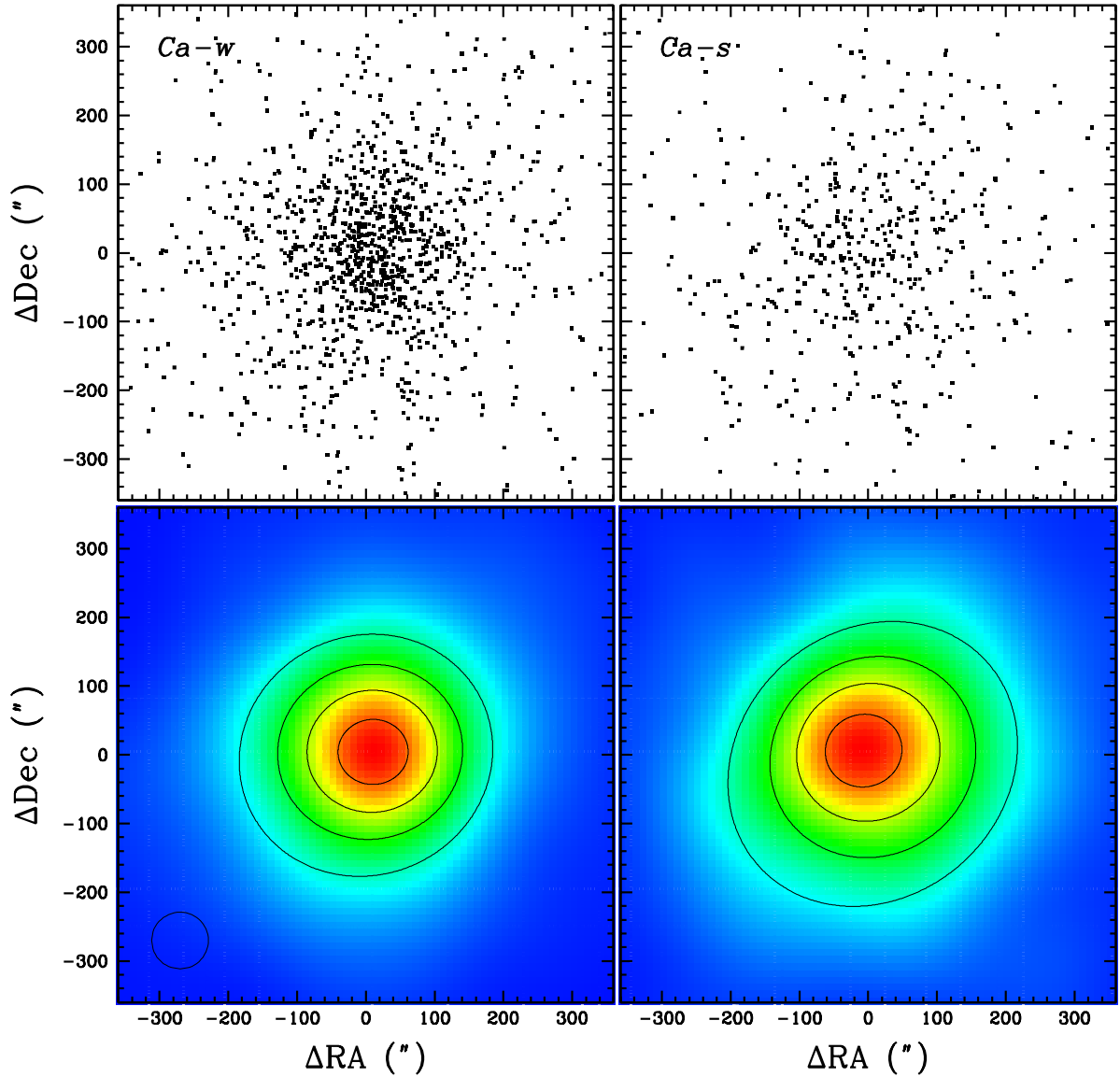}
\caption{Spatial distributions of the \caw\ and the \cas\ RGB stars in M22 are
shown in the upper panels and the smoothed contour maps using 
the fixed Gaussian kernel are shown in the lower panels, where
we show the iso-density contour lines for 90, 70, 50, and 30\% 
of the peak values for both populations. We also show the FWHM of our adopted 
Gaussian kernel in the lower left panel of the figure. Note that 
the distribution of the \cas\ RGB stars is more elongated.}\label{fig:density}
\end{figure}

\begin{figure}
\epsscale{1}
\figurenum{25}
\plotone{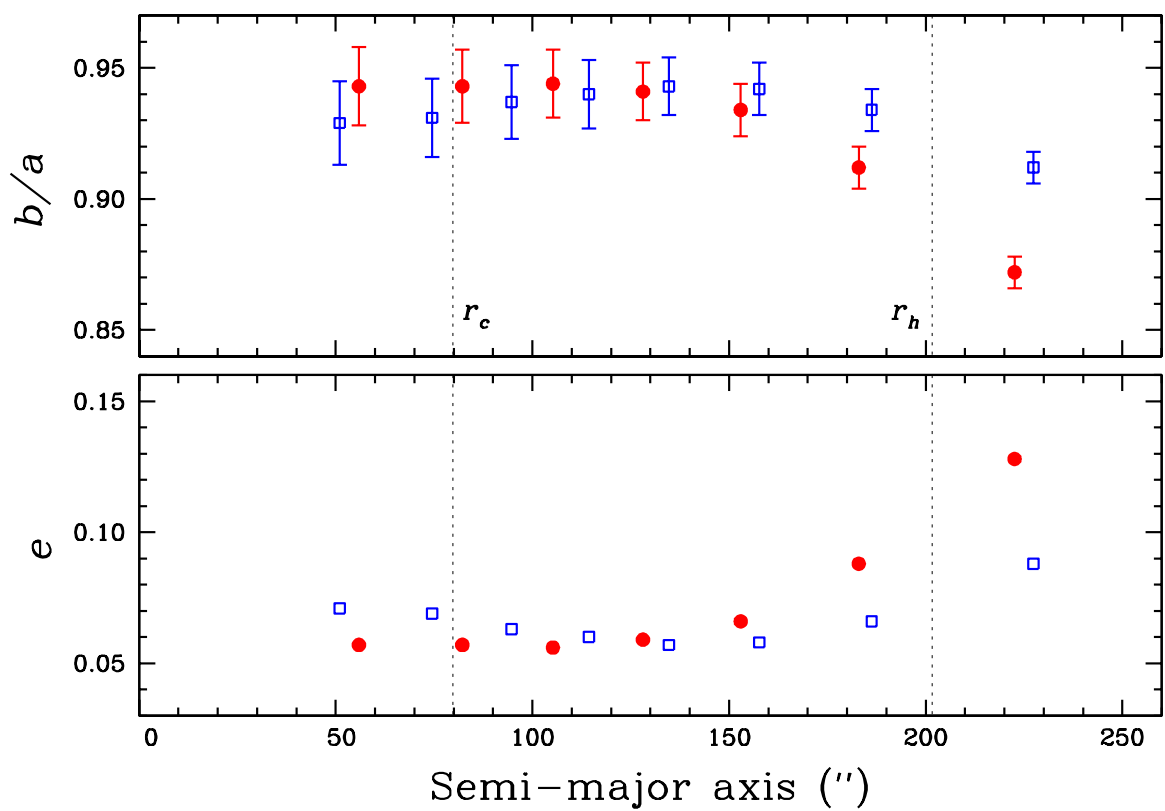}
\caption{The run of the axial ratio, $b/a$, the ellipticity, $e$ ($= 1 - b/a$)
of the \caw\ (blue) and the \cas\ (red) groups 
against the semi-major axis, $a$.
The radial distributions for each group bifurcate at the radial distance larger
than $a$ $\gtrsim$ 150 arcsec, in the sense that the \cas\ group 
is more elongated.
}\label{fig:axis}
\end{figure}

\begin{figure}
\epsscale{1}
\figurenum{26}
\plotone{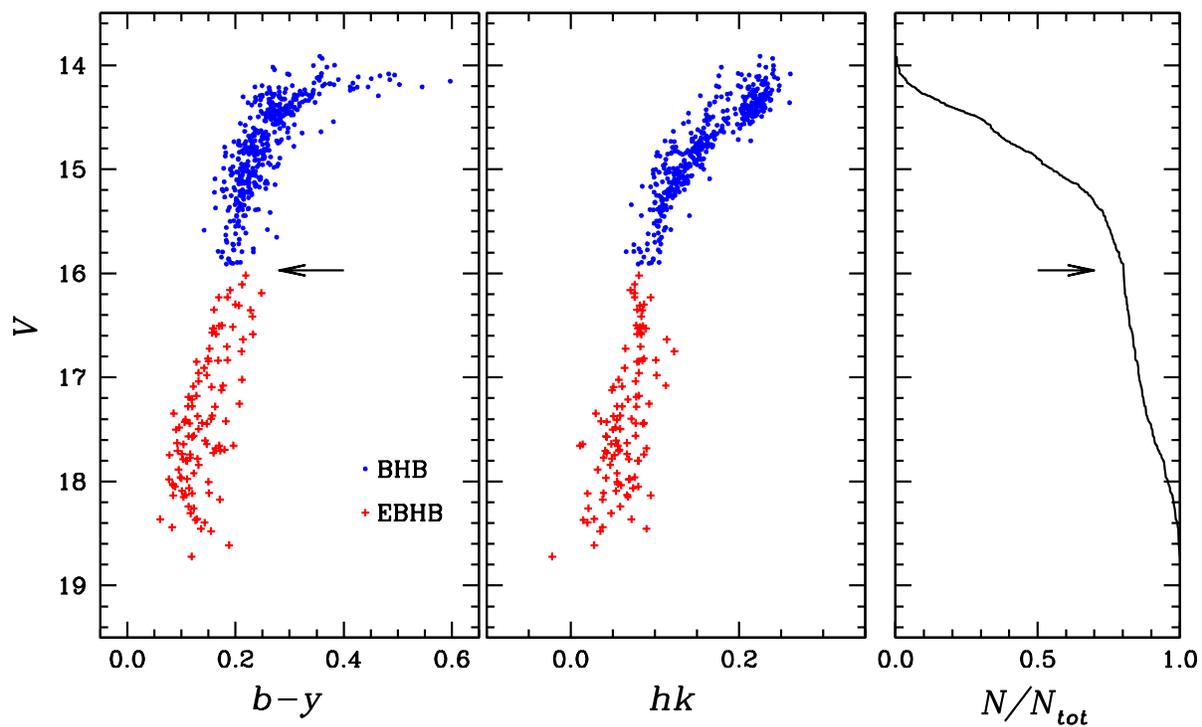}
\caption{CMDs of the HB region of M22 and the cumulative LF of HB stars. 
The HB gap at $V$ = 15.97 mag, with the zero slope in the cumulative LF of
HB stars, is denoted by arrows.
The BHB and EBHB stars are denoted by
blue filled circles and red plus signs, respectively.}\label{fig:hbcmd}
\end{figure}

In our photometric data, the observed number ratio between the BHB and
the EBHB stars becomes \nhb\ = 80:20 ($\pm$ 9), marginally in agreement with
those of the RGB populations, \nrgb\ = 69:31 ($\pm$ 3),
and the SGB populations, \nsgb\ = 69:31 ($\pm$ 6), without the correction
for the evolutionary effects.
With the correction of the evolutionary effect,
the HB number ratio does not agree with that of SGB, \nsgb\ = 62:38 ($\pm$ 6).
Note that our HB number ratio includes the completeness correction that shown
in Figure~\ref{fig:complete}. 
We calculated the average completeness fraction as a function of $V$ magnitude
using those from the central and outer part of the cluster 
as shown in Figure~\ref{fig:complete}, and we applied the correction factor
to individual HB stars by interpolating the average completeness fraction
against $V$ magnitude.
As discussed above, our photometry is complete down to $\approx$ 18.5 mag.
Therefore, only the the EBHB population is insignificantly affected 
($\approx$ 0.6 \%) by the completeness correction.

\begin{figure}
\epsscale{1}
\figurenum{27}
\plotone{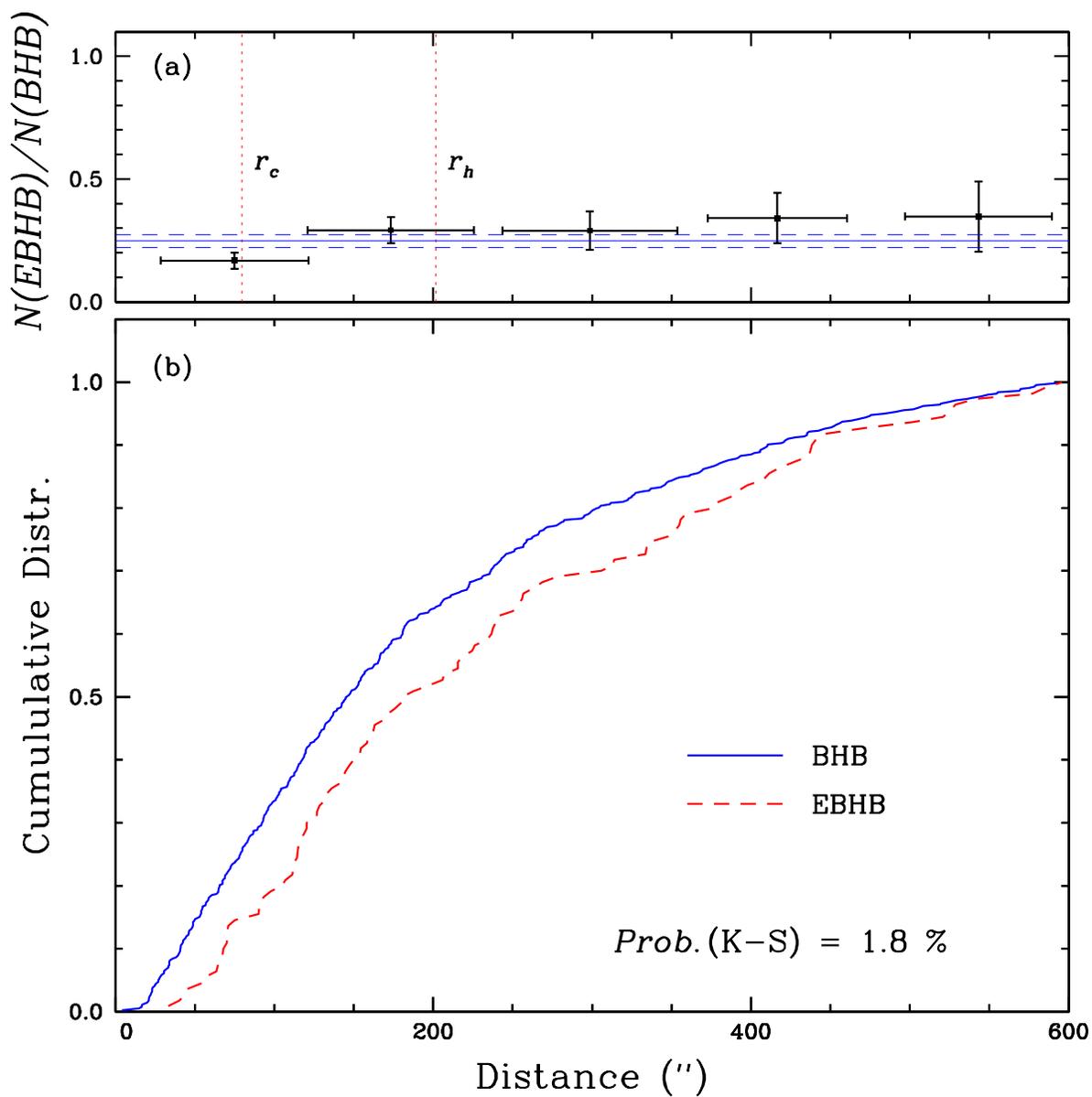}
\caption{Same as Figure~\ref{fig:rgbdist}, but for the BHB and EBHB stars in M22.
Similar to the \caw\ RGB population, the BHB population is more
centrally concentrated.
}\label{fig:hbratio}
\end{figure}

In the upper panel of Figure~\ref{fig:hbratio}, we show the number ratios
of the EBHB to the BHB populations against the radial distance from the center. 
Although, the HB number ratios against the radial distance
are in agreement within measurement errors 
up to $\approx$ 3 times of the half-light radius of the cluster,
the mean HB number ratio appear to have a weak gradient against
the radial distance in the sense that the BHB stars 
are slightly more centrally concentrated
than the EBHB stars, similar that can be seen in the RGB stars
in Figure~\ref{fig:rgbdist}.
In the bottom panel of the figure, we show the cumulative distributions
of the BHB and the EBHB populations against the distance from the center.
The BHB population is more centrally concentrated than the EBHB population,
qualitatively consistent with that can be seen in the RGB populations.
The K-S test indicates a probability of 1.8\% that two HB populations 
are drawn from the same parent population,
suggesting that they have different parent populations 
as can be seen in the RGB stars.

As have done for the RGB and SGB populations, 
due to different masses and chemical compositions,
different lifetimes of individual HB stars  should be taken into account.
We compared the lifetimes between the BHB and the EBHB populations
using the HB tracks by \citet{joo13}
and obtained the correction factor in terms of the number of HB stars,
\nhb\ = 56.5:43.5 ($\pm$ 4.0).
Because the EBHB stars live longer, one naturally expects to observe
a larger number of EBHB stars if initially even numbers of stars
are distributed both on the BHB and on the EBHB locations.

In addition, the number of HB stars evolved from the \caw\ and the \cas\
RGB populations will be slightly different since the RGB-tip masses 
between the two groups will be different.
If the \cas\ population has the higher helium abundance
by $\Delta Y$ = 0.09 \citep{joo13}, the evolution of stars in the \cas\ population
would be faster than those in the \caw\ population due to increasing
mean molecular weight, resulted in smaller RGB-tip masses
in the \cas\ population.
With the HB mass dispersion adopted by \citet{joo13}, 
$\sigma_M$ = 0.023$M_\odot$ for both populations,
we calculated the number of stars within 4$\times\sigma_M$ from the RGB-tip
masses\footnote{This mass range is rather arbitrary, but the final result 
does not significantly depend on the adopted mass range if the same mass
dispersions for both groups are maintained.}
of both isochrones using the same evolutionary population synthesis model
that we constructed previously.
For this purpose, we populated 10$^7$ artificial stars in each group
using the Salpeter's IMF and we generated 50 different sets for both groups.
We obtained the correction factor, \nhb\ = 58.9:41.1 ($\pm$ 0.1),
in the sense that the observed number of HB stars that evolved from
the \cas\ population gets larger 
since the RGB-tip mass of the \cas\ is smaller owing to its faster
evolution from the MS through the RGB phases.
If we take these effects into consideration, the number ratio between
the BHB and the EBHB stars becomes \nhb\ = 88.2:11.8 ($\pm$ 9.8),
significantly different from those of the RGB and the SGB populations.
It should be noted that this result is based on the assumption that
the BHB stars are the progeny of the \caw\ RGB stars while
the EBHB stars are the progeny of the \cas\ RGB stars 
as proposed by \citet{joo13}.

Our result may suggest that the connection between the RGB and HB populations 
may not be as clear as that \citet{joo13} claimed.
Note that \citet{joo13} adopted the number ratio between the BHB and EBHB 
of 70:30 in their synthetic HB model construction and they claimed that 
their HB number ratio does not disagree with the previous estimates 
of RGB or SGB number ratios between the two populations by others.
It is thought that \citet{joo13} overestimated the EBHB fraction in their
calculation compared to our observation.
As \citet{joo13} suggested, if the EBHB population is the only progeny of
the second generation of the RGB population (i.e.\ the \cas\ population),
the discrepancy in the number ratios between the two populations
in the HB and SGB/RGB is difficult to explain.
For instance, it is evident that the $R$ value, the number ratio of the HB 
to the RGB stars \citep{caputo87}, of each population is against
the theoretical expectation.
The observed $R$ value of a simple stellar population is 
the measure of the helium abundance in the sense that
the $R$ value increases with the helium abundance.
We define the {\em modified} $R$ value which includes
RGB stars from $V_{\mathrm HB}$ ( =  14.15 mag for M22) to 12.0 mag 
(not the tip of the RGB sequence), since we already selected 
bright RGB stars with 12 mag $\leq V \leq$ 16.5 mag
in both populations as discussed above.
Then we have the {\em modified} $R$ values of 1.73 $\pm$ 0.14 for
for the BHB versus the \caw\ RGB populations and 
1.21 $\pm$ 0.17 for the EBHB versus the \cas\ RGB populations.
If the EBHB and the \cas\ RGB populations are the second generation of the stars 
in the cluster and they formed out of interstellar gas enriched in helium 
by the first generation of stars, one would naturally expect to have 
a larger $R$ value for the EBHB and the \cas\ RGB populations.
However, the observed $R$ value for the EBHB and the \cas\ RGB populations
is smaller than that for BHB and the \caw\ RGB populations and 
the observe $R$ values are opposite of the theoretical expectation.
Our result may suggest that, like other lighter elemental abundances,
there may exist a spread in the helium abundance of the second generation of stars
and at least about the half of the \cas\ RGB stars must have evolved into 
the BHB sequence (which requires no or little helium enrichment)
and a significant fraction of the \cas\ RGB stars failed to become EBHB stars 
(which requires the helium enrichment of
$\Delta Y \approx$ 0.09, see \citealt{joo13}) in M22.

It would be very interesting to point out the recent result by \citet{gratton14}.
They carried out the spectroscopic survey of the BHB stars in M22
and found a significant fraction ($\approx$ 18\%)
of the metal-rich helium-enhanced BHB stars in their sample 
(Group 3 designated by \citeauthor{gratton14}).
If we consider that the Group 3 BHB stars by \citet{gratton14} 
are the progeny of the \cas\ RGB stars, the number ratio
between the second generation and the first generation of HB stars
becomes 34:66 and this number ratio is in good agreement
with those from the RGB and the SGB stars.
Perhaps, this is also hinted by the separate Na-O distribution of M22 RGB stars,
indicative of internal He spread in both the \caw\ and the \cas\ groups.
If so, the \cas\ population of M22 may be exactly what we observe in NGC~6752, 
which contains both the BHB and the EBHB populations maintaining 
a rather monotonous metallicity distribution.

\subsection{Radial velocity and velocity dispersion}\label{sec:4.5}
Recently, \citet{lane09} performed a wide-field spectroscopic survey
of bright RGB stars in M22 using AAOmega mounted 
on the 3.9-m Anglo-Australian Telescope. 
Their primary concern was to test Newtonian gravity by investigating
velocity profiles of several globular clusters, including M22, at larger radii.
We made use of their radial velocity data, which were kindly provided
by Dr.\ Kiss, in order to investigate the differences in kinematics
of the multiple stellar populations in M22.

We compared our coordinates of \caw\ and \cas\ RGB stars 
with those of \citet{lane09}
and we selected common stars matched within a radius of 2\arcsec.
During our source matching process, we excluded stars having
uncertainties in the radial velocity measurement larger than 10 \kms\
tagged by \citet{lane09}. Through this process, 
we have 208 \caw\ and 87 \cas\ RGB stars.

\begin{figure}
\epsscale{1}
\figurenum{28}
\plotone{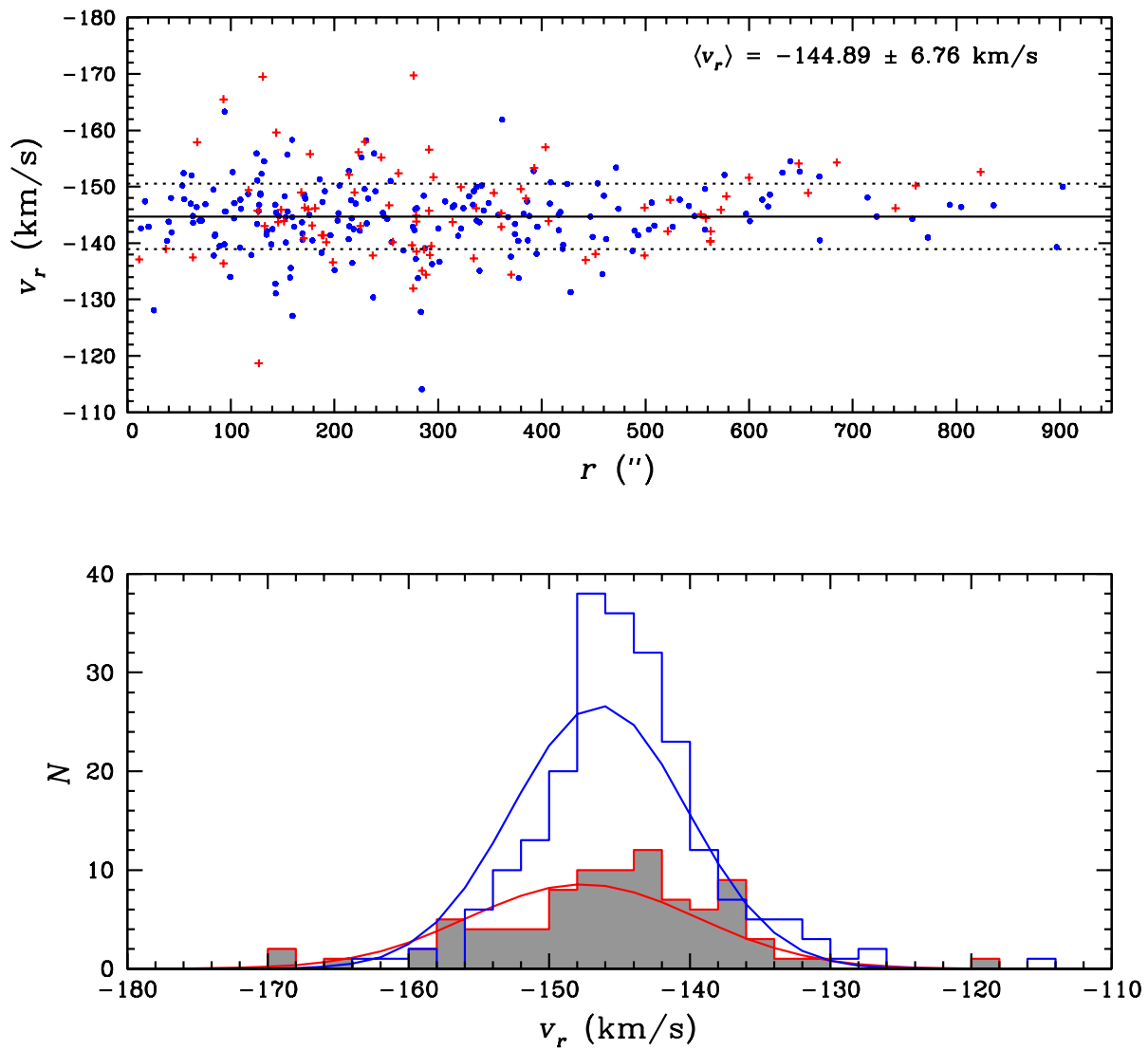}
\caption{
(Upper) The distribution of radial velocities of individual stars against 
the radial distance from the center.The blue color represents the \caw\ group
and the red color the \cas\ group. No radial gradient can be seen.
(Lower) Histograms for the radial velocities of each group with gaussian fits.
The mean radial velocities and the velocity dispersions are very similar
between the two groups.}\label{fig:rv}
\end{figure}

We show the radial velocity of individual stars
in each group against the radial distance from the center
in the upper panel of Figure~\ref{fig:rv}.
We obtained the mean radial velocity of the cluster,
$\langle v_r\rangle = -144.89 \pm 6.76$ \kms.
The distributions of the radial velocity between the two populations are in
good agreement. 
Also the distribution of the radial velocity
of each population does not vary with the radial distance from the center.

Using these stars, we calculate the average radial velocities 
and velocity dispersions for \caw\ and \cas\ groups as given by \citet{pm93},
assuming that each velocity $v_i$ is drawn from the normal distribution,
\begin{equation}
 f(v_i) = \frac{1}{\sqrt{2\pi(\sigma^2_c + \sigma^2_{e,i})}}
 \exp{\left[-\frac{(v_i-v_r)^2}{2(\sigma^2_c + \sigma^2_{e,i})}\right]},
\end{equation}
where $\sigma_{e,i}$ is the known measurement uncertainty of $v_i$ and
$v_r$, $\sigma_c$ are the average radial velocity and
the intrinsic velocity dispersion of each group.
For our calculations, we used a FORTRAN program kindly provided 
by Prof.\ Pryor, obtaining
$v_r$ = $-$144.59 $\pm$ 0.40 \kms\ and
$\sigma_v$ = 4.71 $\pm$ 0.35 \kms\ for the \caw\ group and
$v_r$ = $-$145.77 $\pm$ 0.89 \kms\ and  
$\sigma_v$ = 7.55 $\pm$ 0.66 \kms\ for the \cas\ group.
Our results are shown in the bottom panel of Figure~\ref{fig:rv}.
The mean velocities of both groups are in good agreement,
while the velocity dispersion for the \cas\ group appears to be
slightly larger than that of the \caw\ group. 
Note that \citet{marino14b} studied the velocity dispersions between
the two SGB in NGC~1851 and they did not find any difference.

\subsection{Projected Rotations}
\citet{pc94} first found a clear evidence of the rotation of M22 from both 
proper motions and radial velocities of bright stars in the cluster,
with the amplitude of the mean rotation of about from $\approx$ 6 \kms\ for
$1\arcmin \leq r \leq 3\arcmin$ to $\approx$ 3 \kms\ for
$3\arcmin \leq r \leq 7\arcmin$.
Their result is an unique example of obtaining 
the rotational velocity of GCs using proper motion data.
Later \citet{lane09,lane10} confirmed that M22 has a substantial rotation 
among other GCs that they studied.
In this section, we investigate the mean rotation of RGB stars of 
the double populations in M22 using three methods.

\subsubsection{Radial velocity versus position angle}\label{sec:4.6.1}
\begin{figure}
\epsscale{1}
\figurenum{29}
\plotone{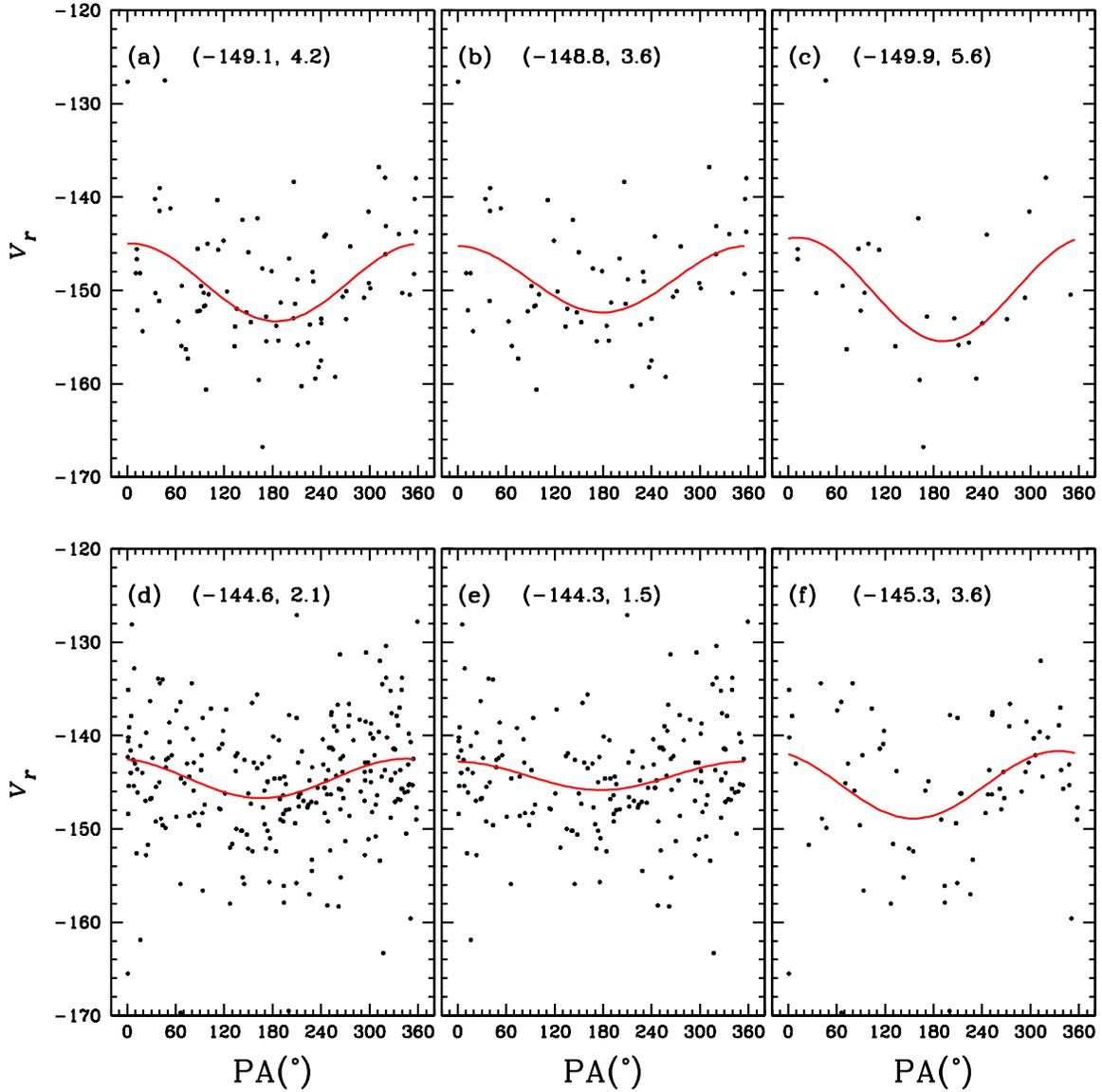}
\caption{Radial velocities of individual RGB stars in M22 against
their position angles. The two numbers in the parentheses in each plot denote
the mean radial velocity and the amplitude of the mean rotational velocity.
The red solid lines are sinusoidal fits to the data.
(a) All RGB stars; (b) \caw\ stars; and (c) \cas\ stars 
with radial velocities measured by \citet{pc94}.
(e) -- (f) As (a) -- (c) but using stars measured by \citet{lane09}.
It is evident that the \cas\ group rotates 
faster than the \caw\ group does.}\label{fig:pavr}
\end{figure}

First, we estimated the mean rotation of the cluster by using the radial
velocity distribution against the position angle of individual stars \citep{pc94}.
In Figure~\ref{fig:pavr}, we show radial velocities of individual stars 
in each group against their position angles measured from North.
We chose the RGB stars\footnote{\citet{pc94} also studied the HB stars 
in the cluster, of which we do not make use in this study. 
Therefore, the mean velocity of the cluster and the number of stars being used 
in this study are different from those in \citet{pc94}.} 
with the proper motion membership probability larger than 90\% from 
\citet{pc94} 
and we merge their radial velocity measurements with our photometric data.
We show radial velocities from \citet{pc94} against
the position angle for individual RGB stars in Figure~\ref{fig:pavr}-(a).
We performed a sinusoidal fit to the data and we obtained
the amplitude of the mean rotation of all RGB stars is about 4.2 \kms\,
which is comparable to that from \citet{pc94}, 3.8 \kms\ \citep{mh97}.

Next, we investigate the mean rotational velocity of the \caw\ and 
the \cas\ groups. We divided RGB stars from \citet{pc94} into two groups 
based on the $hk$ index of individual stars in Figure~\ref{fig:rgbpop} and
we show plots of radial velocities against the position angle
of RGB stars in the \caw\ and the \cas\ groups
in Figure~\ref{fig:pavr}-(b) and -(c).
The amplitude of the sinusoidal fit to the \caw\ group is
about 3.6 \kms, while that to the \cas\ group is about 5.6 \kms.
The amplitude of the projected rotation of the \cas\ group is larger
than that of the \caw\ group.

We carried out the same procedure using the radial velocities by \citet{lane09}.
As shown in the bottom panels of Figure~\ref{fig:pavr},
similar results can be found from the radial velocity data by \citet{lane09},
although the mean radial velocities and the amplitudes of the mean rotation
are different from those from \citet{pc94}.
We obtained the amplitude of mean rotation for the \caw\ group is
about 1.5 \kms, while that for the \cas\ group is about 3.6 \kms.
Again, the amplitude of the mean rotation for the \cas\ group is larger than
that for the \cas\ group.

It should be noted that the difference in the mean radial velocities
of the cluster using the data from \citet{pc94}, $-$149.1 \kms, 
and \citet{lane09}, $-$144.6 \kms, may suggest that 
two sets of radial velocity data are heterogeneous.
Therefore, we did not attempt to merge two sets of data together to
investigate the rotation of the cluster.
The amplitude of the mean rotation using the data by \citet{lane09} is smaller
than that by \citet{pc94}.
Since the number of RGB stars being used is smaller for the case of 
using data by \citet{pc94}, it is suspected that a few stars with large
deviations from the mean velocity can affect the results.
In the upper panel of Figure~\ref{fig:pavr}, the radial velocities of 
three RGB stars deviate more than 3 $\sigma$ level from the mean value;
IV-20 ($-$166.78 \kms), 2-73 ($-$127.65 \kms), and I-27 ($-$127.51 \kms).
If we exclude these three stars in our calculations,
the amplitudes of the mean rotations become 2.7 \kms, 2.5 \kms\ and 3.3 \kms\,
for all RGB stars, the \caw\ and the \cas\ groups, respectively.
These values become comparable to those using data from \citet{lane09}.

\subsubsection{Radial velocity difference between two hemispheres}
\label{sec:4.6.2}
Next, we estimated the amplitude of the mean rotation of the cluster 
using the method described by \citet{lane09}.
Assuming an isothermal rotation, the mean rotation can be measured by dividing
the cluster in half at a given position angle and calculating differences
between the average velocities in the two halves.
We repeated this calculation by increasing the position angle of the boundary
of the two halves by 10$^\circ$.
Then the net rotation velocity is the half of the amplitude 
of the sinusoidal function in the differences between the average velocities
in the two halves.

\begin{figure}
\epsscale{1}
\figurenum{30}
\plotone{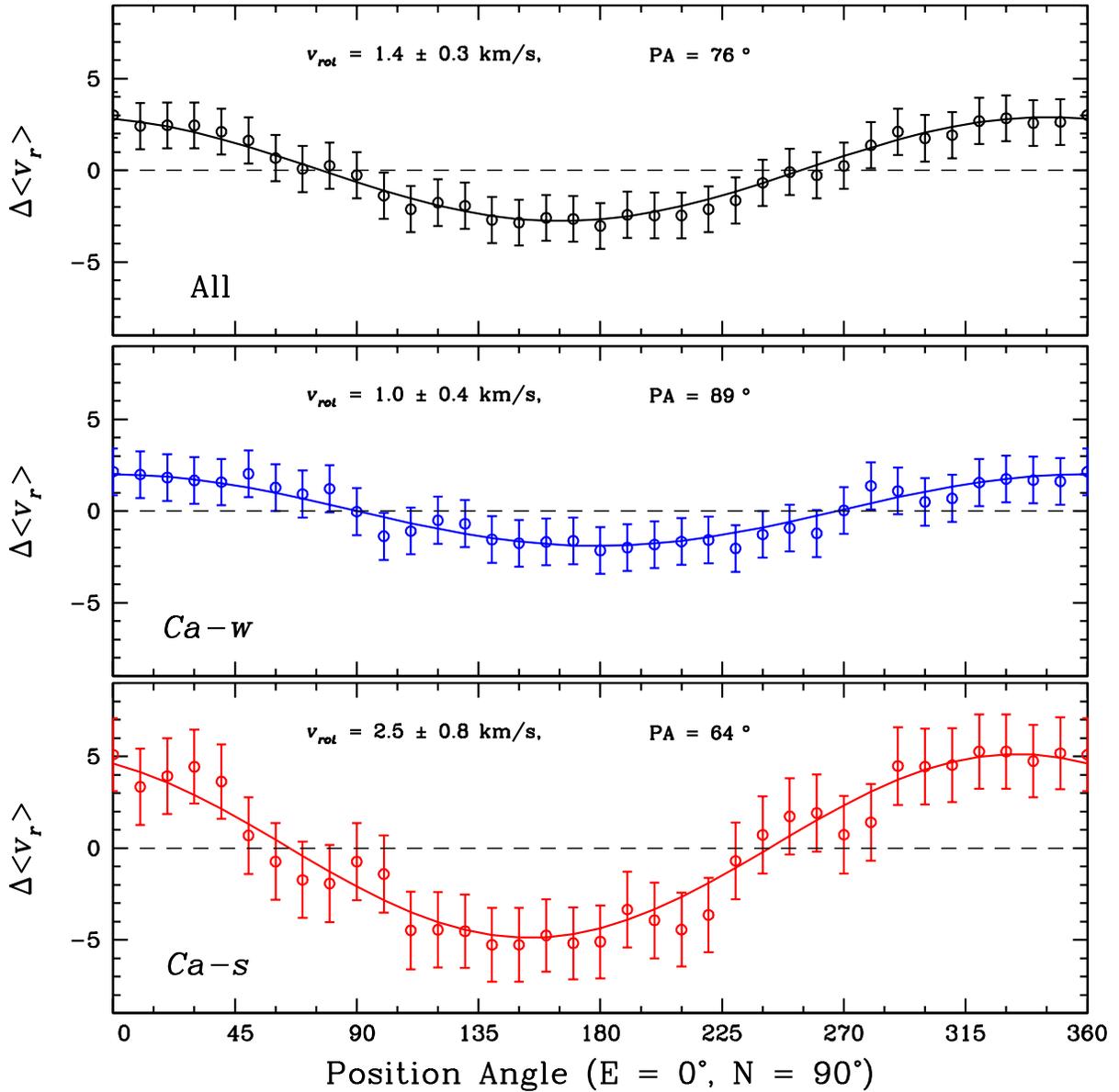}
\caption{The difference in the mean radial velocities between both hemispheres
against its position angle. Again, it is evident that the \cas\ group 
appear to rotate faster than the \caw\ group does.
The position angle refers to that of the equator of the rotation, i.e.\ 
the axis perpendicular to the rotation axis.
The error bar is the error of the mean.
}\label{fig:rotiso}
\end{figure}

We show the differences in the mean radial velocities as a function of
the position angle (East = 0$^\circ$ and North = 90$^\circ$)
along with the best-fitting sine function in Figure~\ref{fig:rotiso}.
Our result for all RGB stars in the cluster is shown in the top panel
of the figure.
We obtained the mean rotation velocity of 1.4 $\pm$ 0.3 \kms\ and 
the position angle of the equator (i.e.\ the axis perpendicular 
to the rotation axis) of the rotation of 81$^\circ$ 
(i.e.\ $\approx$ North), which are very consistent with 
those of previous studies by \citet{lane09,lane10},
1.5 $\pm$ 0.4 \kms\ and approximately North-South, respectively.

In the middle and the bottom panels of Figure~\ref{fig:rotiso},
we show plots of differences in the mean radial velocities for the \caw\
and the \cas\ groups. 
We obtained the mean rotation of 1.0 $\pm$ 0.4 \kms\ with 
the position angle of 86$^\circ$  for the \caw\ group and
2.5 $\pm$ 0.8 \kms\ with the position angle of 73$^\circ$ for the \cas\ group.
Again, the amplitude of the mean rotation of the \cas\ group is 
slightly larger than that of the \caw\ group is.

Note that the mean rotations of this method are smaller than those of 
the sinusoidal fit to individual stars on the radial velocity versus
position angle. 
As we have demonstrated above, the sinusoidal fit to the individual stars
suspects to be vulnerable to stars with large velocity deviations from
the mean value. 
This may explain the difference in the amplitudes of the mean rotation 
between the two methods.

\subsubsection{Radial velocity profile}\label{sec:4.6.3}
\begin{figure}
\epsscale{1}
\figurenum{31}
\plotone{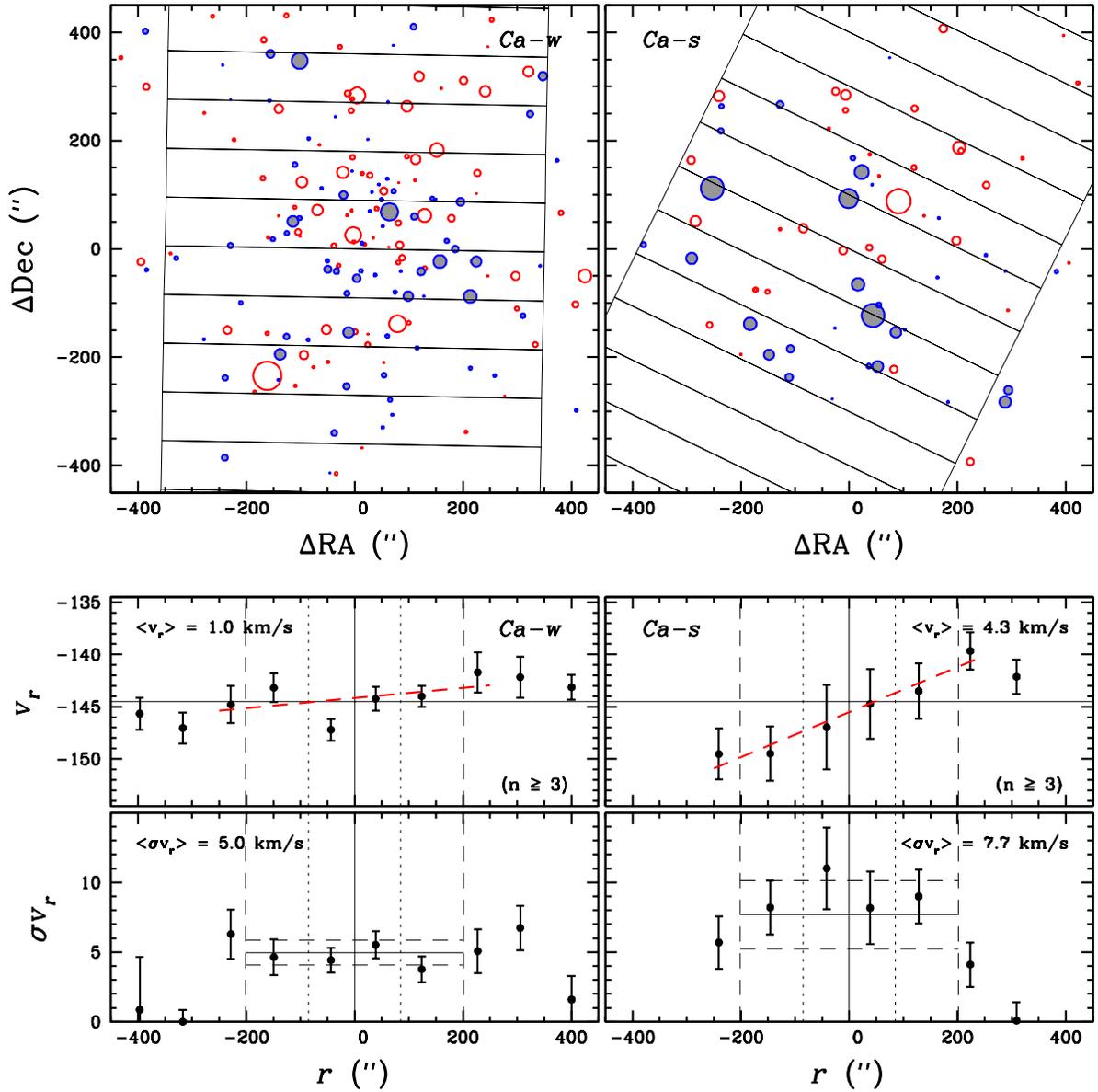}
\caption{The upper panels show the distributions of RGB stars with radial
velocity measured by \citet{lane09}. The blue color denotes the blue
shift and the red color the red shift. The size of each circle indicates
the difference between the velocities of individual stars and the mean velocity.
The bottom panels show the radial velocities and the velocity dispersions
against the projected distance on the equator of the rotation.
In the Figure, the position angles of 86\degr\ and 73\degr\ 
for the \caw\ and the \cas\ groups, respectively, are used.
In the middle panels, red dotted lines represent the mean rotation.
The maximum rotation velocities at the half-light radius
are 1.0 \kms\ for the \caw\ and 4.3 \kms\ for the \cas\ groups.
The velocity dispersion of the \cas\ group is slightly larger but
those of both groups are in agreement within the errors.
}\label{fig:rot}
\end{figure}

Finally, we examine the rotation of the cluster by using the radial velocity
profile along the axis perpendicular to the mean rotation.
In the top panel of Figure~\ref{fig:rot}, we show the distributions  of 
RGB stars in the \caw\ and \cas\ groups with the radial velocity measurements.
In the figure, blue and red colors denote blue-shift and red-shift 
from the mean radial velocity of the cluster, respectively, and
the size of each circle represents the velocity deviation from 
the mean value of the cluster.
We also show grids for calculating the group velocity and 
the velocity dispersion with
rectangles with the width of 800 arcsec and the height of 100 arcsec.
We adopt the tilted angle of the grid rectangles 
that we obtained in \textsection\ref{sec:4.6.2},
89$^\circ$  for the \caw\ group and 64$^\circ$ for the \cas\ group.

We calculated the the mean radial velocity and the velocity dispersion
in each grid rectangle.
Our results are shown in the bottom panels of Figure~\ref{fig:rot}. 
Inside the half-radius of the cluster, the velocity dispersion of
the \cas\ population appears to be slightly larger, consistent
with our previous result presented in \textsection\ref{sec:4.5}.
Also, the radial velocity profile against 
the projected distance perpendicular to the rotation axis are different.

We performed a least square fit to the line of sight velocity versus
the projected distance perpendicular to the rotation axis.
We found that the maximum velocity of 1.0 \kms\ for the \caw\ group 
and 4.3 \kms\ for the \cas\ group at the half-light radius of the cluster.
Again, the RGB stars in the \cas\ group appear to rotate faster
than those in the \caw\ group do, consistent with our previous results.

We examined the anisotropy parameter of the cluster 
\citep[see, for example,][]{bender91},
\begin{equation}
\left(\frac{V_\mathrm{rot}^\mathrm{max}}{\sigma_\mathrm{m}}\right)^* =
\left(\frac{V_\mathrm{rot}^\mathrm{max}}{\sigma_\mathrm{m}}\right)\sqrt{(1-e)/e}
\end{equation}
where $\sigma_\mathrm{m}$ is the mean velocity dispersion and $e$ is 
the ellipticity of the cluster. 
Note that \vsigstar\ $\ll$ 1
indicate strong anisotropy.
Using the relation above, we obtained the anisotropy parameter 
\vsigstar\ = 0.707 for the \caw\ group and 1.62 for the \cas\ group,
where we adopted $e$ = 0.074, \vmax\ = 1.0 \kms\ and  
$\sigma_\mathrm{m}$ = 5.0 \kms\  for the \caw\ group and 
$e$ = 0.106, \vmax\ = 4.3 \kms\ 
and $\sigma_\mathrm{m}$ = 7.7 \kms\ for the \cas\ group.
For ellipticity, we used the simple mean of those given in
Table~\ref{tab:fit} within about 210 arcsec from the center 
(i.e.\ about the half-light radius of the cluster).
The large values of anisotropy parameters for both populations suggest that
the flattening observed in M22 is most likely originated from 
the presence of internal rotation.

\section{SUMMARY}
We presented new wide-field ground-based \caby\ photometry 
and the HST WFC3 photometry of the peculiar GC
M22 (NGC~6656), confirming our previous results that 
M22 has a distinctive RGB split in the $hk$ index, 
mainly due to the bimodal distribution in heavy elemental abundances, 
especially in calcium.

Our ground-based photometry is complete down to $V$ $\approx$ 18.5 mag
both in the central and the outer part of the cluster and incompleteness
at the fainter magnitude regime does not affect our results presented in this paper.
Although M22 is located toward the Galactic bulge, the contamination
from the off-cluster field populations appears to be negligible
when our cluster membership star selection scheme based on 
our multi-color photometry is applied.

Similar to our previous work \citep{jwlnat}, we defined the \caw\ and the \cas\
RGB groups based on the $hk$ index values at a given $V$ magnitude.
We obtained the apparent number ratio of \nrgb\ = 69:31 ($\pm$ 3), 
without the correction for the differential evolutionary effects due to 
the heterogeneous elemental abundances
and slightly different ages between the two groups \citep{marino11,joo13}.
After applying such correction, the intrinsic or the true number ratio becomes
\nrgb\ = 70:30 ($\pm$ 3).

Using the HST/WFC3 data of the central part of the cluster, 
we obtained the apparent SGB number ratio of \nsgb\ = 69:31 ($\pm$ 6) 
without the correction for the differential  evolutionary effects and 
the true SGB number ratio of 62:38 ($\pm$ 6)  with the correction, 
which is in excellent agreement with previous result
from the HST observations in other passbands by \citet{marino09}.

We showed that the RGB split in the $m1$ index in M22 is not mainly due to
the difference in the CNO abundances, but in the heavy metal abundances.
Remarkable resemblance in the HB morphology and the double RGB sequences 
both in the $m1$ and the $hk$ indexes between M22 and the composite GC
using M55 and NGC~6752 (each of which has similar metal abundances 
as the \caw\ and the \cas\ groups in M22) may support our idea.
Therefore, it is thought that the origin of the RGB split in the $m1$ index
in M22 is the same as that in the $hk$ index: 
the difference in the mean heavy metal abundances.

Comparisons with the chemical abundance measurements of RGB stars 
from previous studies showed that the $hk$ index indeed a measure
of calcium abundance at a given luminosity. 
Our $hk$ measurements are well correlated with \caii\ H and K lines, 
infrared \caii\ triplet lines and visual \ion{Ca}{1} lines by others.
Therefore, the $hk$ index can provide an efficient means to distinguish 
the difference in the calcium abundance and furthermore 
the multiple populations in GCs and nearby dwarf spheroidal galaxies.

Using homogeneous CN and CH band strength measurements by \citet{smolinski11},
we showed that the gradients in $\delta$CN and $\delta$CH against metallicity
are naturally expected to occur, when the common lower envelope for GCs with
heterogeneous metallicity is adopted.
In this regard, we showed that the CN-CH positive correlation in M22 
reported by \citet{lim15} is most likely due to metallicity effects 
between the two RGB groups with heterogeneous metallicities in M22.
However, when corrected with its own lower envelope,
each RGB group keeps maintaining its own CN-CH anti-correlation, 
showing a well-known characteristics of normal GCs.
This may suggest that the merger of two GCs with different chemical compositions
would be the most simplest explanation.

Our $cy$ index measurements for each RGB group in M22 confirmed
that there exists a significant spread in the nitrogen abundance among RGB stars
in both groups. Quantitatively, the $cy$ index appears to over-estimate 
the nitrogen abundance when compared with the results from
the high-resolution spectroscopy by \citet{marino11}.

We found that the mean RGB bump $V$ magnitude, $V_{\mathrm bump}$,
of the \caw\ group is about 0.15 mag brighter than that of the \cas\ group,
which is consistent with the idea that the difference in $V_{\mathrm bump}$
between the two groups is mainly due to the difference in metallicity.
Furthermore, the tilted $V_{\mathrm bump}$ magnitude level 
against $\Delta hk$ for the \caw\ group can be seen our data,
suggesting that the $hk$ index is indeed a good measure of metallicity
and the RGB bump luminosity decreases with increasing metallicity.
It is an widely accepted fact that the formation of the EBHB stars may
require a helium enhancement \citep[e.g.][]{dantona02}.
If the \cas\ group has an enhanced helium abundance by $\Delta Y$ = 0.09
as \citet{joo13} proposed, the brighter $V_{\mathrm bump}$ magnitude 
by $\Delta V$ = 0.08 mag is expected for the \cas\ group.
But this is not the case and the difference 
in the $V_{\mathrm bump}$ magnitude within the context
of the enhanced helium abundance of the \cas\ group
is difficult to explain.

With the carefully selected sample of RGB stars,
we derived the coordinates of the center of both RGB populations
using three different methods, finding insignificant differences
in the coordinates of the center of both populations.
The coordinates of the center of each group appear to
be slightly different, but the difference is not substantially large
to claim that the centers of two groups are distinctively different.
On the other hand, the radial and spatial distributions between
the two groups are different.
The \caw\ group appears to be slightly more centrally concentrated while
the number ratio between the \caw\ and the \cas\ groups are
almost flat within the measurement errors up to $\approx$ 3 -- 4 half-mass radii,
a reminiscence of the peculiar GC NGC~1851.
The spatial distribution of the \cas\ group is very similar as 
that of the \caw\ group up to $\approx$ 1.5 core radii,
while the more elongated spatial distribution of the \cas\ group
at larger radii can be seen.

The similarity and the difference between the RGB and the HB populations 
could be the critical puzzle to understand the formation of the EBHB stars.
The BHB stars in M22 are thought to be the canonical progeny of 
the metal-poor \caw\ RGB population with a normal helium abundance while
the EBHB stars are thought to be that of the metal-rich \cas\ RGB population 
with an enhanced helium abundance \citep{joo13}.
Our result showed that the BHB population is more centrally
concentrated than the EBHB population,
qualitatively consistent with the result from the RGB populations.
However, the number ratio between the BHB and the EBHB  populations is
significantly different from that of the RGB and the SGB populations,
\nhb\ = 80:20 ($\pm$ 10)  without the correction for the differential 
evolutionary effects and \nhb\ = 88:12 with such correction.
With this HB number ratio, we showed that the comparison of 
the modified $R$ values of the two groups is against
the theoretical expectation that the \cas\ + EBHB population
(G2 designated by \citeauthor{joo13})  is more helium enhanced
than the \caw\ + BHB population (G1) as \citet{joo13} proposed.

Our results for the HB number ratio and the modified $R$ values
may indicate that the EBHB is not the only progeny of the \cas\ RGB group.
In this regard, the recent spectroscopic study of  the HB stars in M22 
by \citet{gratton14} is very intriguing as we have mentioned previously.
They carried out the spectroscopic survey of the BHB stars in M22
and found a significant fraction ($\approx$ 18\%)
of the metal-rich helium-enhanced BHB stars in their sample 
(Group 3 designated by \citeauthor{gratton14}).
If we consider the Group 3 BHB stars by \citet{gratton14} 
as the progeny of the \cas\ RGB stars, the apparent HB number ratio
equivalent to the \caw\ and the \cas\ RGB groups
becomes 66:34 and this number ratio between the two different HB groups
is in good agreement with those from the RGB and the SGB stars.\footnote{Recall 
that the \cas\ group in M22 may be equivalent to NGC~6752, which contains 
both the BHB and the EBHB stars.}
However, this does not explain why some of the metal-rich helium-enhanced
RGB stars end up with the BHB stars, while others end up with the EBHB stars.
Perhaps, may the discrepancy in the number counts indicate that there exists
a non-canonical channel of the evolution of the EBHB stars without
invoking the helium enhancement as proposed by \citet{moehler11}?
Or, may this indicate that there exist the substantial spread or 
the bimodal distribution in helium abundance of the second generation of stars?
Or, perhaps may the dispersion in the mass loss during the RGB phase
be much larger than that of the canonical channel?
These are the questions should be answered in the future.

We measured the mean radial velocities and velocity dispersions of 
of each RGB population using the results from a wide-field spectroscopic
survey by \citet{lane09}.
We obtained
$v_r$ = $-$144.59 $\pm$ 0.40 \kms\ and
$\sigma_v$ = 4.71 $\pm$ 0.35 \kms\ for the \caw\ RGB population and
$v_r$ = $-$145.77 $\pm$ 0.89 \kms\ and  
$\sigma_v$ = 7.55 $\pm$ 0.66 \kms\ for the \cas\ RGB population.
The mean velocities of both populations are in good agreement,
while the velocity dispersion of the \cas\ population is slightly larger.

Finally, we derived the projected rotation velocities for each population
using three different methods.
Our rotation velocity measurement for all RGB stars is in good agreement
with previous measurements by others, ranging from 1.4 \kms\ to 2.1 \kms\
depending on the methods, suggesting that the elongated structure of M22 
is most likely due to the presence of internal rotation.
However, detailed comparisons between the \caw\ and the \cas\ groups
revealed that they have different kinematic properties.
The \cas\ RGB population (2.5 -- 4.3 \kms) appears to rotate faster 
than the \caw\ population (1.0 -- 1.5 \kms), 
with different position angles of the axis of the rotation,
89$^\circ$  for the \caw\ group and 64$^\circ$ for the \cas\ group.

\section{DISCUSSIONS}
The up-to-date observational evidences strongly suggest that the formation
history of M22 must have been different from that of normal GCs
in our Galaxy.
Two different formation scenarios have been frequently explored
in order to elucidate the nature of M22:
the self-enrichment and the merger scenarios
\citep[e.g.][]{jwlnat,marino11,marino12, joo13,lim15}.
Each hypothesis has pros and cons to explain the detailed
photometric and spectroscopic characteristics observed in M22.

In an attempt to delineate the chemical evolution between the two groups,
at least three or four sources for specific elemental abundance changes
should be involved:
(i) Core-collapsed SNe: Heavy metals including Ca and Fe, but not affect 
the $r$-process elements.
(ii) Intermediate-mass AGBs (IMAGBs) or FRMSs: Na-O, C-N anti-correlations.
The He and overall C and N abundance enhancements.
(iii) Low-mass AGBs (LMAGB): $s$-process elements.
As extensively discussed by \citet{marino11,marino12} and \citet{roederer11},
the self-enrichment formation hypothesis may require unlikely
complex fine-tuning assumptions.
The formation scenario of M22 in the context of the self-enrichment hypothesis
can be described in the following steps\footnote{The self-enrichment hypothesis
by \citet{joo13} is slightly different from that of \citet{marino11},
but it does not affect our discussions made in this paper.}
 \citep[e.g][]{marino11};
(1) The formation of the first generation (FG) of stars of the $s$-process-poor 
(or \caw) group (FGSP);
(2) SNe II explosions, which expelled material far from the center;
(3) IMAGBs of the FGSP polluted the intra-cluster medium, which went into 
the central region via cooling flow;
(4) The formation of the second generation (SG) of stars of the $s$-poor group (SGSP);
(5) LMAGBs of the FG expelled the ejecta with enhanced $s$-process elements;
(6) A second cooling flow, containing the ejecta from the LMAGBs and SNeII,
into the central part of the cluster;
(7) The formation of the FG of the $s$-process-rich (or \cas) group (FGSR);
(8) IMAGBs of the FGSR polluted the intra-cluster medium,
which went into the central region via cooling flow again;
(9) The formation of the SG of the $s$-rich group (SGSR);
(10) Loss of stars only in the FGSP and the FGSR
due to gas expulsion induced by SNe explosions.

The first problem with such self-enrichment hypothesis is the number ratios
between the FG and the SG in both the $s$-poor and $s$-rich groups.
The most conspicuous feature can be seen in the recent spectroscopic study
of the majority of normal GCs is that the SG is the major component of 
normal GCs, generally $\approx$ 50-80 \% \citep[e.g.][]{dantona08,carretta09}.
This cannot be easily explained with the chemical evolution with variant IMFs,
which led \citet{dercole08} to propose the loss of the vast amount of 
the FG stars due to the gas expulsion induced by SNe Ia explosions 
in the early phase of the GC evolution.
This must be the case for both groups of stars in M22.
We estimate that the fractions of the SG for both groups in M22 are very
similar, about 40-50\%, based on the [O/Fe] versus [Na/Fe] diagram
of \citet{marino11}, 
using the definition for the primordial component by \citet{carretta09}.
This means that the FGs in both groups are loosely bounded
to the same extent with respect to each SG, 
so that the similar fractions of the FG stars in both groups 
must have been lost during the phase (10) above, 
which appears to be difficult to explain.\footnote{The loss of the FGSP 
can occur during the phases (4-5), but this cannot explain the similar 
fractions of the FG in both groups and the number ratio between the two groups.}

The second problem is the number ratio between the two groups.
Given the current theoretical framework of the normal GC formation 
which requires extensive mass loss of the previous generations of stars, 
it may be expected that the fraction of the $s$-rich (or the \cas) stars 
is supposed to be larger than that of the $s$-poor (\cas) stars, 
in sharp contrast to the result presented here,  \nrgb\  $\approx$ 70:30.
Initially, if M22 were much more massive and 
located in an isolated place, such as NGC~2419, 
it could be able to retain the $s$-poor (or \caw) stars 
expelled during the phase (10) above \citep[e.g.][]{caloi11}. 
In fact, the recent observations showed that 
the metal-poor population is the major component of NGC~2419
\citep{beccari13,ywlee13}.
If so, isn't it natural to expect that the FGSP and the FGSR are 
the major components for both groups?

The third problem is the almost flat number ratio against the radial distance,
with a hint of the  slightly more centrally concentrated structure
of the \caw\ group in M22.
One of the general characteristics of normal GCs is the central concentration
of the SG population \citep[e.g.][]{lardo11}, 
including NGC~2419 \citep{beccari13}.
As mentioned earlier, the $N$-body numerical simulations by \citet{vesperini13}
suggested that the time required to have a flat number ratio 
between the FG and the SG of normal GCs is likely much larger than 
a Hubble time. 
If their result is applicable to the two groups of stars in M22, 
the flat number ratio against the radial distance is not likely
due to the dynamical evolution of the cluster with multiple stellar populations
but it must have been set initially \citep[e.g.][]{larsen15}.

The last problem is the difference in the velocity dispersions.
As discussed by \citet{bekki10}, the SG formed gaseous ejecta of the FG
in the central region of the cluster can have larger rotational amplitudes 
but smaller velocity dispersions than the FG.
He suggested that the dynamically cold nature of the SG is largely  due 
to gaseous dissipation during the gas accretion onto the cluster center.
Thus the lower velocity dispersion of the SG
is the key signature that can be imprinted.
If the \cas\ group were formed out of gaseous material accreted onto the center
within the self-enrichment scenario,
the lower velocity dispersion is expected.
However this is not consistent with our result,
showing that the velocity dispersion of the \cas\ group is slightly larger than
that of the \caw\ group.
Therefore, the difference in the velocity dispersions between the two groups
does not appear to be well understood  within the current frame of 
the dynamical evolution of the GCs with multiple stellar populations.

It is interesting to note that the heavy elemental abundance pattern
between the two groups of M22 can be reproduced by
the merger of two mono-metallic normal GCs, 
M5 and M4 \citep{roederer11}.
In addition, we showed that the photometric characteristics of M22 
can also be successfully reproduced by the two mono-metallic normal GCs, 
M55 and NGC~6752. 
As mentioned earlier, the observed CN and CH distributions in M22 RGB stars
can be simply explained with the merger scenario.
The differences in the spatial distributions and the kinematic properties
between the two groups in M22 can not be well understood
in the context of the self-enrichment hypothesis
and they appear to favor the merger scenario.
Also, the rather high flattened structure and rotation of M22 can be induced by 
the merging of a GC pair.
We proposed that the plausible explanation of the formation of M22
would be that M22 formed via a merger of two GCs, preferentially, 
in a dwarf galaxy environment where the GC merger events are much 
more probable \citep[e.g.][]{thurl02,bekkiyong12}, 
and then accreted later to the Milky Way
\citep[see also][]{carretta10,marino11}.

\acknowledgements
J.-W. L. acknowledges financial support from the Center for Galaxy Evolution Research
through the National Research Foundation of Korea (grant no. 2010-0027910).
The author thank Dr.\ Kiss for providing AAOmega data,
Prof.\ Pryor for a fortran code that calculates velocity dispersion,
Dr.\ Joo for providing model isochrones and the anonymous referee 
for constructive comments.
Finally, special thanks should go to Prof. Bruce Carney, who first
invented the $Ca$ filter decades ago and provided invaluable guidance to this work.

\clearpage

\begin{deluxetable}{ccrrrccl}
\tablecaption{Journal of observations.\label{tab:obslog}}
\tablenum{1}
\tablewidth{0pc}
\tablehead{
\multicolumn{1}{c}{date} &
\multicolumn{1}{c}{} &
\multicolumn{3}{c}{Int. Exp. Time (s)} &
\multicolumn{1}{c}{} &
\multicolumn{1}{c}{Tel.} &
\multicolumn{1}{c}{Weather}\\
\cline{3-5}
\colhead{(yy/mm/dd)} & \colhead{} & \multicolumn{1}{c}{$y$} & 
\multicolumn{1}{c}{$b$} & \multicolumn{1}{c}{$Ca$} & 
\colhead{} & \colhead{(CTIO)} & \colhead{} }
\startdata
11/03/27 && 180 &   310 &  1,500 && 1.0m & Photometric \\
11/03/28 && 340 &   580 &  2,400 && 1.0m & Non-photometric \\
11/03/29 && 340 &   580 &  2,400 && 1.0m & Photometric \\
11/03/30 && 340 &   580 &  1,800 && 1.0m & Non-photometric \\
11/03/31 && 160 &   540 &  2,400 && 1.0m & Photometric \\
11/04/01 && 600 & 1,350 &  4,200 && 1.0m & Non-photometric \\
11/04/02 && 275 &   600 &  2,100 && 1.0m & Photometric \\
11/04/03 && 400 &   900 &  3,000 && 1.0m & Photometric \\
11/04/07 && 200 &   450 &  1,500 && 1.0m & Photometric \\
11/08/19 && 100 &   200 &  1,200 && 1.0m & Photometric \\
11/08/20 && 200 &   400 &  1,500 && 1.0m & Non-photometric \\
11/08/21 && 350 &   600 &  2,400 && 1.0m & Photometric \\
11/08/22 && 300 &   500 &  2,400 && 1.0m & Photometric \\
11/08/31 && 310 &   240 &    900 && 1.0m & Non-photometric \\
11/09/04 && 100 &   200 &    900 && 1.0m & Photometric \\
12/04/29 &&  60 &   180 &  1,200 && 0.9m & Non-photometric \\
12/07/12 && 250 &   330 &  1,200 && 0.9m & Non-photometric \\
12/07/14 && 200 &   600 &  1,200 && 0.9m & Non-photometric \\
12/07/16 && 100 &   300 &  1,500 && 0.9m & Non-photometric \\
12/07/18 && 200 &   500 &  1,500 && 0.9m & Non-photometric \\
12/07/19 && 100 &   250 &  1,500 && 0.9m & Non-photometric \\
12/07/21 && 200 &   500 &  1,500 && 0.9m & Non-photometric \\
12/07/22 &&  60 &   180 &  2,400 && 0.9m & Non-photometric \\
12/07/23 &&  50 &   150 &  1,200 && 0.9m & Photometric \\
12/07/24 && 130 &   390 &  1,200 && 0.9m & Photometric \\
12/10/10 && 140 &   420 &  3,000 && 1.0m & Photometric \\
12/10/11 && 120 &   360 &  2,400 && 1.0m & Photometric \\
13/04/07 &&  80 &   200 &  1,200 && 1.0m & Photometric \\
13/04/08 && 185 &   475 &  1,500 && 1.0m & Non-photometric \\
13/04/11 && 180 &   500 &  1,500 && 1.0m & Photometric \\
13/04/12 && 115 &   310 &  1,500 && 1.0m & Photometric \\
13/04/13 && 100 &   250 &  1,200 && 1.0m & Photometric \\
13/05/14 && 100 &   250 &  1,500 && 1.0m & Non-photometric \\
13/07/27 &&  80 &   200 &  1,200 && 1.0m & Non-photometric \\
13/07/28 && 180 &   260 &  1,200 && 1.0m & Non-photometric \\
13/07/31 &&  95 &   240 &  1,200 && 1.0m & Non-photometric \\
13/08/02 && 180 &   500 &  1,500 && 1.0m & Non-photometric \\
14/05/23 && 150 &   450 &  1,800 && 1.0m & Non-photometric \\
14/05/25 && 300 &   900 &  3,900 && 1.0m & Non-photometric \\
14/05/26 && 300 &   900 &  3,900 && 1.0m & Photometric \\
14/05/27 && 240 &   600 &  3,600 && 1.0m & Photometric \\
14/05/30 && 320 &   800 &  3,600 && 1.0m & Photometric \\
14/05/31 && 615 & 1,545 &  8,700 && 1.0m & Photometric \\
14/06/01 && 120 &   360 &  1,500 && 1.0m & Photometric \\

\hline\hline
total    && 9,145 & 20,930 & 90,900 &&& \\
\enddata
\end{deluxetable}

\clearpage

\begin{deluxetable}{ccc}
\tablecaption{Comparisons of photometric results.\label{tab:comp}}
\tablenum{2}
\tablewidth{0pc}
\tablehead{
\multicolumn{1}{c}{} &
\multicolumn{1}{c}{Anthony-Twarog et al.} &
\multicolumn{1}{c}{Richter et al.} }
\startdata
No. of stars & 198 & 60 \\
$\Delta V$ & $+0.024 \pm 0.051$ & $-0.038 \pm 0.027$ \\
$\Delta (b-y)$ & $+0.034 \pm 0.050$ & $+0.003 \pm 0.037$ \\
$\Delta m1$ & $-0.028 \pm 0.063$ & $+0.021 \pm 0.037$ \\
$\Delta hk$ & $-0.029 \pm 0.079$ & \nodata \\
\enddata
\end{deluxetable}

\clearpage

\begin{deluxetable}{rccccc}
\tablecaption{Color-Magnitude Diagram data.\label{tab:cmd}}
\tablenum{3}
\tablewidth{0pc}
\tablehead{
\multicolumn{1}{c}{ID} &
\multicolumn{1}{c}{RA} &
\multicolumn{1}{c}{Dec}&
\multicolumn{1}{c}{$V$} &
\multicolumn{1}{c}{$b-y$} &
\multicolumn{1}{c}{$hk$}}
\startdata
     96 &  279.502014  & -24.007305 &  11.436 &   0.849  &  1.280 \\
    104 &  279.061249  & -24.014750 &  11.895 &   0.717  &  0.901 \\
    106 &  279.059052  & -24.033777 &  11.756 &   0.927  &  0.954 \\
    130 &  278.959198  & -24.261583 &  11.649 &   0.895  &  1.178 \\
    146 &  279.027954  & -24.347834 &  12.278 &   0.528  &  0.463 \\
    163 &  279.502747  & -24.352249 &  11.570 &   0.836  &  1.186 \\
    197 &  279.064087  & -23.805361 &  12.211 &   0.885  &  0.815 \\
    252 &  279.312927  & -23.539223 &  11.941 &   0.308  &  0.409 \\ 
    285 &  279.184082  & -23.980139 &  12.530 &   0.888  &  0.842 \\
    297 &  279.138306  & -23.940722 &  12.655 &   0.942  &  0.830 \\
    301 &  278.767181  & -24.275888 &  13.256 &   0.202  &  0.221 \\
\enddata
\tablecomments{
Table \ref{tab:cmd} is presented in its entirety in the electronic
edition of the Astrophysical Journal Supplement. A portion is shown here
for guidance regarding its form and content.}
\end{deluxetable}

\clearpage

\begin{deluxetable}{cccc}
\tablecaption{Number ratios.\label{tab:ratio}}
\tablenum{4}
\tablewidth{0pc}
\tablehead{
\multicolumn{1}{c}{} &
\multicolumn{1}{c}{Before Correction} &
\multicolumn{1}{c}{After Correction} &
\multicolumn{1}{c}{Error} 
}
\startdata
\nrgb & 69:31 & 70:30 & $\pm$ 3 \\
\nsgb & 69:31 & 62:38 & $\pm$ 6 \\
\nhb  & 80:20 & 88:12 & $\pm$ 10 \\
\enddata
\end{deluxetable}

\clearpage

\begin{deluxetable}{ccrrcrrcrr}
\tablecaption{Centers of M22.\label{tab:cnt}}
\tablenum{5}
\tablewidth{0pc}
\tablehead{
\multicolumn{1}{c}{} &
\multicolumn{1}{c}{} &
\multicolumn{2}{c}{Simple mean} &
\multicolumn{1}{c}{} &
\multicolumn{2}{c}{Half-sphere} &
\multicolumn{1}{c}{} &
\multicolumn{2}{c}{Pie-slice }\\
\cline{3-4}\cline{6-7}\cline{9-10}
\colhead{} & \colhead{} & 
\colhead{$\Delta$X(\arcsec) } & \colhead{$\Delta$Y(\arcsec)} &
\colhead{} &
\colhead{$\Delta$X(\arcsec) } & \colhead{$\Delta$Y(\arcsec)} &
\colhead{} &
\colhead{$\Delta$X(\arcsec) } & \colhead{$\Delta$Y(\arcsec)} }
\startdata
All    & & 2.9 & $-$0.1 &&    5.5 &    1.1 &&    7.1 &    1.0 \\
\caw\  & & 3.8 &    0.4 &&    7.5 &    1.6 &&    7.7 & $-$2.8 \\
\cas\  & & 0.8 &    5.6 && $-$1.4 & $-$1.4 && $-$4.9 &    6.5 \\
\enddata
\end{deluxetable}

\clearpage

\begin{deluxetable}{rrrrrrr}
\tablecaption{Ellipse fitting parameters.\label{tab:fit}}
\tablenum{6}
\tablewidth{0pc}
\tablehead{
\multicolumn{1}{c}{Pop.} &
\multicolumn{1}{c}{grid} &
\multicolumn{2}{c}{Ellipse center} &
\multicolumn{1}{c}{$\theta$} &
\multicolumn{1}{c}{$b/a$ } &
\multicolumn{1}{c}{$e$}\\
\cline{3-4}
\colhead{} & \colhead{} & \colhead{$\Delta$X(\arcsec) } &
\colhead{$\Delta$Y(\arcsec)} &\colhead{} &
\colhead{} & \colhead{}}
\startdata
\caw\   & 0.9 &  10.5 $\pm$  0.4 &   4.2 $\pm$  0.4 &   1.6 $\pm$  1.4 &  0.929 $\pm$  0.016 &  0.071 \\
        & 0.8 &  10.1 $\pm$  0.6 &   4.6 $\pm$  0.6 &   3.2 $\pm$  1.3 &  0.931 $\pm$  0.015 &  0.069 \\
        & 0.7 &   9.2 $\pm$  0.7 &   5.2 $\pm$  0.7 &   4.9 $\pm$  1.2 &  0.937 $\pm$  0.014 &  0.063 \\
        & 0.6 &   7.9 $\pm$  0.7 &   5.0 $\pm$  0.8 &   7.3 $\pm$  1.1 &  0.940 $\pm$  0.013 &  0.060 \\
        & 0.5 &   6.2 $\pm$  0.8 &   4.3 $\pm$  0.8 &  11.2 $\pm$  1.0 &  0.943 $\pm$  0.011 &  0.057 \\
        & 0.4 &   3.5 $\pm$  0.8 &   2.7 $\pm$  0.8 &  17.4 $\pm$  0.8 &  0.942 $\pm$  0.010 &  0.058 \\
        & 0.3 &   0.2 $\pm$  0.7 &  -0.6 $\pm$  0.8 &  24.3 $\pm$  0.7 &  0.934 $\pm$  0.008 &  0.066 \\
        & 0.2 &  -2.3 $\pm$  0.7 &  -5.2 $\pm$  0.7 &  31.4 $\pm$  0.5 &  0.912 $\pm$  0.006 &  0.088 \\
\hline\hline 
\cas\   & 0.9 &  -6.6 $\pm$  0.4 &   6.2 $\pm$  0.4 &  15.3 $\pm$  1.3 &  0.943 $\pm$  0.015 &  0.057 \\
        & 0.8 &  -3.3 $\pm$  0.6 &   5.4 $\pm$  0.6 &  18.9 $\pm$  1.2 &  0.943 $\pm$  0.014 &  0.057 \\
        & 0.7 &   0.1 $\pm$  0.7 &   3.6 $\pm$  0.7 &  24.1 $\pm$  1.1 &  0.944 $\pm$  0.013 &  0.056 \\
        & 0.6 &   3.8 $\pm$  0.8 &   0.8 $\pm$  0.7 &  29.2 $\pm$  1.0 &  0.941 $\pm$  0.011 &  0.059 \\
        & 0.5 &   6.8 $\pm$  0.8 &  -3.2 $\pm$  0.8 &  35.4 $\pm$  0.8 &  0.934 $\pm$  0.010 &  0.066 \\
        & 0.4 &   8.1 $\pm$  0.8 &  -8.0 $\pm$  0.8 &  42.7 $\pm$  0.7 &  0.912 $\pm$  0.008 &  0.088 \\
        & 0.3 &   6.6 $\pm$  0.7 & -13.2 $\pm$  0.8 &  41.8 $\pm$  0.6 &  0.872 $\pm$  0.006 &  0.128 \\
        & 0.2 &   2.4 $\pm$  0.7 & -13.5 $\pm$  0.7 &  37.6 $\pm$  0.4 &  0.837 $\pm$  0.005 &  0.163 \\
\enddata
\end{deluxetable}

\clearpage

\clearpage

\clearpage

\clearpage

\clearpage

\clearpage

\clearpage

\clearpage

\clearpage

\clearpage

\clearpage

\clearpage

\clearpage

\clearpage

\clearpage

\clearpage

\clearpage

\clearpage

\clearpage

\clearpage

\clearpage

\clearpage

\clearpage

\clearpage

\clearpage

\clearpage

\clearpage

\clearpage

\clearpage

\clearpage

\clearpage

\end{document}